\RequirePackage[2020-02-02]{latexrelease} 

\documentclass[%
 reprint,
 amsmath,amssymb,
prx,
]{revtex4-2}

\usepackage{graphicx}
\usepackage{dcolumn}
\usepackage{bm}
\usepackage{hyperref}
\usepackage{cleveref}
\usepackage{xcolor}

\newcommand{\ie}{\textit{i.e.}}
\newcommand{\cf}{\textit{c.f.}}
\newcommand{\eg}{\textit{e.g.}}
\newcommand{\etc}{\textit{etc.}}

\begin{document}

\title{Synthesis of parametrically-coupled networks}

\author{Ofer Naaman}
\email{ofernaaman@google.com}
\affiliation{Google Quantum AI, Santa Barbara, CA 93117 USA}

\author{Jos\'e Aumentado}
\email{jose.aumentado@boulder.nist.gov}
\affiliation{National Institute of Standards and Technology, Boulder, CO 80305 USA.}

\date{\today}

\begin{abstract}

We show that a common language can be used to unify the description of parametrically-coupled circuits\textemdash parametric amplifiers, frequency converters, and parametric nonreciprocal devices\textemdash with that of band-pass filter and impedance matching networks. This enables one to readily adapt network synthesis methods from microwave engineering in the design of parametrically-coupled devices having prescribed transfer characteristics, \textit{e.g.}, gain, bandwidth, return loss, and isolation. We review basic practical aspects of coupled mode theory and filter synthesis, and then show how to apply both, on an equal footing, to the design of multi-pole, broadband parametric and non-reciprocal networks. We supplement the discussion with a range of examples and reference designs.
\end{abstract}
\maketitle

\tableofcontents

\section{Introduction}
\label{sec:introduction}
The success of Josephson parametric amplifiers in enabling high-fidelity readout of superconducting qubits has led to a flurry of research into parametrically-coupled networks, including amplifiers, frequency converters, and parametric non-reciprocal networks~\cite{aumentado2020superconducting, vijay2011observation, ranzani2019circulators}. While much of the recent parametric amplifier activity is aimed at increasing amplifier saturation power and bandwidth~\cite{macklin2015near, mutus2014strong, roy2015broadband, naaman2019high, frattini2018optimizing}, a parallel research path has focused on generating nonreciprocal frequency conversion and amplification\textemdash an effort motivated by the need to improve qubit isolation from noise in the measurement chain, with the ultimate goal of minimizing the reliance on bulky ferrite circulators~\cite{lecocq2017nonreciprocal, sliwa2015reconfigurable, abdo2019active, lecocq2021efficient, peterson2017demonstration, chapman2017widely}. Recent work on parametric conversion has additionally expanded into electro-optomechanical systems~\cite{andrews2014bidirectional,mirhosseini2020superconducting, han2020cavity}, specifically facilitating the use of mechanical modes to mediate electrical to optical transduction.  

Despite significant progress, we make the following observations: a) there is currently no methodology in place to engineer the transfer characteristics (\eg, bandwidth, ripple, gain, return loss, \etc) of parametrically-coupled devices to arbitrary (physically realisable) specifications, and b) there is no unified language for including  parametrically-coupled devices on an equal footing with the electrical, mechanical, optical, or hybrid circuit in which they may be embedded. 

In this tutorial, we will show that the problem of designing parametrically-coupled circuits can be mapped onto a band-pass network synthesis problem. In doing so, we establish a common language to describe parametric interactions as circuit elements\textemdash this allows one to design and simulate general circuits that involve both parametric and passive coupling between multiple resonant modes, while also bringing band-pass network synthesis methods to bear on the design of more complicated parametrically-coupled devices. To be clear, this approach is applicable to any system that can be described by a system of linear coupled mode equations, including both resonant and parametric coupling on equal footing. 

This tutorial is organized as follows. In Section \ref{sec:matrix} we briefly summarize the coupling-graph approach and associated coupling-matrix formalism described in Refs.~\onlinecite{ranzani2015graph, lecocq2017nonreciprocal, peterson2020parametric_THESIS}, which provide a convenient way to visualize parametrically-coupled circuits and calculate their S-parameters. In Section \ref{sec:immittance} we discuss parametric couplers and show that they function as generalized admittance- or impedance-inverters\textemdash common in microwave-engineering circuit design as critical elements in the design of filter and impedance matching networks. Next, we briefly outline in Section \ref{sec:bandpass} some basic concepts and methods in band-pass network synthesis, and in Section~\ref{sec:all_together} we unify the microwave-engineering language of band-pass network synthesis with that of conventional coupled mode theory. We apply the concepts developed throughout the preceding sections to design a range of parametrically-coupled devices: a broadband parametric converter, a broadband parametric circulator, and a broadband non-degenerate Josephson parametric amplifier, all combining both passive (resonant) and parametric couplings to achieve a specific target response.

\section{Coupling matrices and graphs}
\label{sec:matrix}

\subsection{Coupling matrix formalism}\label{sec:formalism}
In this section we will sketch a formalism for the analysis and visualization of coupled-mode networks, define terminology, and establish notational conventions. The formalism that we will use is based on Ref. \onlinecite{ranzani2015graph}, which was later employed in Refs.~\onlinecite{lecocq2017nonreciprocal, peterson2020parametric_THESIS,lecocq2020microwave,lecocq2021efficient}. We will show how to describe an arbitrary network of resonantly- or parametrically-coupled modes by a coupled-modes equation of motion (EoM) matrix $\mathbf{M}$, and how the input/output boundary condition~\cite{vool2017introduction} can be used to derive a generalized multi-port scattering matrix (S-parameters). Although this approach was developed in the context of parametrically-coupled networks, it is equally adept at describing networks with constant (passive) coupling, providing a flexible description that can be used to combine physical matching networks with parametric frequency conversion and amplification. 

We consider a set of resonant modes having natural frequencies $\{\omega_k\}$. Each mode is characterized by a complex mode amplitude $a_k$, such that $a^*_ka_k=n_k$ is the number of photons in mode $k$. For example, in an electrical circuit we may define~\cite{gao2021practical}:
\begin{equation}\label{eqn:normalmode_defn}
    a_k=\frac{1}{\sqrt{2\hbar Z_k}}L_kI_k+i\sqrt{\frac{Z_k}{2\hbar}}C_kV_k
\end{equation}
where $L_k$ ($C_k$) is the inductance (capacitance) of the mode, $Z_k=\sqrt{L_k/C_k}$ is the mode impedance, $I_k$ is the mode current and $V_k$ is the mode voltage. Note that here $a_k$ is classical mode amplitude and the choice to scale to the square root photon (or phonon) number is a convenient correspondence to its formal annihilation operator counterpart. We note that the coupling description given here yields the same linear scattering matrix one would derive with a more rigorous quantum approach (\cf, Ref.\,\cite{peterson2020parametric_THESIS}).

The equations of motion of an arbitrary linear coupled mode system can then be cast in terms of these complex amplitudes $a_j(t)$, 
\begin{equation}\label{eq:mode_diff_eqn}
    \dot{a}_j=-i\left(\omega_j-i\frac{\gamma_j}{2}\right)a_j - i\sum_{k\neq j}c_{jk}\left(a_k+a^*_k\right)+\sqrt{\gamma^\mathrm{ext}_j}a^{in}_j,
\end{equation}
where we have included the drive term $a_j^\mathrm{in}$, and $j$ spans the set of resonances/resonators. The dissipation rate $\gamma_j=\gamma^\mathrm{int}_j+\gamma^\mathrm{ext}_j$ is the sum of the mode's internal dissipation, $\gamma^\mathrm{int}_j$, and external loading, $\gamma^\mathrm{ext}_j$, due to coupling out through the signal ports. In an electrical circuit, the dissipation rate is simply $\gamma=1/RC$, where $R$ is the equivalent shunt resistance seen by a parallel resonant circuit (in recent literature, this quantity is often denoted as $\kappa$, but we choose to maintain here the notation of \cite{ranzani2015graph}). The drive term corresponds to incident propagating waves with a rate $|a_j^\mathrm{in}|^2$ quanta (\eg, photons or phonons) per second. The coefficients $c_{jk}$  are coupling rates (often denoted as $g$, but we reserve this symbol to later signify filter prototype coefficients), and can be constant, modulated (time-varying), or some combination of the two. For instance, in an all passive electrical network these couplings might be realized by mutual inductances or coupling capacitors between resonators, while in a parametric amplifier or frequency converter, these elements can be varactors (tunable capacitors)~\cite{tucker1964circuits} or Josephson junctions~\cite{aumentado2020superconducting}. 

\subsubsection{The equations of motion matrix, $\mathbf{M}$} 
To determine the driven response of this coupled system, it is more convenient to Fourier transform the equations of motion (EoM), Eq.~(\ref{eq:mode_diff_eqn}). In a parametrically-coupled multi-mode system, one finds solutions that correspond to oscillations in each resonator, often at several different frequencies because of mixing terms generated by coupling modulation. If we relabel the Fourier frequency variable $\omega\rightarrow\omega_j^s$ to correspond to one of the input drives into the $j$-th resonator at frequency $\omega_j^s$ (the superscript $s$ is for \textit{signal} or \textit{stimulus}), the system can respond both at $\omega_j^s$ as well as mixing products generated by the coupling modulation. The family of resonator response amplitudes, at all possible mixing product frequencies, comprises the set of possible steady-state solutions to the driven equations of motion Eq.~(\ref{eq:mode_diff_eqn}). We can write the \textit{internal} mode amplitudes in the Fourier domain as the set $\vec{v}=\{a_j[\omega_{j}^s]\}$. In general, one may annotate the mode amplitudes according to the resonator wherein they reside, and an additional label to indicate a specific mixing product within that resonance; we will suppress these additional bookkeeping details here for clarity. The vector $\vec{v}$ can additionally include conjugate mode amplitudes. For each of the amplitudes in $\vec{v}$, we can assign a corresponding drive, and the set of all drive terms can be represented by  $\vec{v}_\mathrm{in}=\{a_j^\mathrm{in}[\omega_{j}^s]\}$. In this basis we can write Eq.~(\ref{eq:mode_diff_eqn}) in a compact matrix form,
\begin{equation}\label{eq:FT_EOM}
    -i\gamma_0\mathbf{M}\vec{v}=\mathbf{K}\vec{v}_\mathrm{in},
\end{equation}

The matrix $\mathbf{M}$ encapsulates the frequency-domain equations of motion, describing how energy is coupled between oscillating fields and their mixing products. The matrix $\mathbf{K}$ describes external dissipation for all modes in the mode basis, $\mathbf{K}\equiv \mathrm{diag}(\{\sqrt{\gamma_{j}^\mathrm{ext}}\})$. The prefactor $\gamma_0$ is an overall normalization\textemdash a characteristic rate in the system\textemdash whose form is chosen to suit different physical problems as we discuss below.

It is important to note that, in general, the complete mode basis can be quite large, yet many of these mode amplitudes are far off-resonant, and do not contribute to the steady state dynamics. It is therefore customary to reduce the mode basis $\vec{v}$ by performing rotating wave approximations (RWA), eliminating modes whose frequencies are fast with respect to their corresponding host resonance natural frequency, \ie, if $|\omega_j^s - \omega_j|\gg \gamma_j$.

The structure of $\mathbf{M}$ has a very simple general form,
\begin{equation}\label{eq:coupling_matrix}
    \mathbf{M}=
    \begin{bmatrix}
        \Delta_1 & \beta_{12} & \cdots & \beta_{1N} & & \beta_{11^*} & \cdots & \beta_{1N^*}\\
        \beta_{21} & \Delta_2 & \cdots & \beta_{2N} & & \beta_{21^*} & \cdots & \beta_{2N^*}\\
        \vdots &              & \ddots & \vdots & & \vdots & \ddots & \vdots \\
        \beta_{N1} & \cdots &       & \Delta_N  & & \beta_{N1^*} & \cdots & \beta_{NN^*} \\
        & & & & & & & & \\
        \beta_{1^*1} &  \cdots & & \beta_{1^*N} & & -\Delta^*_1 & \cdots & \beta_{1^*N^*} \\
        \vdots & &  & \vdots & & \vdots & \ddots & \vdots \\
        \beta_{N^*1} & \cdots & & \beta_{N^*N} & & \beta_{N^*1^*} & \cdots & -\Delta^*_N
    \end{bmatrix},
\end{equation}
where we define the diagonal detuning terms
\begin{equation}\label{eq:Delta_def}
    \Delta_k \equiv \frac{1}{\gamma_0}\left(\omega^s_k-\omega_k+i\frac{\gamma_k}{2}\right)
\end{equation}
and the off-diagonal coupling terms
\begin{equation}\label{eq:beta_def}
    \beta_{jk} = \frac{c_{jk}}{2\gamma_0}.
\end{equation}
The normalized coupling rates $|\beta_{jk}|^2$ are related to the cooperativity parameter in cavity QED \cite{kimble1998strong}, circuit QED \cite{clerk2020hybrid}, or optomechanics \cite{aspelmeyer2014cavity}.

There are four main blocks in $\mathbf{M}$ in Eq.~(\ref{eq:coupling_matrix}): the block-diagonal represents passive or frequency-conversion (difference-frequency) coupling within the mode manifold (upper left $N\times N$ block) and anti-conjugate-mode manifold (lower right block); the off-diagonal $N\times N$ blocks represent parametric amplification coupling (sum-frequency) between the two manifolds. Often, the connectivity of the network will result in a total matrix $\mathbf{M}$ that can be reduced to separate redundant matrices with no coupling between them\textemdash we will henceforth work with the minimal non-redundant subset of modes and sub-matrix that describes the coupled network of interest. As a concrete minimal example, one could describe the driven response of a single resonator with a two-mode basis, $\vec{v} = (a[\omega_A^s], a^*[-\omega_A^s])^T$, and a 2$\times$2 $\mathbf{M}$ matrix. In the absence of parametric pumping, the two modes are uncoupled ($\beta_{AA^*} = \beta_{A^*A} = 0$) and it is sufficient to describe the system with a single equation of motion, $-i\gamma_A\Delta_A a[\omega_A^s] = \sqrt{\gamma_A^\mathrm{ext}}a^\mathrm{in}[\omega_A^s]$, since the other equation of motion is just the conjugate of this equation. If, on the other hand, the resonator is parametrically pumped with $\omega_P\simeq 2\omega_A$, the modes $a[\omega_A^s]$ and $a[\omega_A^s - \omega_P]$ are coupled and the full 2$\times$2 $\mathbf{M}$ matrix is required. We refer the reader to Ref.~\onlinecite{ranzani2015graph} for a more detailed treatment.

The coupling coefficients $\beta_{jk}$ follow a particular pattern that is determined by the coupling mechanism \cite{ranzani2015graph}. If the coupling is passive, they are symmetric and real, $\beta_{jk} = \beta_{kj} \in \mathbb{R}$. If the coupling corresponds to frequency conversion, that is, the physical coupling element is modulated at the difference frequency between modes $j$ and $k$, the coupling coefficients are related by conjugation, $\beta_{jk} = \beta_{kj}^*$. Lastly, if the coupling is modulated at the sum frequency, generating amplification, the coefficients are related by anti-conjugation, $\beta_{jk} = -\beta_{kj}^*$ . 

The simple structure of $\mathbf{M}$ allows one, with experience, to identify the relevant mode basis quickly, and assign the appropriate conjugation and signs to the coupling, building a picture of the dynamics within a system. 

\subsubsection{Generalized scattering} 
The input and output drive amplitudes, $\vec{v}_\mathrm{in}$ and $\vec{v}_\mathrm{out}$, obey a boundary condition at each port, which connects them to the internal mode amplitudes \cite{yurke2004input,vool2017introduction}, $\vec{v}_\mathrm{in} + \vec{v}_\mathrm{out} = \mathbf{K}\vec{v}$. From this, together with Eq.~(\ref{eq:FT_EOM}), one can compute the full scattering matrix,
\begin{equation}\label{eq:S_matrix_def}
    \mathbf{S}=i\frac{1}{\gamma_0}\mathbf{K}\mathbf{M}^{-1}\mathbf{K}-\mathbb{I}.
\end{equation}
It is useful to write this in component form,
\begin{equation}\label{eq:s_params_components}
    S_{jk}=\frac{a^\mathrm{out}_j}{a^\mathrm{in}_k}=i\frac{\sqrt{\gamma_j^\mathrm{ext}\gamma_k^\mathrm{ext}}}{\gamma_0}\left[\mathbf{M}^{-1}\right]_{jk}-\delta_{jk}.
\end{equation}

When all coupling in the network is passive, there is no frequency translation and Eq.~(\ref{eq:s_params_components}) describes scattering between physical input/output ports. It is then essentially identical to its electrical engineering counterpart\textemdash the usual S-parameters of the network. However, the presence of time-varying, modulated coupling generates scattering between frequencies, both within single resonators as well as between different physical resonators and Eq.~(\ref{eq:s_params_components}) is broadly defined to include scattering between mode amplitudes rotating at different frequencies. One other notable difference between our definition above and the usual S-parameters is that here, scattering is defined as the ratio of output to input mode amplitudes that we have chosen to normalize to energy quanta, $\hbar\omega_k^s$, see Eq.~(\ref{eqn:normalmode_defn}). By contrast, scattering parameters in electrical engineering are commonly defined by the ratio of the equivalent voltage amplitudes, $S_{jk} \equiv V_j^\mathrm{out}/V_k^\mathrm{in}$ \cite{pozar2009microwave}. A consequence of this difference in definition is that a process having unity gain in our convention, will be associated with a gain factor in the common electrical engineering definition of the scattering parameters, stemming from the Manley-Rowe relations \cite{manley1956some}. An example of this difference is discussed in Section \ref{sec:all_together_converter}.

\subsubsection{Simplifications for network synthesis}\label{sec:simplifications} 
In the above, we have cast the problem of mode coupling in a linear system as generally as possible, following the prescription in \cite{ranzani2015graph}. However, for the purpose of the present discussion, we can perform some simplifications to this model to focus on the specific problem of network synthesis and amplifier/converter design. Namely, we will assume that only a subset of the modes are coupled to signal ports, and further consider the ideal case where there is no internal loss. 

As an example, a simple 2-port microwave $N$-pole cavity filter will have $N$ cavities, each with extremely low internal loss, but since it is connected to input/output ports at the ends, $\gamma_1$ and $\gamma_N$ present the only available dissipation channels. In other words, the total dissipation rates of the `internal' modes can be set to zero, $\{\gamma_2,\ldots,\gamma_{N-1}\} = 0$, and the subset of scattering elements, 
\begin{equation}
\mathbf{S} =
\begin{bmatrix}
    S_{11} & S_{1N}\\
    S_{N1} & S_{NN}
\end{bmatrix}
\end{equation}
is sufficient to describe the filter response. To this end, we will mark a subset of the modes as \textit{ports}, $\mathcal{P} \subset  \{ v_j \}$, through which energy can be injected and extracted, and set all mode dissipations and input drives to zero, $\gamma_k = 0, v_k^\mathrm{in} = 0$ for everything that is not a port. Likewise, since the total dissipation of a resonator $k$ is the sum of its internal and external losses, $\gamma_k = \gamma_k^\mathrm{int} + \gamma_k^\mathrm{ext}$, the `no internal loss' assumption means that $\gamma_k = \gamma_k^\mathrm{ext}$ and we can drop the `$\mathrm{int}/\mathrm{ext}$' superscripts in the remainder of this tutorial. In this same spirit, we choose a form for the normalization rate that is the geometric mean of the dissipations of all connected ports,
\begin{equation}\label{eq:gamma0_def}
    \gamma_0\equiv\sqrt[\leftroot{-3}\uproot{5}N_{\mathcal{P}}]{\prod_{k\in \mathcal{P}} \gamma_k},
\end{equation}
where $N_\mathcal{P}$ is the number of ports. For the 2-port example above, $\gamma_0 = \sqrt{\gamma_1 \gamma_N}$. Modes that are marked as ports can additionally be characterized with a finite quality factor
\begin{equation}\label{eq:port_gamma_q}
    Q_j=\frac{\omega_j}{\gamma_j}.
\end{equation}

Since in the rest of this tutorial we will focus on the problem of network synthesis primarily through the lens of electrical circuit design, it will be convenient to connect the notion of input admittance to the general picture presented thus far. These are connected by inverting the usual formula for the reflection coefficient in terms of input admittance, $Y_\mathrm{in} = Y_0 (1 - S_{kk})/(1 + S_{kk})$. One can then show that the admittance, looking into resonator $k$ at mode frequency $\omega_k^s$ is 
\begin{equation}\label{eq:port_admittance}
Y_\mathrm{in} = -Y_0 \left[2i \frac{\gamma_0}{\gamma_k}\frac{1}{[\mathbf{M}^{-1}]_{kk}} + 1\right],
\end{equation}
where $Y_0$ is the reference admittance of the environment connected to port $k$, usually taken to be $(50\;\Omega)^{-1}$.

\subsubsection{Internal dissipative losses}\label{sec:losses}
We will proceed for the remainder of this tutorial with the ``zero internal loss" assumption. This assumption is reasonable for implementations based on superconducting circuits and microwave cavities, however, in optomechanical circuits internal losses play a more central role \cite{aspelmeyer2014cavity}. Furthermore, any accurate analysis of quantum noise in parametric networks must include all internal dissipation sources.

The topic of coupled-resonator network design techniques in the presence of internal dissipative losses has received considerable attention in the microwave engineering literature, and we point the reader to, \eg, Refs.~\cite{cohn1959dissipation, dishal1949design, hunter2005passive}. As we will see throughout this tutorial, network synthesis techniques from microwave engineering can be readily adopted in parametrically-coupled circuits. 

Noise properties of coupled-mode systems can be calculated using the EoM matrix formalism described in this section by including all internal losses as effective ports \cite{peterson2020parametric_THESIS, andersson2020squeezing}. The simplifications made in Sec.~\ref{sec:simplifications} are useful in network synthesis, and the resulting ideal networks can then be re-annotated to include finite temperature dissipation sources for the purpose of noise calculations. This general noise analysis approach is beyond the scope of this tutorial.

\subsection{Coupled-mode graph representation}\label{sec:graphs}
The formalism described in Section~\ref{sec:formalism} provides a method for calculating S-parameters for driven modes within an arbitrary system of coupled resonators, where ports are generalized to both physical and frequency-space and the couplings can be complex, carrying the phases of the pumps driving them. This allows one to include both parametric coupling processes (amplification and frequency conversion) and passive coupling structures in a straightforward way, but requires careful bookkeeping. In multiple-resonator systems, the addition of even a few parametric couplings can generate a surprising increase in the mode basis size (see, \eg, Refs.~\cite{lecocq2020microwave, lecocq2015quantum, peterson2017demonstration}). While this accounting can, with care, be automated in a numerical calculation, one can also utilize a graph representation \cite{ranzani2015graph, lecocq2017nonreciprocal, peterson2020parametric_THESIS} that facilitates bookkeeping and serves as an aid in both analysis and design/synthesis. 

\begin{figure*}[htb]
    \centering
    \includegraphics[width=16cm]{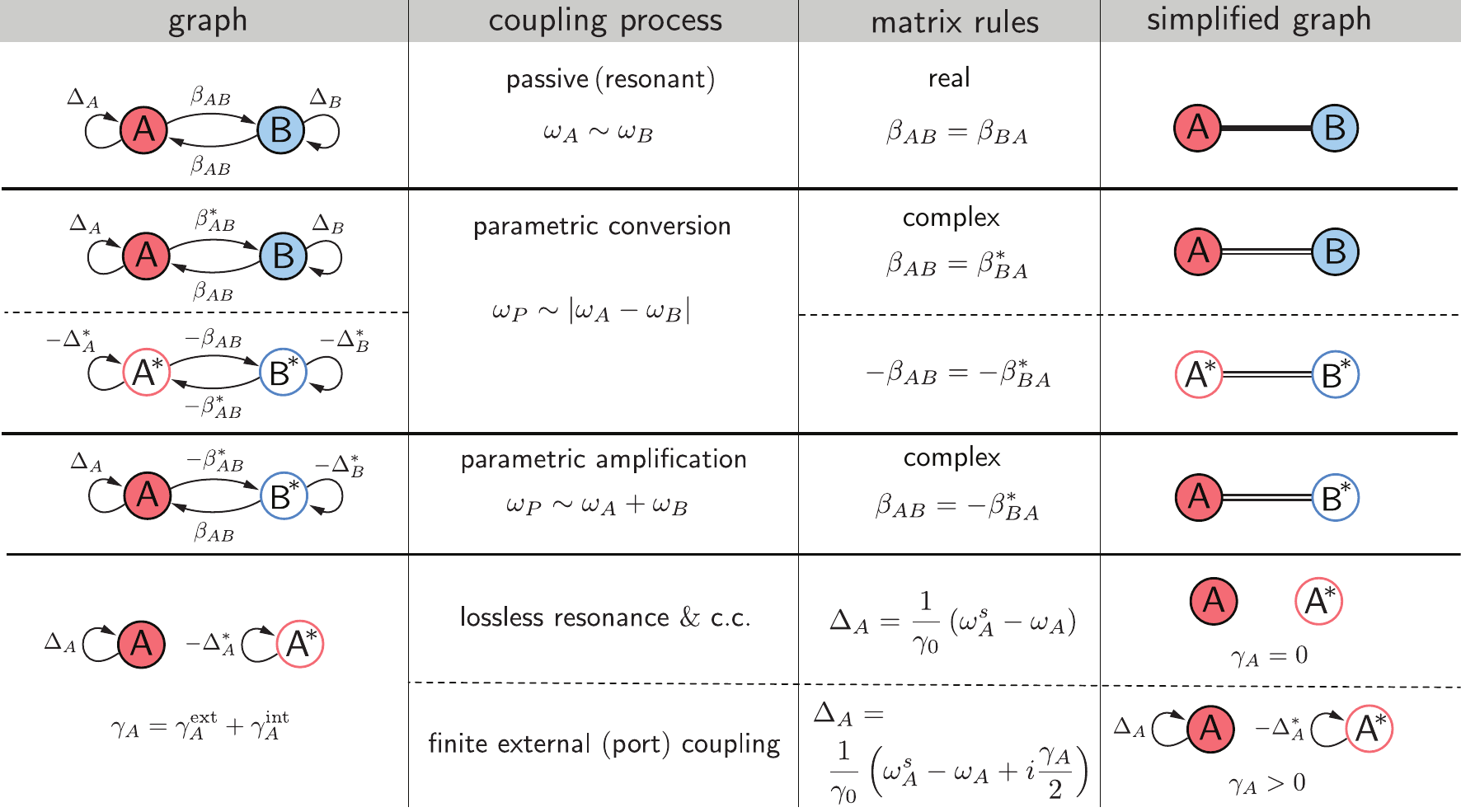}
    \caption{Rules for writing the EoM matrix elements of a system from its graph representation. Left column: graph primitives using the notation of \cite{ranzani2015graph}, second column: corresponding coupling process, third column: matrix elements and their conjugation relations, right column: abbreviated notation used in this tutorial.\label{fig:matrix_rules}}
\end{figure*}

Since the coupled-modes EoM matrix $\mathbf{M}$ is square, it can be represented as a directed graph \cite{brualdi1991combinatorial,greenman1976graphs}, where each of the modes can be represented as a node. In this picture, we represent the diagonal elements, $\Delta_k$, as self-loops, connecting a node to itself, while all off-diagonal couplings $\beta_{jk}$ are represented by directed edges (arrows) that connect one node to another. Above, we outlined rules for relating $\beta_{jk}$ to $\beta_{kj}$ based on the nature of the coupling. With these rules, one can derive the EoM of a linear coupled mode system quickly, simply by drawing a graph and identifying all of the coupling signs and conjugations correctly by inspection. 

Graph primitives for 1- and 2-mode processes are given in the left-hand column in  Figure\,\ref{fig:matrix_rules} along with the rules for writing their corresponding matrix elements. After identifying the graph representation, one can immediately obtain the EoM matrix $\mathbf{M}$, and from it scattering parameters via \Cref{eq:s_params_components}. From these primitives, one can construct more complex networks such as multimode parametric amplifiers \cite{ranzani2015graph,sliwa2015reconfigurable,lecocq2017nonreciprocal,lecocq2020microwave,lecocq2021efficient} and hybrid optomechanical circuits \cite{peterson2017demonstration}, which rely on the topology of the network to produce directionality and phase-sensitive amplification.

Since there are only three possible mechanisms for coupling\textemdash resonance (passive coupling), parametric frequency conversion (modulated at the difference-frequency), and parametric amplification (modulated at the sum-frequency)\textemdash with well-defined relationships between forward and backward couplings, we can simplify the graph representation to show all directed edge pairs ($\beta_{jk}$ and $\beta_{kj}$) as single (resonant) or double lines (parametric). Parametric frequency conversion and amplification are further distinguished by whether they connect co-rotating or counter-rotating (conjugated) modes, indicated by mode labels and face color for each node. Figure \ref{fig:matrix_rules} summarizes the translation between the original directed graph notation of~\cite{ranzani2015graph} and the abbreviated notation used in this tutorial, along with the corresponding matrix coupling element conjugation rules. 

In the simplified graph representation, we only draw self-loops for modes that are connected to ports and therefore have non-vanishing dissipation.

\subsection{Examples}\label{sec:graph_examples}
Practical use of the tools outlined above is best illustrated through simple canonical examples of parametrically coupled networks. In the examples below, we will highlight a design flow that uses the network graph to extract its EoM matrix, calculate its S-parameters, and derive design parameters based on performance requirements.

\subsubsection{Parametric frequency converter}\label{sec:examples_FC}
\begin{figure}[th]
    \centering
    \includegraphics[width=\columnwidth] {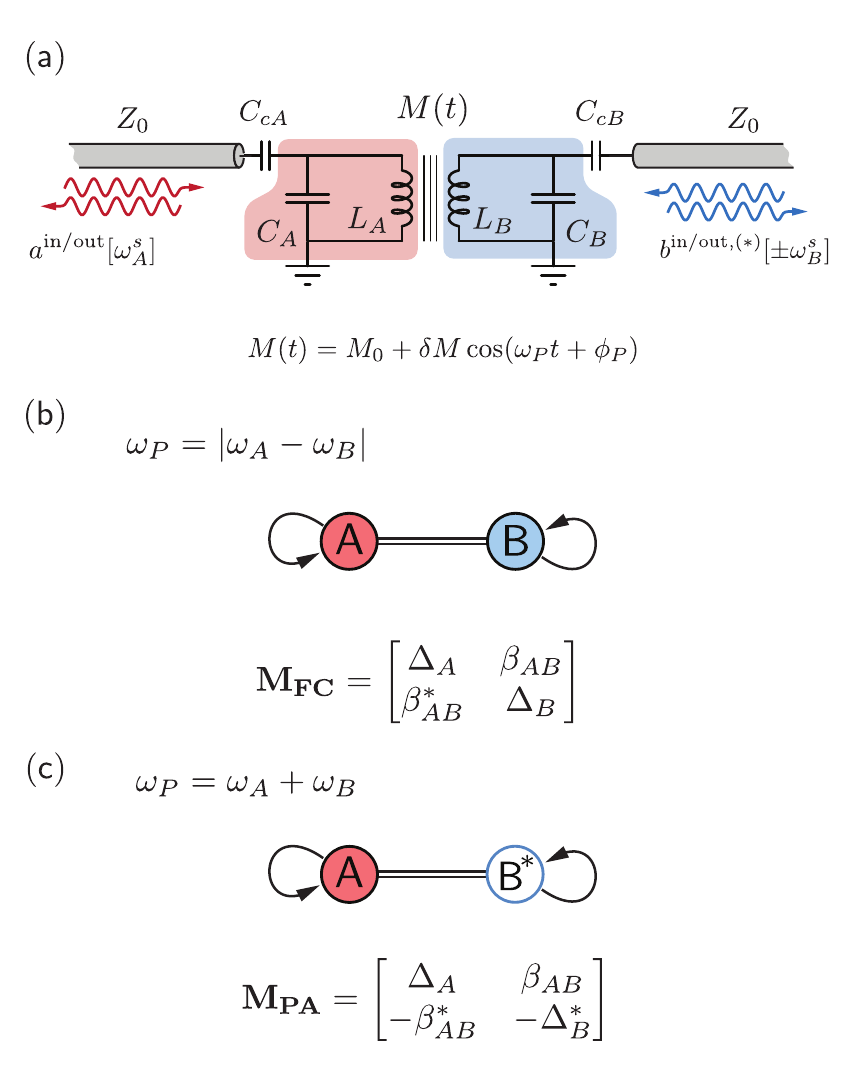}
    \caption{(a) Schematic of a circuit containing two resonant modes coupled through a time-varying mutual inductance pumped at a frequency $\omega_P$. (b) Coupled-modes graph of a 2-mode parametric converter and the corresponding EoM matrix, written in the basis $\vec{v}=\left(a\left[\omega^s_A\right],\,b\left[\omega^s_B\right]\right)^T$. (c) Coupled-modes graph of a 2-mode parametric amplifier and the corresponding EoM matrix, written in the basis $\vec{v} = (a[\omega_A^s],\,b^*[-\omega_B^s])^T$.\label{fig:FCPA_graph_matrix}}
\end{figure}

The 2-mode parametric frequency converter is the simplest non-trivial parametric device. As a concrete example, one may construct a circuit comprised of two $LC$ resonators, coupled by a modulated mutual inductance (Figure \ref{fig:FCPA_graph_matrix}(a)). The resonators have different natural frequencies, $\omega_A$ and $\omega_B$, and are coupled to external transmission lines with characteristic impedance $Z_0$ through capacitors $C_{cA}$ and $C_{cB}$. Propagating electromagnetic waves ($a^\mathrm{in/out}[\omega_A^s]$ and $b^\mathrm{in/out}[\omega_B^s]$) can propagate in and out of A/B resonators with the rates:

\begin{eqnarray}
    \gamma_A &=& \frac{\omega^2_A C_{cA}^2 Z_0}{C_A}\\
    \gamma_B &=& \frac{\omega^2_B C_{cB}^2 Z_0}{C_B}.
\end{eqnarray}

If the natural frequencies differ by more than the bandwidths, $|\omega_A - \omega_B|\gg\sqrt{\gamma_A\gamma_B}$, energy exchange is greatly reduced, but can be recovered if the mutual inductance is pumped (sinusoidally modulated) at the difference frequency, $\omega_P\simeq|\omega_A - \omega_B|$, resulting in frequency conversion\textemdash power is transferred from $a^\mathrm{in}$ to $b^\mathrm{out}$, and from $b^\mathrm{in}$ to $a^\mathrm{out}$.

The graph representation of the 2-mode frequency converter is shown in Fig.~\ref{fig:FCPA_graph_matrix}(b). Using the rules in Fig.~\ref{fig:matrix_rules}, we can immediately write the EoM matrix $\mathbf{M}_\mathrm{FC}$ as shown in Fig.~\ref{fig:FCPA_graph_matrix}(b). Here, the EoM matrix is written in the driven mode basis $\vec{v} = (a[\omega_A^s],\,b[\omega_B^s])^T$. The detuning terms, including the respective port dissipation rates, are written according to Eq.~(\ref{eq:Delta_def}) as:
\begin{align}
    \Delta_A&=\frac{1}{\gamma_0}\left(\omega^s_A-\omega_A+i\frac{\gamma_A}{2} \right)\\
    \Delta_B&=\frac{1}{\gamma_0}\left(\omega^s_A+\omega_P-\omega_B+i\frac{\gamma_B}{2} \right),\label{eq:FC_det_term}
\end{align}
where we have already substituted  for $\omega^s_B$ via the pump frequency, $\omega^s_B = \omega^s_A+\omega_P$.

Observe that since the pump is tuned to the difference frequency, we also have that $\omega_P-\omega_B=-\omega_A$, so we could have instead written in Eq.~(\ref{eq:FC_det_term}) $\Delta_B=\frac{1}{\gamma_0}\left(\omega^s_A-\omega_A+i\frac{\gamma_B}{2}\right)$. In other words, the fact that the modes are at different frequencies drops out of the equations. We will use this insight repeatedly here, keeping in mind that deviation of the pump frequency from perfect tuning (while beyond the scope of the present discussion) can open additional useful design space in certain applications.

We will want to design the network to be impedance-matched to the 50$\;\Omega$ ports. Using Eq.~(\ref{eq:s_params_components}), we can write the reflection off of port A:
\begin{equation}
    S_{AA}=i\frac{\gamma_A}{\gamma_0}\frac{\Delta_B}{\Delta_A\Delta_B-|\beta_{AB}|^2}-1.
\end{equation}
At zero detuning, $\omega^s_A-\omega_A=0$, we have
\begin{equation}
    S_{AA}=\frac{1-4|\beta_{AB}|^2}{1+4|\beta_{AB}|^2},
\end{equation}
and requiring $S_{AA}=0$ for perfect matching results in $|\beta_{AB}|=0.5$. Further, from the transmission $S_{BA}$ we can calculate $\Delta\omega$, the 3-dB bandwidth of the network, and find that the bandwidth is optimized when $\gamma_A=\gamma_B$, where it is equal to $\Delta\omega=\sqrt{2}\gamma_0$.

\begin{figure}
    \centering
    \includegraphics[width=\columnwidth] {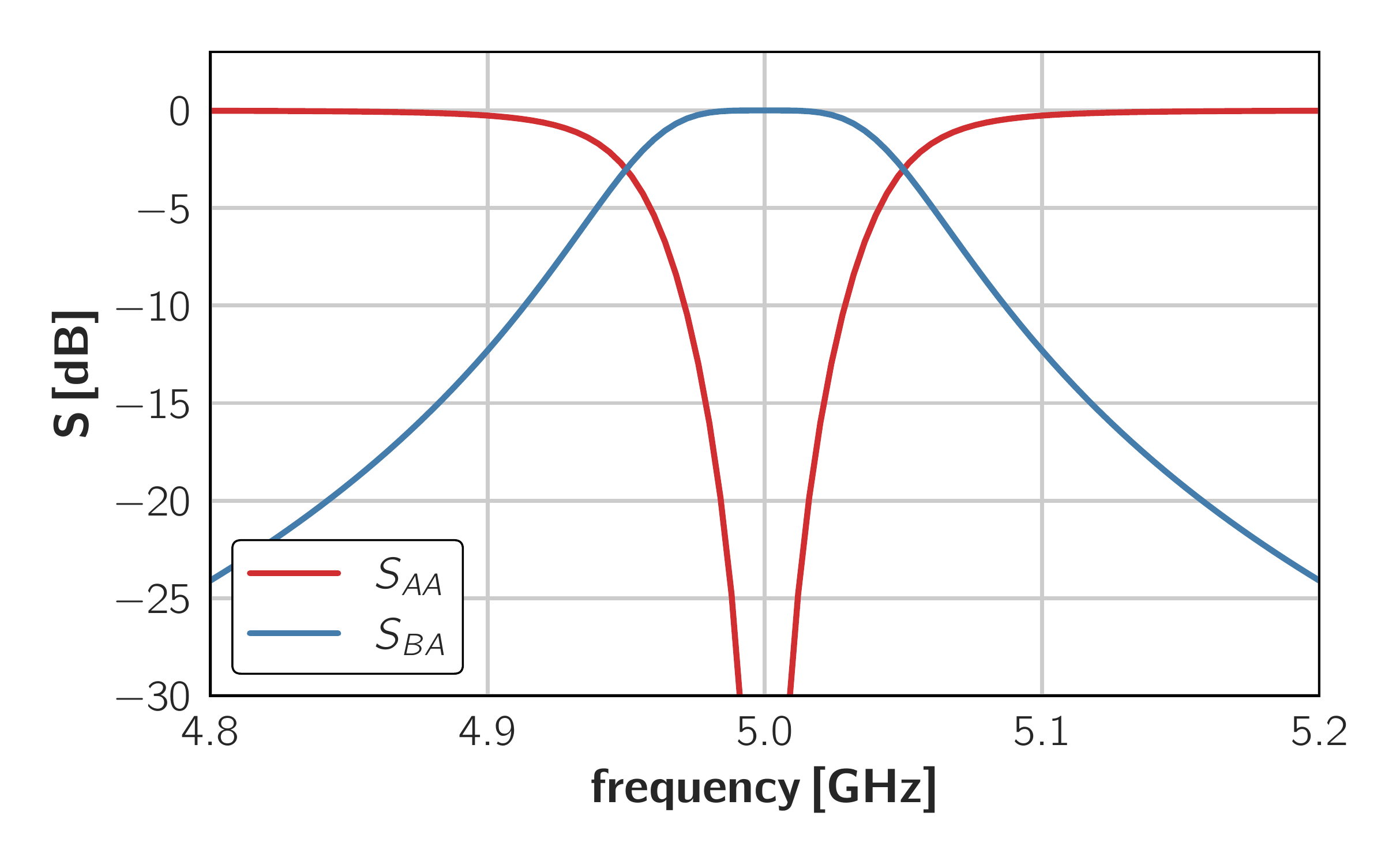}
    \caption{S-parameters of the 2-mode frequency converter calculated using Eq.~(\ref{eq:s_params_components}). The frequency axis is referenced to the signal port: $S_{AA}$ is the reflection off of the input port at the indicated frequency, $\omega^s_A/2\pi$; $S_{BA}$ is the conversion gain\textemdash output photon flux at port B at $\omega^s_A+\omega_P$ normalized to input flux at port A at frequency $\omega^s_A$.\label{fig:FC_graph_sparams}}
\end{figure}

Fig.~\ref{fig:FC_graph_sparams} shows S-parameters in dB vs.~signal frequency of a parametric frequency converter designed to meet the following requirements: $\omega_A/2\pi=5$~GHz, $\omega_B/2\pi=7$~GHz, and 3-dB bandwidth of $\Delta\omega/2\pi=100$~MHz. According to the discussion above, we can calculate the port coupling rate $\gamma_0/2\pi=70.7$~MHz (Eq.~\ref{eq:gamma0_def}), and parametric coupling rate $c_{AB}/2\pi=2\gamma_0\beta_{AB}/2\pi=70.7$~MHz (Eq.~\ref{eq:beta_def}). For each signal frequency in Fig.~\ref{fig:FC_graph_sparams}, we invert the matrix in Fig.~\ref{fig:FCPA_graph_matrix}(b) and calculate the S-parameters using Eq.~(\ref{eq:s_params_components}). Note that the inductor and capacitor values aren't specified here, and this approach is, instead, cast in terms of frequencies, dissipation, and coupling rates. Specific circuit parameter values can be designed to realize these while accommodating realistic design and fabrication constraints of a chosen technology.

In Section~\ref{sec:all_together} we will see that the converter we designed here exactly implements a 2-pole max-flat (Butterworth) response, as Fig.~\ref{fig:FC_graph_sparams} already hints. 

In Fig.~\ref{fig:FC_graph_sparams}, and all subsequent figures that show S-parameters in this tutorial , we will use the fact that when the pump frequency is exactly tuned to the difference- (conversion) or sum- (amplification) frequency, the problem has a single independent reference frequency variable\textemdash indicated on the frequency axis in these figures. Curves that represent S-parameters for frequency-translating processes, should be understood in relation to that reference frequency where either the input or the output frequency or both correspond to translating the reference frequency by the pump frequency for the corresponding process. Specifically, in Fig.~\ref{fig:FC_graph_sparams}, the curve labeled $S_{BA}$ represents the output photon flux at port B at $\omega^s_A+\omega_P$ normalized to input photon flux at port A at frequency $\omega^s_A$, where $\omega^s_A/2\pi$ also serves as the independent frequency variable indicated on the x-axis.

\subsubsection{Parametric amplifier}
The coupled-mode graph of a 2-mode parametric amplifier is shown in Fig.~\ref{fig:FCPA_graph_matrix}(c). The same graph can describe both a degenerate reflection amplifier such as the JPA~\cite{yurke1989observation,mutus2013design} when $\omega_A=\omega_B$ and the same physical resonator hosts both modes, and a non-degenerate amplifier such as the JPC~\cite{abdo2013nondegenerate} or FPJA~\cite{lecocq2017nonreciprocal}, where the modes are hosted in physically separate resonators or resonances similar to Fig.~\ref{fig:FCPA_graph_matrix}(a). In this case, the EoM matrix $\mathbf{M_{PA}}$ is written in the basis $\vec{v} = (a[\omega_A^s],\,b^*[-\omega_B^s])^T$. Observe that the mode $B$ is now conjugated as indicated by its open face shading and the anti-conjugation of corresponding detuning term $-\Delta^*_B$ in the EoM matrix. We have again used the rules in Fig.~\ref{fig:matrix_rules} to fill in the off-diagonal elements with the appropriate anti-conjugation of $\beta_{BA}=-\beta^*_{AB}$ for a parametric process driven by a sum-frequency pump $\omega_P=\omega_A+\omega_B$. The detuning terms are:
\begin{align}
    \Delta_A&=\frac{1}{\gamma_0}\left(\omega^s_A-\omega_A+i\frac{\gamma_A}{2} \right)\\
    -\Delta^*_B&=\frac{1}{\gamma_0}\left(\omega^s_A-\omega_P+\omega_B+i\frac{\gamma_B}{2} \right),
\end{align}
where we have replaced $\omega^s_B=-\omega^s_A+\omega_P$. Since it is also the case that $-\omega_P+\omega_B=-\omega_A$ we could have written instead $-\Delta^*_B=\frac{1}{\gamma_0}\left(\omega^s_A-\omega_A+i\frac{\gamma_B}{2}\right)$, demonstrating again that under resonant pump condition ($\omega_P$ perfectly tuned to the sum-frequency), the system has only a single independent frequency variable.

The reflection off of the signal mode $A$ using Eq.~(\ref{eq:s_params_components}) is 
\begin{equation}\label{eq:paramp_graph_sparams}
    S_{AA}=i\frac{\gamma_A}{\gamma_0}\frac{\Delta^*_B}{\Delta_A\Delta^*_B-|\beta_{AB}|^2}-1,
\end{equation}
and at the center frequency of the amplifier we get \begin{equation}\label{eq:paramp_graph_gain}
    S_{AA}=\sqrt{G}=\frac{1+4|\beta_{AB}|^2}{1-4|\beta_{AB}|^2},
\end{equation}
where $G$ is the signal power gain. From Eq.~(\ref{eq:paramp_graph_gain}) we can extract $\beta_{AB}$ needed to get the desired gain, for example, 20 dB gain will require $|\beta_{AB}|=0.452$.

For high gain and small detuning, Eq.~(\ref{eq:paramp_graph_sparams}) approximates a Lorentzian whose bandwidth is 
\begin{equation}\label{eq:paramp_graph_bandwidth}
    \Delta\omega=\frac{2}{\sqrt{G}}\frac{\gamma_A\gamma_B}{\gamma_A+\gamma_B},
\end{equation}
which is maximal when $\gamma_A=\gamma_B$, and inversely proportional to the amplitude gain.

The two driven modes of the parametric amplifier are often identified as \textit{signal} and \textit{idler}. This terminology is anchored in historical jargon\textemdash early nondegenerate parametric amplifiers terminated the idler tank circuit (\ie, the resonator that did not host the incident signal mode) \cite{louisell1960coupled,blackwell1961semiconductor}. The idler circuit provides a critical internal degree of freedom necessary for amplification, but might have been historically viewed as an ancillary mode.

\begin{figure}[th]
    \centering
    \includegraphics[width=\columnwidth] {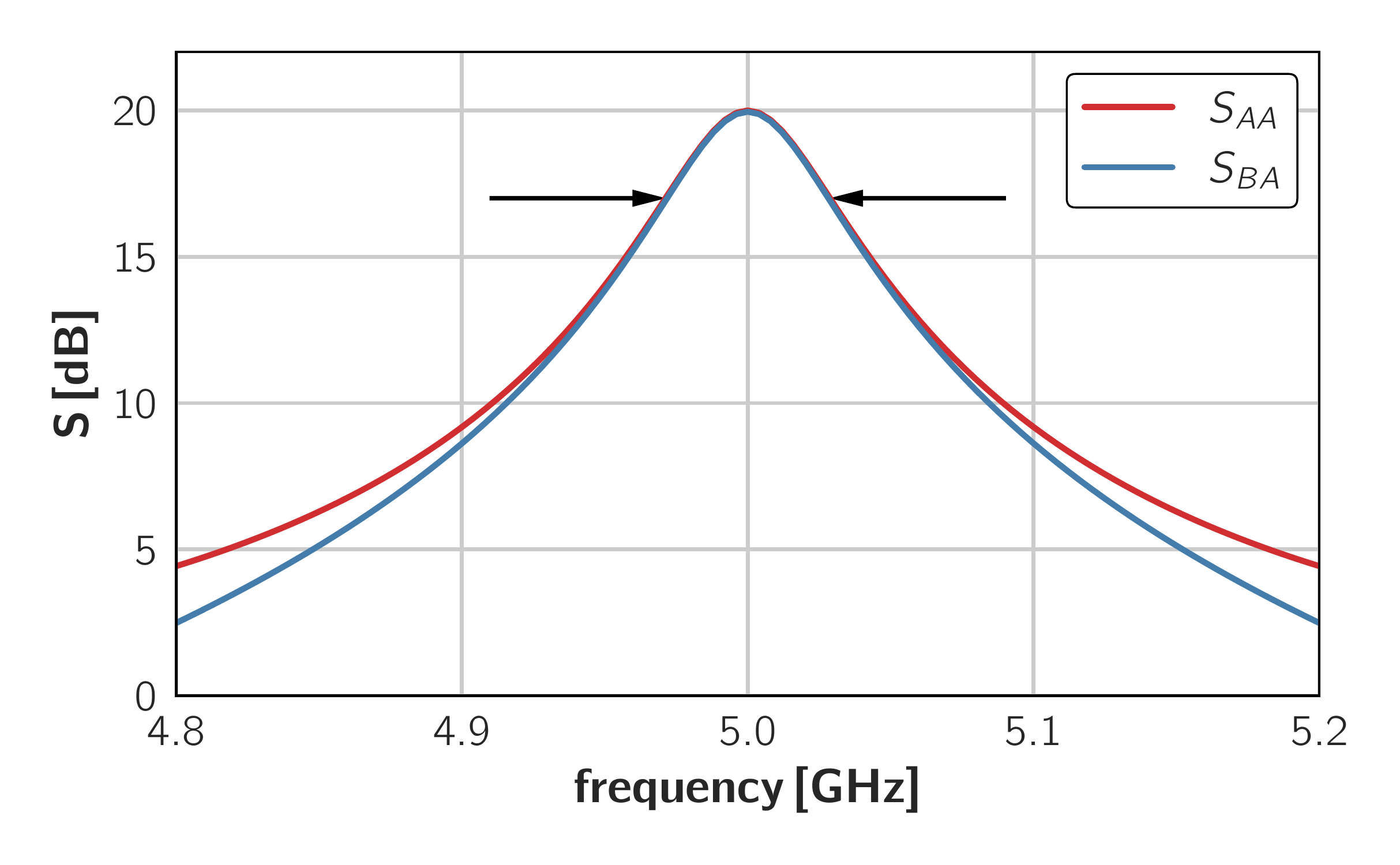}
    \caption{S-parameters of the 2-mode parametric amplifier calculated using the EoM matrix in Fig.~\ref{fig:FCPA_graph_matrix}(c) and Eq.~(\ref{eq:s_params_components}). The frequency axis is referenced to the signal port. $S_{AA}$ is the signal reflection gain and $S_{BA}$ is the idler gain. Arrows indicates the 3 dB bandwidth according to Eq.~(\ref{eq:paramp_graph_bandwidth}).\label{fig:PA_graph_sparams}}
\end{figure}

Fig.~\ref{fig:PA_graph_sparams} shows the signal ($S_{AA}$) and idler ($S_{BA}$) gain profiles of an amplifier pumped with the $\beta_{AB}$ we calculated above to give $G=20$~dB at $\omega_0/2\pi=5$~GHz, and with $\gamma_A/2\pi=\gamma_B/2\pi=600$~MHz. The arrows indicate the 3 dB bandwidth calculated using Eq.~(\ref{eq:paramp_graph_bandwidth}), 60 MHz in this case. The bandwidth of the amplifier can be increased rather dramatically by embedding the 2-mode primitive of Fig.~\ref{fig:FCPA_graph_matrix}(c) in a passive matching network~\cite{Matthaei1961, henoch1963new, mutus2014strong, roy2015broadband, naaman2019high}. Section~\ref{sec:all_together_paramp} shows how to engineer these matching networks to obtain prescribed gain characteristics. 

\subsubsection{Parametric circulator}\label{sec:graph_example_circulator}
The parametric converter and parametric amplifier are represented by simply-connected graphs (there are no loops) and therefore the phase of the pump only adds an overall phase in the S-parameters but otherwise does not affect the behavior of the device. Therefore, in the examples above we could have taken the coupling terms to be real without loss of generality. When the network's graph is multiply-connected, the relative phases of the coupling terms matter. The (anti-)conjugate symmetry of the parametric coupling coefficients, combined with interference from closed loops in these geometries can generate synthetic non-reciprocal scattering~\cite{ranzani2015graph}.

\begin{figure}[th]
    \centering
    \includegraphics[width=3.0in] {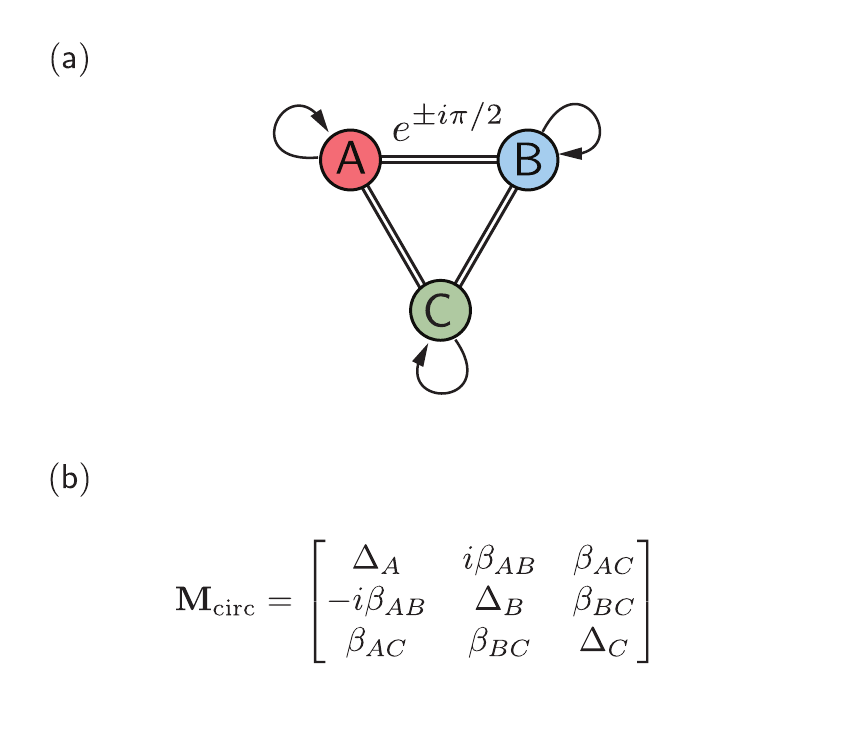}
    \caption{(a) Coupling graph of a 3-mode parametric circulator, indicating the relative pump phase, $\pm\pi/2$, of the A-B edge. (b) Corresponding EoM matrix for a pump phase of $\pi/2$.\label{fig:cir_graph_matrix}}
\end{figure}

Fig.~\ref{fig:cir_graph_matrix}(a) shows the coupled-mode graph of the 3-mode circulator~\cite{lecocq2017nonreciprocal}, one of the simplest devices to implement parametric non-reciprocity. The device has three resonant modes, coupled pair-wise with parametric conversion processes (difference-frequency pumps), where one of the couplings (the A-B edge in Fig.~\ref{fig:cir_graph_matrix}) carries a phase of $\pm\pi/2$ to affect circulation, and the circulation direction depends on the sign of that phase. In the figure, the modes are colored to represent three different mode-frequencies; note, however, that the minimal construction would require only mode B to be parametrically coupled and thus at a different frequency, while modes A and C can have the same frequency with a passive coupling between them.   

Fig.~\ref{fig:cir_graph_matrix}(b) shows the EoM matrix where we have included the $\pi/2$ phase of the A-B edge ($i\beta_{AB}$ in the upper triangle of the matrix, explicitly factoring out the phase so that $\beta_{AB}$ is real) and used the conjugation rule in Fig.~\ref{fig:matrix_rules} to write $\beta_{BA}=-i\beta_{AB}$ in the lower triangle. All other couplings are assumed to be real.

\begin{figure}
    \centering
    \includegraphics[width=\columnwidth] {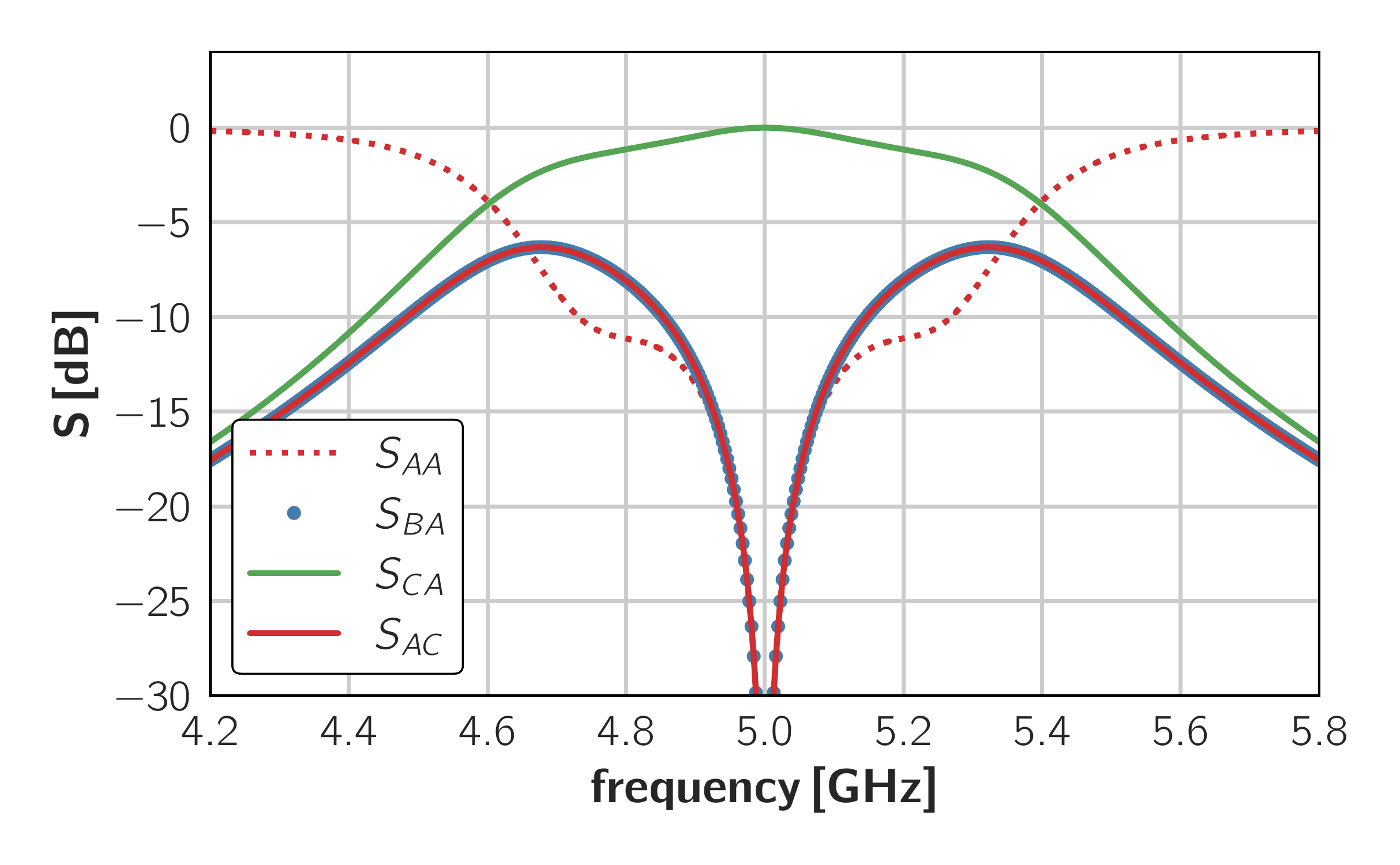}
    \caption{S-parameters of the 3-mode parametric circulator calculated using Eq.~(\ref{eq:s_params_components}). The frequency axis is referenced to the signal port A. \label{fig:circ_graph_sparams}}
\end{figure}

If the port coupling is equal for all ports, $\gamma_A=\gamma_B=\gamma_C=\gamma_0$, then from Eq.~(\ref{eq:s_params_components}) we see that the condition for perfect isolation from port C to port A at zero detuning, $S_{AC}(\omega^s_A=\omega_A)=0$, requires $\beta_{AB}\beta_{BC}=\beta_{AC}/2$, and similarly isolation from port A to port B, $S_{BA}=0$, requires $\beta_{AB}\beta_{AC}=\beta_{BC}/2$. These two conditions give $\beta_{AB}=\beta_{BC}=\beta_{AC}=0.5$. This set of parameters also results in match at each of the ports at zero detuning, \eg, $S_{AA}(\omega^s_A=\omega_A)=0$.

Figure~\ref{fig:circ_graph_sparams} shows the S-parameters calculated using Eq.~(\ref{eq:s_params_components}) and the coupling matrix of Fig.~\ref{fig:cir_graph_matrix}(b), with all port couplings set to $\gamma_0/2\pi=400$~MHz, and with all $\beta$'s equal to 0.5. At the center frequency of the device we get unity conversion ($S_{CA}$, green) of signals entering port A near $\omega_A$ to signals leaving port C near $\omega_C$, while in the backwards direction (from $A$ to $B$, $S_{BA}$ blue, or from $C$ to $A$, $S_{AC}$ red, solid) we get zero conversion. In addition, all ports are matched (\eg, $S_{AA}$, red, dots), as is the case for an ideal circulator.

\subsection{Graph Reduction}
\label{sec:graphreduction}
The graph of a coupled-mode system can be simplified by eliminating nodes from the graph, in a process known as Kron reduction~\cite{dorfler2012kron}. The edges and self-loops of any remaining node that was previously connected to the eliminated node will be re-scaled to account for change in the connectivity of the graph. When a node $k$ is eliminated, the new, resulting graph is represented by a new EoM matrix $\mathbf{M}'$, such that the new matrix components can be calculated from those of the old matrix $\mathbf{M}$ via~\cite{peterson2017demonstration} (no summation implied):
\begin{equation}\label{eq:kron_def}
    M'_{ij} = M_{ij} - \frac{M_{ik}M_{kj}}{M_{kk}}. 
\end{equation}
For example, referring to the parametric converter in Fig.~\ref{fig:FCPA_graph_matrix}(b), if we are only interested in the reflection off of mode $A$, we can eliminate mode $B$ from the graph\textemdash in this case the resulting graph will have only one mode, and the EoM matrix reduces to a single element $\Delta'_A$:
\begin{equation}
    \Delta'_A=M'_{AA}=M_{AA}-\frac{M_{AB}M_{BA}}{M_{BB}}=\Delta_A-\frac{|\beta_{AB}|^2}{\Delta_B}.
\end{equation}

\begin{figure}[h]
    \centering
    \includegraphics[width=\columnwidth] {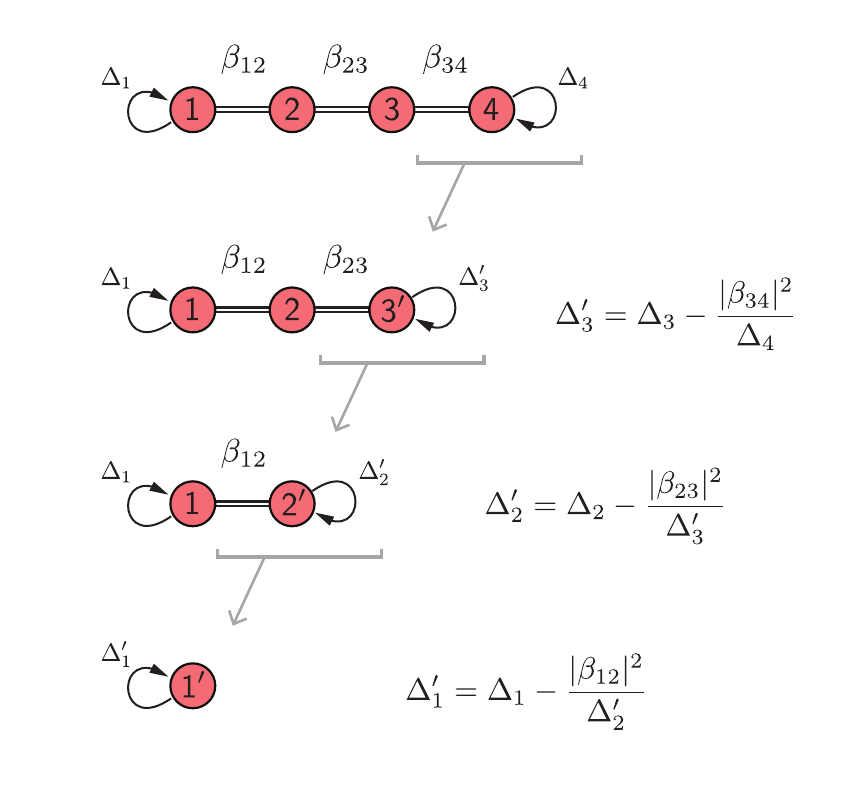}
    \caption{Graph reduction of a four-mode circuit with nearest-neighbor connectivity, where couplings are either passive or via parametric frequency conversion processes.\label{fig:reduction_example}}
\end{figure}

For the purpose of driving toward our main goal in this tutorial, we are interested in the admittance seen at a certain port, and we will use graph reduction to write
Eq.~(\ref{eq:port_admittance}) in the form:
\begin{equation}\label{eq:admittance_reduced}
    \frac{Y_k}{Y_0}=-2i\frac{\gamma_0}{\gamma_k}\Delta'_k-1,
\end{equation}
where $\Delta'_k$ is the overall result of successively applying Eq.~(\ref{eq:kron_def}) to eliminate all other modes, as shown for the one-dimensional nearest-neighbor coupled four-mode circuit in Fig.~\ref{fig:reduction_example}.

Carrying out the successive substitutions as indicated in Fig.~\ref{fig:reduction_example}, we can write:
\begin{equation}\label{eq:kron_cfrac}
    \Delta'_1=\Delta_1-\cfrac{|\beta_{12}|^2}{\Delta_2-\cfrac{|\beta_{23}|^2}{\Delta_3-\cfrac{|\beta_{34}|^2}{\Delta_4}}}.
\end{equation}
Note that the dissipation due to the port connected to mode 4 in Fig.~\ref{fig:reduction_example}, is embedded in Eq.~(\ref{eq:kron_cfrac}) via its dependence on $\Delta_4$. Therefore $\Delta'_1$ will have an effective `internal' dissipation included in its nonvanishing imaginary part.

The pattern that emerges from Eqs.~(\ref{eq:admittance_reduced}) and~(\ref{eq:kron_cfrac}) is that the admittance function of a 1D-connected graph can be written as a continued-fraction expression, where each of the terms is linear in signal frequency. Recall (Sec.~\ref{sec:graph_examples}) that for parametric processes under a resonant pump condition, the different frequencies of the modes all get projected to the signal band, and all detuning terms will have the same frequency dependence $\propto (\omega-\omega_0)$.

In Appendix~\ref{apx:kron_to_cauer} we show that the continued fraction expression of Eq.~(\ref{eq:kron_cfrac}) can be used to write the admittance in the Cauer form \cite{aatre1986network}, which is used in network synthesis to extract the network's normalized low-pass prototype coefficients. 

\section{Parametric coupling as a circuit element}
\label{sec:immittance}
In the previous section we developed implementation-agnostic methods to analyze parametrically-coupled circuits using the coupling matrix formalism~\cite{ranzani2015graph}. To bring the discussion down to the device level, we next examine how we can understand parametric couplings as functional circuit elements. We will develop this understanding by considering a sinusoidally-modulated inductance in a microwave circuit. This model is a generalization of parametrically modulated elements that are commonly used today, including single Josephson junctions, superconducting quantum interference devices (SQUIDs) (see Appendix \ref{apx:JJSQUID}), and kinetic inductance-based implementations \cite{parker2021near}. The discussion here is inspired by much earlier work on modulated capacitors (varactors) \cite{Getsinger1963}.

\begin{figure*}
    \centering
    \includegraphics[width=\textwidth]{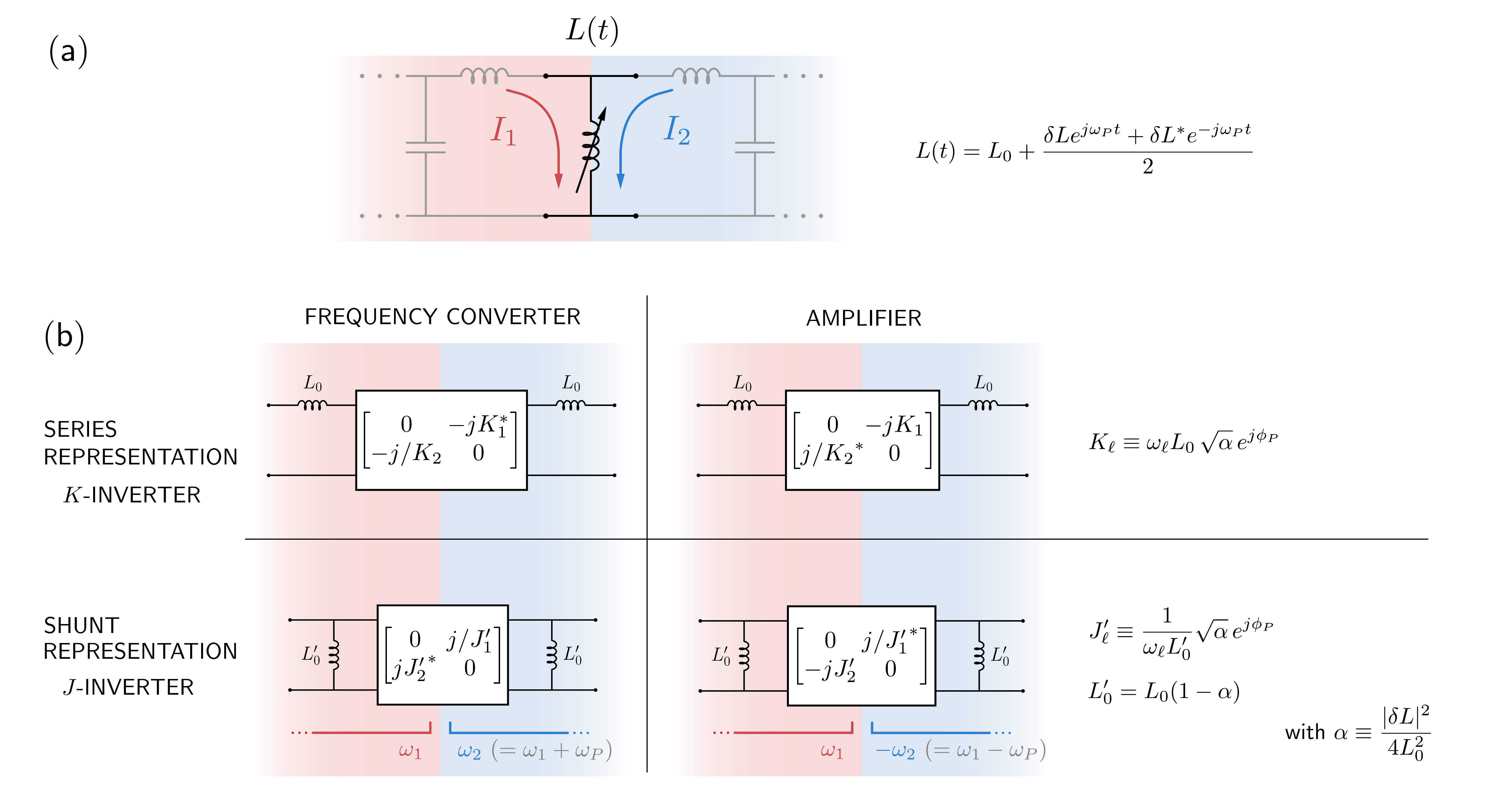}
    \caption{(a) Modulated inductor model with signal and idler branch currents $I_1$ and $I_2$. An example embedding circuit is shown (grayed-out) to provide context. (b) Equivalent 2-port circuit models for the \textit{series} and \textit{shunt representations} of the modulated inductor. The boxed elements represent impedance or admittance inverters, with the corresponding $ABCD$ matrices shown. The constants $K_\ell$, $J_\ell^\prime,$ and $L_0^\prime$ are defined in the figure.}
    \label{fig:LequivInverters}
\end{figure*}

We consider an inductor whose value is sinusoidally modulated, 
$L(t)=L_0+|\delta L|\cos{\omega_Pt}$,  as shown in Fig.~\ref{fig:LequivInverters}(a). The modulated inductor may be embedded in a microwave circuit as shown in the figure, and can participate in the total inductance that defines one or more $LC$ resonators. If this element carries two branch currents, $I_1$ and $I_2$, oscillating at two different frequencies, the voltage across this modulated inductor is
\begin{eqnarray}
    \label{eq:body_vflux}
    V(t) &=& \frac{d\Phi(t)}{dt}\\
    \label{eq:body_VofTexpanded}
     &=& \frac{d}{dt}\biggl[\biggl(L_0 + \frac{\delta L e^{j\omega_Pt} + \delta L^* e^{-j\omega_Pt}}{2}\biggr)  \nonumber \\
     & &\times \sum_{k=1,2}\frac{I_k e^{j\omega_kt} + I_k^* e^{-j\omega_kt}}{2}\biggr].
\end{eqnarray}

We can identify signals oscillating at $\pm\omega_1$ as the \textit{signal}, and those oscillating at $\pm\omega_2$ as the \textit{idler} \cite{aumentado2020superconducting}. We assume that the circuit embedding the modulated inductor is designed such that oscillations at any other frequency are effectively ``shorted out": $V_k=0$ for $k\notin \{1,2\}$, and we can therefore ignore other mixing products~\cite{tucker1964circuits}. In physics jargon, elimination of these other modes is a form of \textit{rotating wave approximation}. In non-degenerate amplifier and frequency conversion circuits, we will further assume same-frequency isolation between the signal and idler circuits, meaning that the idler network presents an open-circuit (in shunt representation) or short-circuit (in series representation) at the signal frequency, and \textit{vice versa}. Below, we take the convention that all frequencies $\omega_k>0$ and $\omega_2>\omega_1$.

If the pump is driven at $\omega_{P,\mathrm{FC}}= \omega_2 - \omega_1$ we couple the voltage and current components oscillating at $\omega_1$ to those oscillating at $\omega_2$, corresponding to a parametric frequency conversion process. If the pump frequency alternatively satisfies  $\omega_{P,\mathrm{PA}}= \omega_1 + \omega_2$ we couple components oscillating at $\omega_1$ to those at 
$-\omega_2$, corresponding to a parametric amplification process. When $\omega_1\neq\omega_2$ (non-degenerate parametric amplification) we get additional frequency conversion. 

\subsection{Parametric conversion}
When the inductance in Fig.~\ref{fig:LequivInverters}(a) is pumped at the difference frequency, $\omega_{P,\mathrm{FC}} = \omega_2 - \omega_1$, only those voltage components that oscillate at $\omega_1$ and $\omega_2$ survive in Eq.~(\ref{eq:body_VofTexpanded}). We can therefore write: 
\begin{eqnarray}
    V_1 &= j\omega_1 L_0 I_1 + j\omega_1 \frac{\delta L^*}{2} I_2\\
    V_2 &= j\omega_2 \frac{\delta L}{2}I_1 + j\omega_2 L_0 I_2,
\end{eqnarray}
yielding the following impedance Z-matrix~\cite{pozar2009microwave}:
\begin{equation}
    \mathbf{Z}_{L,\mathrm{FC}}= 
    \begin{bmatrix}
        j\omega_1 L_0 & j\omega_1\frac{\delta L^*}{2}\\
        j\omega_2 \frac{\delta L}{2} & j\omega_2 L_0
    \end{bmatrix}.
\label{eq:ZLFC}
\end{equation}
The Z-matrix (or its inverse, the admittance Y-matrix) can be converted to an $ABCD$ (transmission) matrix~\cite{pozar2009microwave} (see Appendix \ref{apx:ZY_ABCD}), and the linear time-independent part of the inductance can be factored out as shown in Fig.~\ref{fig:LequivInverters}(b). The remaining, purely parametric, part of the frequency converter $ABCD$ matrix in a series representation becomes
\begin{equation}\label{eq:abcd_fc}
    \mathbf{T}_\mathrm{FC}=
    \begin{bmatrix}
        0 & -jK^*_1 \\
        -j/K_2 &  0
    \end{bmatrix},
\end{equation}
where $K_\ell$ is defined in Fig.~\ref{fig:LequivInverters}(b).

\subsection{Parametric frequency conversion is a generalized impedance/admittance inverter}
What functional role does an element described by the $ABCD$ matrix Eq.~(\ref{eq:abcd_fc}) play in a circuit? Calculating the impedance $Z_1(\omega_1)=V_1/I_1$ seen from the circuit's input when its output port is terminated with an impedance $Z_2(\omega_2)=V_2/I_2$ we get
\begin{equation}\label{eq:fc_impedance_inverter}
    Z_1(\omega_1)=\frac{K^*_1K_2}{Z_2(\omega_2)}.
\end{equation}
Microwave engineers will recognize the function of the parametric coupling, as embodied in Eq.~(\ref{eq:fc_impedance_inverter}), as an impedance $(K)$ inverter~\cite{collin2007foundations}\textemdash the canonical example in passive circuits is the quarter wave transformer with impedance $Z_{\lambda/4}=K$, which transforms a load $Z_\mathrm{L}$ into an input impedance $Z_\mathrm{in}=Z^2_{\lambda/4}/Z_\mathrm{L}$ \cite{pozar2009microwave}. Similarly, the shunt representation in Fig.~\ref{fig:LequivInverters}(b) functions as an admittance $(J)$ inverter, transforming admittances according to $Y_1=J'^*_1J'_2/Y_2$. 

A key insight of this tutorial is that a parametric coupling driven with a difference-frequency pump generalizes the concept of the passive impedance (admittance) inverter, in that it transforms between impedances connected to ports that are not necessarily at the same frequency \cite{Matthaei1961, henoch1963new}. Additionally, the parametric impedance inverter $K$ can be complex, meaning that it carries information about the phase of the pump. Although one can view the modulated inductor as a cross-frequency inverter as we have shown here, it is important to understand this element as physically generating the mode coupling rates $c_{jk}$ in our general coupled mode picture. The inverter constants can be directly related to the normalized coupling rates,
\begin{equation}
    |\beta_{12}|^2 = \frac{1}{\gamma_0^2}\frac{K_1^*K_2}{L_1 L_2} = \frac{1}{\gamma_0^2}\frac{\omega_1\omega_2 |\delta L|^2}{4 L_1 L_2},
    \label{eq:K2beta}
\end{equation}
where $L_1$ and $L_2$ are the total effective inductances (including $L_0$) of the two coupled modes \cite{louisell1960coupled}, see Fig~\ref{fig:LequivInverters}(a).

Passive impedance or admittance  inverters (collectively referred to as `immittance' inverters in electrical engineering), are used extensively in microwave engineering as a tool to construct impedance-matching and filter networks~\cite{collin2007foundations}. We discuss inverters and their use in Sec.~\ref{sec:bandpass}. The close functional similarity they share with parametric conversion processes, invites the designer of parametrically-coupled devices to borrow techniques and methodologies from the existing, vast body of knowledge on filter network design. In return, the additional characteristics of parametric couplers, such as the complex phase, can expand the palette available to the filter network designer to encompass new features such as synthetic non-reciprocity.   

\subsection{Parametric amplification}\label{sec:immittance_paramp}
When the inductance in Fig.~\ref{fig:LequivInverters}(a) is pumped at the sum-frequency, $\omega_{P,PA}=\omega_1+\omega_2$, Eq.~(\ref{eq:body_VofTexpanded}) reduces to:
\begin{align}
    V_1 &= j\omega_1 L_0 I_1 + j\omega_1 \frac{\delta L}{2} I_2^*\\
    V_2^* &= -j\omega_2 \frac{\delta L^*}{2}I_1 - j\omega_2 L_0 I_2^*,
\end{align}
and can be described by the Z-matrix
\begin{equation}
    \mathbf{Z}_{L,\mathrm{PA}} = 
    \begin{bmatrix}
        j\omega_1 L_0 & j\omega_1\frac{\delta L}{2}\\
        -j\omega_2 \frac{\delta L^*}{2} & -j\omega_2 L_0
    \end{bmatrix}.
\label{eq:ZLPA}
\end{equation}
From the resulting $ABCD$ matrix (Fig.~\ref{fig:LequivInverters}(b) and Appendix \ref{apx:ZY_ABCD}) we see that, in the series representation, the parametric amplification process transforms impedances according to
\begin{equation}\label{eq:pa_impedance_inverter}
    Z_1(\omega_1)=-\frac{K'_1K'^*_2}{Z^*_2(\omega_2)},
\end{equation}
and, in the shunt representation, admittances are transformed according to 
\begin{equation}\label{eq:pa_admittance_inverter}
Y_1(\omega_1)=-\frac{J'^*_1J'_2}{Y^*_2(\omega_2)} .
\end{equation}

Eqs.~(\ref{eq:pa_impedance_inverter}-\ref{eq:pa_admittance_inverter}) suggests that the parametric amplification process functions as an `anti-conjugating' immittance inverter, to which there is no passive analogue. The negative sign in Eq.~(\ref{eq:pa_impedance_inverter}) indicates that the input port is presented with an effective impedance whose real part is negative, resulting in gain. A useful concept for practical design is the so-called `pumpistor model' introduced in Ref.~\cite{sundqvist2014negative}, where the input admittance, $Y_\mathrm{in}$, presented to the signal circuit at the terminals of the modulated inductor relates to admittance of the idler circuit as transformed via this anti-conjugating inverter. Using the shunt representation of Fig.~\ref{fig:LequivInverters}(b) and definitions therein, this can be written as
\begin{equation}\label{eq:pumpistor_Y}
    Y_\mathrm{in}(\omega_1)=\frac{1}{j\omega_1L'_0}\left[1 + \frac{\alpha}{j\omega_2L'_0Y^*_\mathrm{idler}(\omega_2)-1}\right],
\end{equation}
where $Y_\mathrm{idler}$ is the admittance of the idler circuit. Eq.~(\ref{eq:pumpistor_Y}) is equivalent to the `pumpistor' admittance derived in Ref.~\cite{sundqvist2014negative}, and models the modulated inductor as an inductance in parallel with a negative resistance.

A relation similar to Eq.~(\ref{eq:K2beta}), relating the inductance modulation strength to the coupling rate in the coupled-modes picture can be derived from the discussion in Sec.~\ref{sec:pump_amplitude}, and is given in Appendix~\ref{apx:para_inverter}.

\subsection{Modulated mutual coupling}

\begin{figure*}
    \centering
    \includegraphics[width=\textwidth]{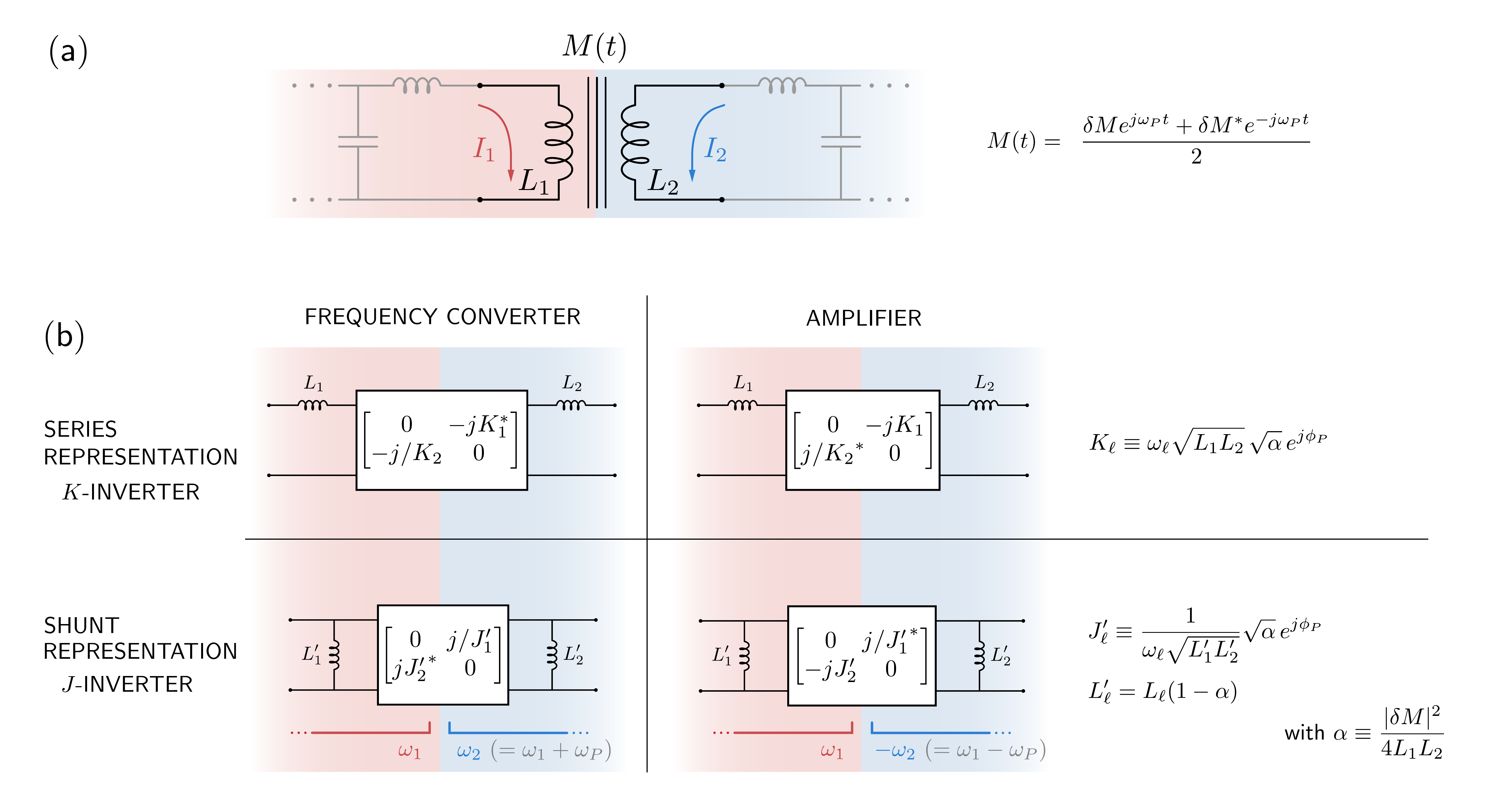}
   \caption{(a) Modulated mutual inductance model with signal and idler branch currents $I_1$ and $I_2$. An example embedding circuit is shown (grayed-out) to provide context. (b) Equivalent 2-port circuit models for the \textit{series} and \textit{shunt representations} of the modulated mutual. The boxed elements represent impedance or admittance inverters, with the corresponding $ABCD$ matrices shown. The constants $K_\ell$, $J_\ell^\prime,$ and $L_\ell^\prime$ are defined in the figure.\label{fig:MequivInverters}}
\end{figure*}

Above we discussed a particular circuit implementation of a parametric coupler, namely, the grounded modulated inductor, which can be implemented by a Josephson junction or a dc-SQUID (Appendix~\ref{apx:JJSQUID}). We note however, that the same reasoning can be applied to other types of parametric couplers, for example, signal and idler circuits that are coupled via a modulated mutual inductance, as was shown in Fig.~\ref{fig:FCPA_graph_matrix}(a),
\begin{equation}\label{eq:modulated_mutual}
    M(t)=M_0+\delta M\cos(\omega_Pt).
\end{equation}
This model is more appropriate for circuits like the JPC~\cite{abdo2013nondegenerate}, the rf-SQUID coupler ~\cite{allman2010rf-squid, chen2014gmon, naaman2016on-chip}, or even dispersive couplers ~\cite{yan2018tunable}. These types of couplers may be advantageous over the grounded SQUID in that they allow nulling of the passive part of the coupling, $M_0=0$, providing same-frequency isolation between the idler and signal circuits, and making the coupling purely parametric.

For a circuit having a signal inductor $L_1$ and an idler inductor $L_2$, which are coupled by the modulated mutual inductance of Eq.~(\ref{eq:modulated_mutual}), and where $M_0$ is nulled (Fig.~\ref{fig:MequivInverters}(a)), we can write the Z-matrices coupling currents and voltages oscillating at $\omega_1$ and $\omega_2$,
\begin{equation}
    \mathbf{Z}_{M,\mathrm{FC}} = 
    \begin{bmatrix}
        j\omega_1 L_1 & j\omega_1\frac{\delta M^*}{2} \\
        j\omega_2\frac{\delta M}{2} & j\omega_2 L_2
    \end{bmatrix},
\end{equation}
and
\begin{equation}
    \mathbf{Z}_{M,\mathrm{PA}} = 
    \begin{bmatrix}
        j\omega_1 L_1 & j\omega_1\frac{\delta M}{2} \\
        -j\omega_2\frac{\delta M^*}{2} & -j\omega_2 L_2
    \end{bmatrix}.
\end{equation}

Comparison with Eqs.~(\ref{eq:ZLFC}) and (\ref{eq:ZLPA}) indicates that this can be considered a generalization of the modulated inductor model with $L_1 = L_2 = L_0$ and, likewise, one can think of the inductance modulation amplitude $\delta L$ as corresponding to a modulated mutual inductance. The equivalent transmission matrices and inverter constant definitions are summarized in Fig.~\ref{fig:MequivInverters}(b).

\section{Band-pass impedance matching and filter networks}
\label{sec:bandpass}
Up to this point we have outlined a graph-based language that facilitates rapid analysis of arbitrary, linearly coupled mode systems, including a direct mapping between scattering parameters and the collections of coupling permutations that generate them. We have seen hints in Sections~\ref{sec:graph_examples} and~\ref{sec:graphreduction} that point to a deeper connection between parametric coupled-mode network design and the problem of filter and matching network design known from microwave engineering. In fact, similar graph-based approaches are already established in the design of microwave frequency cavity-based filters~\cite{cameron2003advanced} and photonic circuits~\cite{liu2011synthesis}. Further, in Section~\ref{sec:immittance} we saw that parametric couplings serve a similar circuit function to that of immittance inverters in microwave filter networks. Before we detail in Section~\ref{sec:all_together} an exact correspondence between the language of coupled-mode theory and that of filter design, we review here a few key concepts from microwave band-pass network synthesis. Filter synthesis techniques are already a very mature topic within electrical engineering \cite{pozar2009microwave,cameron2018microwave,MYJ},  so we will keep the discussion here to the minimum necessary to provide a foundation for Section~\ref{sec:all_together}. Appendix~\ref{apx:cauer_synthesis} gives a more detailed explanation of how network prototype coefficients are calculated.

\begin{figure}
    \centering
    \includegraphics[width=\columnwidth] {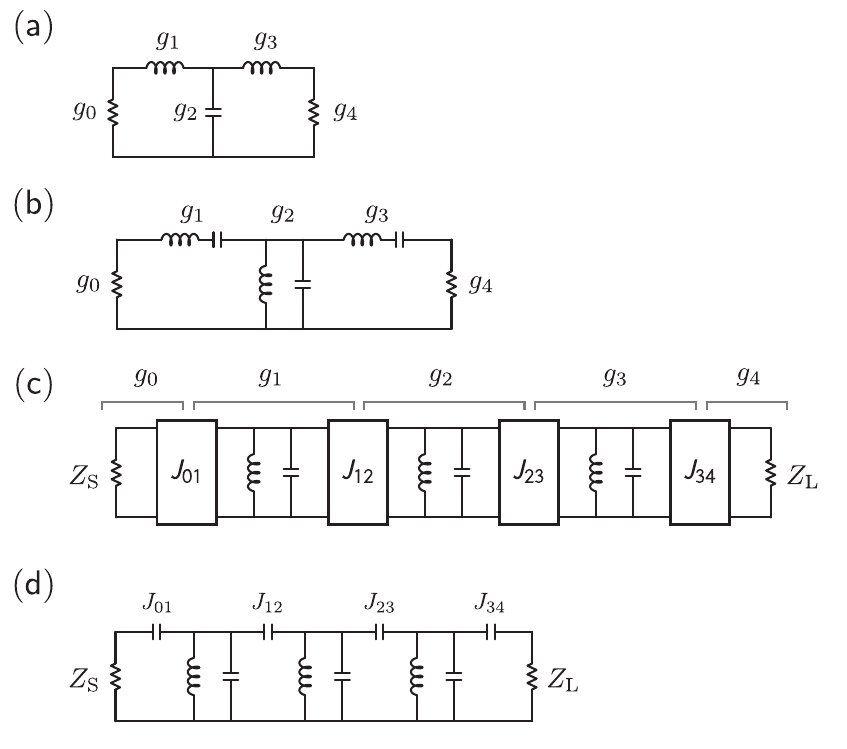}
    \caption{Example of 3-pole band-pass network construction. (a) Low-pass prototype, (b) band-pass transformation, (c) added admittance inverters, and (d) inverters implemented. \label{fig:filter_steps}}
\end{figure}

Band pass filter design begins by selecting a target transmission profile (\eg, ripple and rolloff characteristics), from which we can choose the desired number of filter sections $N$, the filter's center frequency $\omega_0$, bandwidth $\Delta\omega$, and response type (Chebyshev, Butterworth, \etc). We find the corresponding normalized filter coefficients $\{g_i\}$ from tables in \eg,~Refs.~\onlinecite{pozar2009microwave, MYJ}, where for an $N$ section filter we will have $N+2$ zero-indexed coefficients. These coefficients relate to polynomials specifying the input impedance of the network as a function of frequency (Appendix~\ref{apx:cauer_synthesis}). The coefficient $g_0$, representing the conductance of the source, is often omitted from tables as usually $g_0=1$ by definition. The last coefficient, $g_{N+1}$, represents the conductance of the load. The remaining coefficients $g_1\dots g_N$ correspond to the normalized reactances of the elements (alternating capacitors and inductors) that make up the low-pass filter prototype, as in Fig.~\ref{fig:filter_steps}(a).

\subsection{Series/shunt band-pass ladder network design}\label{sec:denorm}
Students are usually taught to start with a low-pass prototype network, Fig.~\ref{fig:filter_steps}(a), then convert each of the reactances to a resonant circuit via a band-pass transformation~\cite{pozar2009microwave}.  The result is a coupled-resonator circuit, Fig.~\ref{fig:filter_steps}(b), that alternates between shunt-connected parallel $LC$ resonators and series-connected series $LC$ resonators.

For a given set of prototype coefficients $\{g_i\}$, a specified reference impedance $Z_0=1/Y_0$, and a specified center frequency $\omega_0$ and bandwidth $\Delta\omega$, the resonators should be constructed such that the admittance $Y_j$ of the parallel resonator corresponding to the prototype coefficient $g_j$ is given by
\begin{equation}\label{eq:denorm_parallel}
    Y_j=g_j\frac{\omega_0}{\Delta\omega}Y_0,
\end{equation}
and the impedance $Z_k$ of the series resonator corresponding to the prototype coefficient $g_k$ is
\begin{equation}\label{eq:denorm_series}
    Z_k=g_k\frac{\omega_0}{\Delta\omega}Z_0.
\end{equation}
The last coefficient, $g_{N+1}$, corresponds to the load immittance, and its denormalization will depend on whether it is preceded by a series section or a parallel section:
\begin{equation}\label{eq:denorm_load}
    Z_\mathrm{L}=\left\{
    \begin{aligned}
        \frac{Z_0}{g_{N+1}};\;\; & N^{th}\mathrm{~section~series}\\
        Z_0\times g_{N+1};\;\; & N^{th}\mathrm{~section~parallel}
    \end{aligned}
    \right.
\end{equation}

Once the resonator impedances $Z_i$ are known, their inductances and capacitances can be calculated in the usual way, $L_i=Z_i/\omega_0$ and $C_i=1/Z_i\omega_0$. The coupling rate between the $j$ parallel resonator and the neighboring $k=j\pm1$ series resonator is a function of their respective impedances:
\begin{equation}
    c_{jk}=\omega_0\sqrt{\frac{Z_j}{Z_k}}.
\end{equation}

This method of circuit construction will be useful in understanding the results of Ref.~\onlinecite{roy2015broadband}, as we show in Sec.~\ref{sec:vijay}.

\subsection{Coupled-resonator method for band-pass network design}\label{sec:design_method}
 The series/shunt ladder network we get by following the procedure in Sec.~\ref{sec:denorm} may be difficult to implement in practice, as it offers little flexibility in component selection, and does not generalize well to circuits that are not based on lumped-element electrical resonators. A more convenient circuit can be realized by replacing the series-connected series-$LC$ resonators with shunt-connected parallel-$LC$ resonators sandwiched between two admittance inverters~\cite{collin2007foundations}, $J_{jk}$, as shown in Fig.~\ref{fig:filter_steps}(c). In this configuration, the inverters are responsible for the coupling between the resonators. Because the inverters can also transform between impedance levels, the designer is now free to choose the resonator impedances to best fit the available technology. This circuit topology of coupled resonators additionally lends itself more naturally to a description based on the coupled-mode language developed in Section~\ref{sec:matrix}.

Coupled-resonator band-pass network design methods are illustrated in detail in Ref.~\onlinecite{MYJ}. Start with a set of $N+2$ prototype coefficients $\{g_j\}$ for a network of $N$ resonators having a resonance frequency $\omega_0$ and impedances $Z_j$, $j=1\dots N$. To implement a network with a fractional bandwidth $w=\Delta\omega/\omega_0$, that operates between a source impedance $Z_\mathrm{S}$ and a load impedance $Z_\mathrm{L}$, and is based on lumped-element shunt-$LC$ resonators, calculate the values $J_{jk}$ of the admittance inverters~\cite{MYJ}:
\begin{align}
    J_{01} & = \sqrt{\frac{w}{g_0g_1Z_\mathrm{S}Z_1}}, \label{eq:J_01}\\
    J_{jk} & = \frac{w}{\sqrt{g_jg_kZ_jZ_k}}, \label{eq:J_jk}\\
    J_{N,N+1} & = \sqrt{\frac{w}{g_Ng_{N+1}Z_NZ_\mathrm{L}}}. \label{eq:J_nn1}
\end{align}

\begin{figure}
    \centering
    \includegraphics[width=\columnwidth] {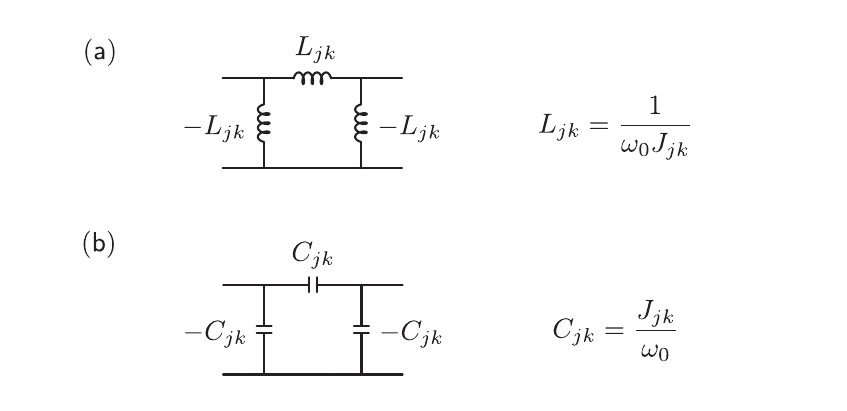}
    \caption{Lumped-element physical implementations of an admittance inverter, using (a) an inductive $\pi$-section, and (b) a capacitive $\pi$-section. \label{fig:inverters_pi}}
\end{figure}

Next, choose a physical implementation for each of the inverters; in a lumped-element circuit we can choose either the inductive circuit of Fig.~\ref{fig:inverters_pi}(a) or the capacitive circuit of Fig.~\ref{fig:inverters_pi}(b), and calculate the inductance or capacitance of the inverter based on the corresponding $J$ value, \ie, $C_{jk}=J_{jk}/\omega_0$ and $L_{jk}=1/\omega_0J_{jk}$. The negative values of the shunt components in Fig.~\ref{fig:inverters_pi} will be absorbed by the neighboring resonators when the inverter is incorporated into the circuit. The resulting circuit will have the general structure of Fig.~\ref{fig:filter_steps}(d) in which capacitive inverters were used. Since the source (load) resistor cannot absorb a negative reactance associated with the first (last) inverter, we have to modify their component values as outlined in Appendix~\ref{apx:inverter_termination} and Ref.~\onlinecite{MYJ}. Finally, calculate $L$ and $C$ of the shunt resonators using $\omega_0$, their respective impedances, and the absorbed inverter component values. 

The physical implementation of immittance inverters, as in Fig.~\ref{fig:inverters_pi}, is always associated with the additional frequency dependence of the inverter itself. For this reason, the methods described herein are typically suitable for designs with fractional bandwidth up to $\approx20$\%. Techniques for designs with higher bandwidth can be found in, \eg, Ref.~\cite{MYJ}.

We illustrated the coupled-resonator design method here with lumped-element parallel $LC$ resonators coupled by admittance inverters. If the preferred resonator type is a series $LC$, a similar procedure that uses impedance- (K-) inverters can be used~\cite{MYJ, collin2007foundations}. Ref.~\onlinecite{MYJ} additionally details expressions similar to Eqs.~(\ref{eq:J_01})-(\ref{eq:J_nn1}) for implementations based on half- and quarter-wave transmission line resonators. A practical design example is given in Section~\ref{sec:filter_example}.

\subsection{Impedance-matching of a resonated load}\label{sec:resonated_load}
We already mentioned that the use of immittance inverters in a filter design offers flexibility in choosing the impedance levels of the constituent resonators. Indeed, it is clear from Eqs.~(\ref{eq:J_01}) and~(\ref{eq:J_nn1}) that the design can also accommodate arbitrary source and load impedances, so that networks designed using the coupled-resonator method can conveniently serve as impedance-matching networks.

In many cases that will be important to our discussion here, we will have to design band-pass networks to match a given `resonated load'\textemdash essentially a resistance shunted by an $LC$ resonator, or more generally, some admittance function $Y(\omega)$ that can be approximated by an $LC$ resonator with a finite quality factor $Q$ at a particular frequency of interest. The latter is a case that we encounter in Section~\ref{sec:all_toghether_circ} when matching a parametric circulator. In the case of Josephson parametric amplifiers (Sec.~\ref{sec:all_together_paramp}), we will be presented with the need to match a load (the parametrically pumped SQUID) that looks like a (negative) resistance in parallel with an inductance; we `resonate' this load by adding a shunt capacitor to the device.

The $Q$ factor of the resonated load can be calculated from its admittance $Y(\omega)$ via~\cite{MYJ}
\begin{equation}\label{eq:loaded_res_Q}
    Q=\frac{\omega_0}{2\mathrm{Re}\left\{Y\left(\omega_0\right)\right\}}\left. \frac{d}{d\omega}\mathrm{Im}\left\{Y\left(\omega\right)\right\}\right\rvert_{\omega=\omega_0},
\end{equation}
where $\omega_0$ is the resonance frequency. For a resistor $R$ shunted by a parallel lumped $LC$ resonator with impedance $Z_\mathrm{res}$, this reduces to the usual $Q=R/Z_\mathrm{res}$. 

Once the $Q$ factor is known, we can proceed with the network design, with the resonated load accounting for the elements labeled as $g_0$ and $g_1$ in Fig.~\ref{fig:filter_steps}(c). The network bandwidth will be constrained by the so-called `decrement' relation~\cite{MYJ}
\begin{equation}\label{eq:decrement}
    w=\frac{g_0g_1}{Q},
\end{equation}
where here, too, $w$ is the fractional bandwidth.

Eq.~(\ref{eq:decrement}) means that if we have a bandwidth requirement to satisfy, we can find the required impedance for the resonator embedding the load. If, instead, we have a given resonator but the load resistance may be adjustable (\textit{e.g.}~via its dependence on a parametric pump amplitude, \textit{c.f.}~Section~\ref{sec:pump_amplitude}), then  Eq.~(\ref{eq:decrement}) lets us set that control knob~\cite{naaman2019high}. If $Q$ of the resonated load is fixed, then we have to design the rest of the network to have the same bandwidth $w$ given by Eq.~(\ref{eq:decrement}).

Note that the decrement condition Eq.~(\ref{eq:decrement}) is synonymous to requiring a transformation of the reference admittance $Y_0=1/Z_0$ through the inverter $J_{01}$ specified by the network prototype, \textit{i.e.}:
\begin{equation}
\mathrm{Re}\{Y(\omega_0)\}=\frac{J^2_{01}}{Y_0}.\nonumber
\end{equation}

\subsection{Coupled-resonator filter design example}\label{sec:filter_example}
As an example, we will design a 3-pole, 500 MHz band-pass filter centered at $\omega_0/2\pi=5$~GHz using a 0.5 dB ripple Chebyshev prototype, in a $Z_0=50\,\Omega$ environment.

\begin{figure}[h]
    \centering
    \includegraphics[width=\columnwidth] {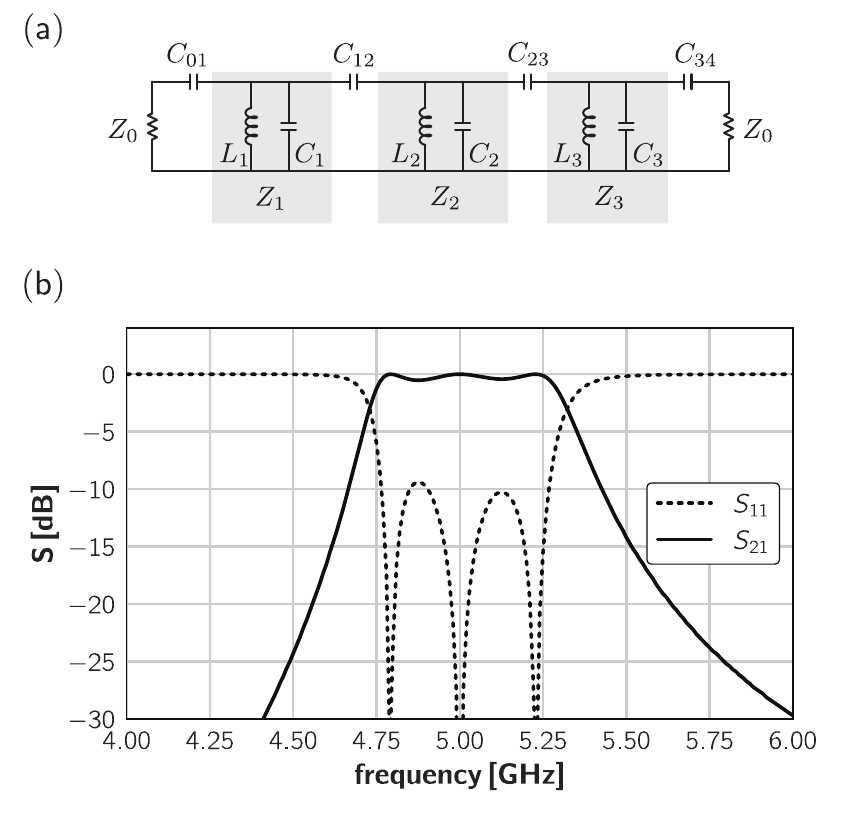}
    \caption{3-pole filter design example. (a) Circuit schematic,  (b) S-parameter simulation in Keysight ADS. \label{fig:filter_example}}
\end{figure}

The prototype coefficients for a 3-pole, 0.5 dB ripple Chebyshev network can be found in \eg,~Ref.~\onlinecite{pozar2009microwave}: $g_i=\{1.0,\,1.5963,\,1.0967,\,1.5963,\,1.0\}$. We will choose (rather arbitrarily for demonstration purposes) the impedances of the three resonators, highlighted in Fig.~\ref{fig:filter_example}(a), to be $Z_1=40\,\Omega$, $Z_2=30\,\Omega$, and $Z_3=40\,\Omega$.

We will implement the admittance inverters using capacitive $\pi$-sections, as in Fig.~\ref{fig:inverters_pi}(b). Using Eqs.~(\ref{eq:J_01})-(\ref{eq:J_nn1}), and the required fractional bandwidth of $w=0.1$, we calculate the inverter values: $J_{01}=J_{34}=0.0056\,\Omega^{-1}$, and $J_{12}=J_{23}=0.0022\,\Omega^{-1}$. From here, the coupling capacitances of the `internal' inverters $C_{jk}=J_{jk}/\omega_0$ can be evaluated (refer to the schematic in Fig.~\ref{fig:filter_example}(a)), $C_{12}=C_{23}=0.069$~pF. For the `external' inverters that couple the filter to the source and load (see Appendix~\ref{apx:inverter_termination}), we use the relations~\cite{MYJ}:
\begin{align}
    C_{01}&=\frac{J_{01}}{\omega_0\sqrt{1-\left(Z_0J_{01}\right)^2}} \\
    C_{34}&=\frac{J_{34}}{\omega_0\sqrt{1-\left(Z_0J_{34}\right)^2}}
\end{align}
to get $C_{01}=C_{34}=0.186$~pF. 

The values of the inductance $L_j$ of resonator $Z_j$ is simply calculated using $L_j=Z_j/\omega_0$: $L_1=L_3=1.27$~nH and $L_2=0.96$~nH. The resonator capacitors have to absorb the negative shunt capacitances of the inverters, so that:
\begin{align}
    C_1 &= \frac{1}{Z_1\omega_0}-C_{01e}-C_{12}=0.56\,\mathrm{pF}, \\
    C_2 &= \frac{1}{Z_2\omega_0}-C_{12}-C_{23}=0.92\,\mathrm{pF}, \\
    C_3 &= \frac{1}{Z_3\omega_0}-C_{23}-C_{34e}=0.56\,\mathrm{pF},
\end{align}
where we have used (Appendix \ref{apx:inverter_termination}, Eq.~(\ref{eq:c01e}), and Ref.~\cite{MYJ}) $C_{01e}=\frac{J_{01}}{\omega_0}\sqrt{1-\left(Z_0J_{01}\right)^2}$, and similarly for $C_{34e}$.

Now that all component values are specified, we can construct the circuit in Fig.~\ref{fig:filter_example}(a) and simulate its response. An S-parameter simulation using Keysight ADS is shown in Fig.~\ref{fig:filter_example}(b), showing that the center frequency, bandwidth, and ripple requirements are met. The slight asymmetry seen in the response is due to the additional frequency dependence introduced by the physical implementation of the admittance inverters as capacitive networks.

\subsection{Filter synthesis methodically determines resonator coupling rates}\label{sec:methodically}
The preceding sections, and the example in Sec.~\ref{sec:filter_example}, demonstrate that band-pass filter synthesis boils down to determining the values of the immittance inverters that are disposed between the filter resonators, and between the resonators and the environment. We have seen that these structures effectively control the coupling rates in the circuit, as their physical implementation as coupling capacitors or inductors would suggest. 

A third key insight in this tutorial is that the art of band-pass network synthesis is concerned with methodically engineering the coupling rates in a system of resonant modes. These methods transcend electrical circuit design and could be applied in diverse areas such as mechanical \cite{rhodes1975modern}, acoustic \cite{kinsler2000fundamentals}, or optical \cite{madsen1999optical, liu2011synthesis} filter design.

\section{Engineering coupled-mode networks}
\label{sec:all_together}
The discussion thus far points toward a deep connection between parametric networks and filter networks. Here, we will build on the insights developed in the preceding sections to unify the language used to describe these systems. 

A key insight from Section~\ref{sec:graph_examples} is that in parametric networks that are driven by pumps that are tuned to either the sum- or the difference-frequency, the fact that the different modes in the circuit have different frequencies essentially drops out of the equations\textemdash the behavior of the circuit can be described using a single reference frequency, usually taken as that of the signal mode. In that respect, the EoM matrices of systems coupled via parametric conversion processes with 1D-connectivity are indistinguishable (up to an overall phase) from those of passively coupled circuits such as band-pass filters. The novelty of parametric coupling only manifests in the off-diagonal elements of the EoM matrix, through non-trivial phases (in the case of multiply connected circuits) and anti-conjugation (in the case of parametric amplification).

The second key insight, arising from Section~\ref{sec:immittance}, is that parametric couplings function as generalized immittance inverters. The use of immittance inverters in the design of filter networks was described in Section~\ref{sec:bandpass}, where we developed an additional insight, that the art of filter synthesis is ultimately concerned with methodically determining coupling rates between resonant modes in the circuit. Immittance inverters therefore should be thought of as the circuit implementation of the off-diagonal elements in the coupled-mode EoM matrices of Sec.~\ref{sec:matrix}.

\subsection{Correspondence of coupled-modes and filter networks}\label{sec:correspondence}
\begin{figure}[h]
    \centering
    \includegraphics[width=\columnwidth]{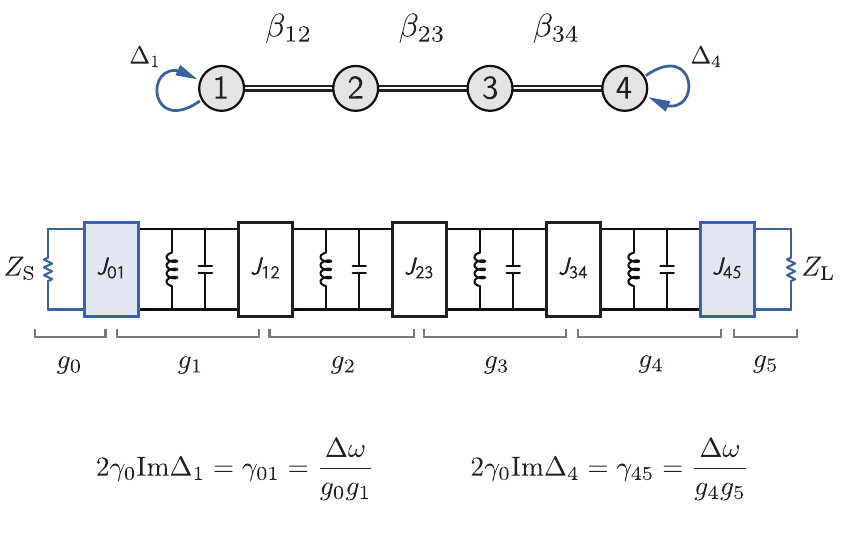}
    \caption{Correspondence between a graph representation of a parametrically coupled network and a coupled-resonator band-pass filter. $\beta_{jk}$ are the coupling elements in the coupled-modes matrix representation, $J_{jk}$ are admittance inverters, and $g_i$ are filter prototype coefficients. The port dissipation rates $\gamma$ are embedded in the imaginary part of the self loops on modes 1 and 4.\label{fig:filter_to_graph}}
\end{figure}

Figure~\ref{fig:filter_to_graph} displays, side by side, a four-mode graph of a coupled-mode system and a fourth-order band-pass electrical network. This graph can represent, for example, an $N$-stage transduction network \cite{wang2022generalized, wang2022quantum}, a multi-mode parametric converter, or matched parametric amplifier \cite{naaman2019high}. Each of the nodes in the graph corresponds to an $LC$ resonator in the filter network, and the ports\textemdash indicated by the self loops in the graph\textemdash correspond to the source and load resistors of the electrical circuit. The graph edges $\beta_{jk}$ (which represent the off-diagonal coupling elements in the circuit's EoM matrix) correspond to the admittance inverters $J_{jk}$, and the port dissipation rate $\gamma_{01}$ (respectively, $\gamma_{45}$) correspond to the admittance inverter $J_{01}$ (respectively, $J_{45}$). 

The coupling rate provided by an admittance inverter $J_{jk}$ disposed between resonant modes $j$ and $k$, in a network design based on prototype with coefficients $\{g_j\}$, is given by~\cite{MYJ}:
\begin{equation}\label{eq:inverter_coupling_rate}
    c_{jk}=\frac{\Delta\omega}{\sqrt{g_jg_k}},
\end{equation}
where $\Delta\omega$ is the bandwidth of the network. The loaded quality factors of the first and last resonators due to their coupling to the environment through $J_{01}$ and $J_{N,N+1}$ is given by~\cite{MYJ}:
\begin{align}\label{eq:inverter_dissipation_rate}
    Q_1&=\frac{\omega_0}{\Delta\omega}g_0g_1\nonumber\\
    Q_N&=\frac{\omega_0}{\Delta\omega}g_Ng_{N+1},
\end{align}
where $\omega_0$ is the network's center frequency.

Comparing Eqs.~(\ref{eq:inverter_coupling_rate}) and~(\ref{eq:inverter_dissipation_rate}) to Eqs.~(\ref{eq:beta_def}) and~(\ref{eq:port_gamma_q}), respectively, we arrive at our main result: that a correspondence between the coupled-mode picture and filter synthesis can be made by the relations
\begin{align}
    \gamma_{01} &= \frac{\Delta\omega}{g_0g_1} \label{eq:correspond_source} \\
    \gamma_{N,N+1} &= \frac{\Delta\omega}{g_Ng_{N+1}} \label{eq:correspond_load}\\
    \beta_{jk} &= \frac{\Delta\omega}{2\gamma_0\sqrt{g_jg_k}}\label{eq:correspond_beta}
\end{align}

Further, comparing to Eqs.~(\ref{eq:J_01})-(\ref{eq:J_nn1}) for lumped-element circuits, we have:
\begin{align}
    \gamma_{01}&=\omega_0Z_\mathrm{S}Z_1\times J^2_{01} \\
    \gamma_{N,N+1}&=\omega_0Z_\mathrm{L}Z_N\times J^2_{N,N+1} \\
    \beta_{jk}&=\frac{\omega_0}{2\gamma_0}\sqrt{Z_jZ_k}\times J_{jk}\label{eq:beta_to_J}
\end{align}
where $Z_\mathrm{S}$ and $Z_\mathrm{L}$ are the source and load impedances, and $Z_j$ is the impedance of mode $j$. Note that when inverters are connected between modes of different frequencies, the normalized bandwidth $w$ should always be calculated with respect to the reference frequency, so that the absolute bandwidth $\Delta\omega$ is fixed throughout the whole circuit.

Below we will demonstrate how we can use the above correspondence to design parametrically coupled circuits with prescribed transfer characteristics.

\subsection{Synthesis of a parametric frequency converter}\label{sec:all_together_converter}
In Section~\ref{sec:examples_FC} we claimed that a 2-mode frequency converter, designed for perfect port match, exactly implements a second-order Butterworth filter. We can now see that indeed, when using the coefficients for the appropriate prototype~\cite{pozar2009microwave}, $g_i=\{1.0,\,1.4142,\,1.4142,\,1.0\}$, in Eqs.~(\ref{eq:correspond_source})-(\ref{eq:correspond_beta}) we obtain the same values found in Section~\ref{sec:examples_FC}, \ie,  $\gamma_{01}=\gamma_{23}=\Delta\omega/\sqrt{2}$ and $\beta_{12}=0.5$.

As a further example, we will design a parametric frequency converter with a signal frequency of $\omega_A/2\pi=5$~GHz, an idler frequency of $\omega_B/2\pi=7$~GHz, and with a bandwidth of $\Delta\omega/2\pi=250$~MHz. To facilitate the broadbanding of the usual 2-mode converter (Sec.~\ref{sec:examples_FC}) we add to the circuit an additional mode at each of the signal and idler frequencies, and design the circuit as a whole to have response characteristics of a 4-pole Chebyshev network with 0.01~dB ripple.

\begin{figure}
    \centering
    \includegraphics[width=\columnwidth]{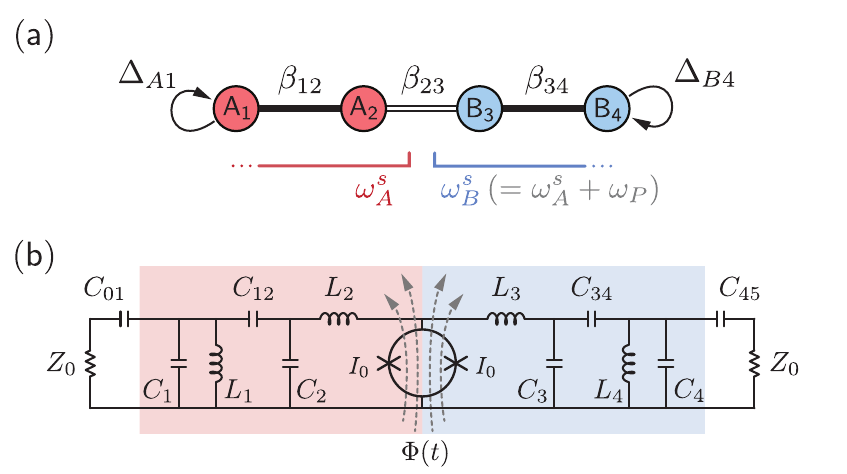}
    \caption{\label{fig:broadbandFC}(a) Coupling graph of a 4-mode matched parametric frequency converter, (b) circuit schematic.}
\end{figure}

The coupled-mode graph for the design is shown in Fig.~\ref{fig:broadbandFC}(a). There are two $A$ modes at frequency $\omega_A$ and two $B$ modes at frequency $\omega_B$. Mode $A_1$ is coupled to an external port with rate $\gamma_{01}$ (indicated by the self-loop, with $\gamma_{01}=2\gamma_0\mathrm{Im}\{\Delta_{A1}\}$), and mode $B_4$ is coupled to an external port with rate $\gamma_{45}$ (indicated by the self-loop, with $\gamma_{45}=2\gamma_0\mathrm{Im}\{\Delta_{B4}\}$). Modes $A_1$ and $A_2$ are at the same frequency and are coupled with a passive element, and similarly for modes $B_3$ and $B_4$. Modes $A_2$ and $B_3$ are coupled by a parametric frequency conversion process with a pump tuned to $\omega_P=\omega_B-\omega_A$.

We can write the EoM matrix associated with the graph in Fig.~\ref{fig:broadbandFC}(a) as:
\begin{equation}\label{eq:broadband_FC_matrix}
    \mathbf{M}=
    \begin{bmatrix}
        \Delta_{A1} & \beta_{12} & 0 & 0 \\
        \beta_{12} & \Delta_{A2} & \beta_{23} & 0 \\
        0 & \beta^*_{23} & \Delta_{B3} & \beta_{34} \\
        0 & 0 & \beta_{34} & \Delta_{B4}
    \end{bmatrix}.
\end{equation}
We use design tables~\cite{pozar2009microwave, MYJ} to find the appropriate prototype coefficients for a 4-pole Chebyshev network with the required 0.01~dB ripple,
\begin{equation}
    g_i=\{1.0,\, 0.7128,\, 1.2003,\, 1.3212,\, 0.6476,\, 1.1007\},\nonumber
\end{equation}
and calculate the values of the coupling terms using Eq.~(\ref{eq:correspond_beta}) to get $\beta_{12}=\beta_{34}=0.385$ for the passive couplings and $|\beta_{23}|=0.283$ for the parametric coupling. The port coupling rates can be calculated using Eqs.~(\ref{eq:correspond_source}) and~(\ref{eq:correspond_load}) to obtain $\gamma_{01}/2\pi=\gamma_{45}/2\pi=351$~MHz.

\begin{figure}[th]
    \centering
    \includegraphics[width=\columnwidth] {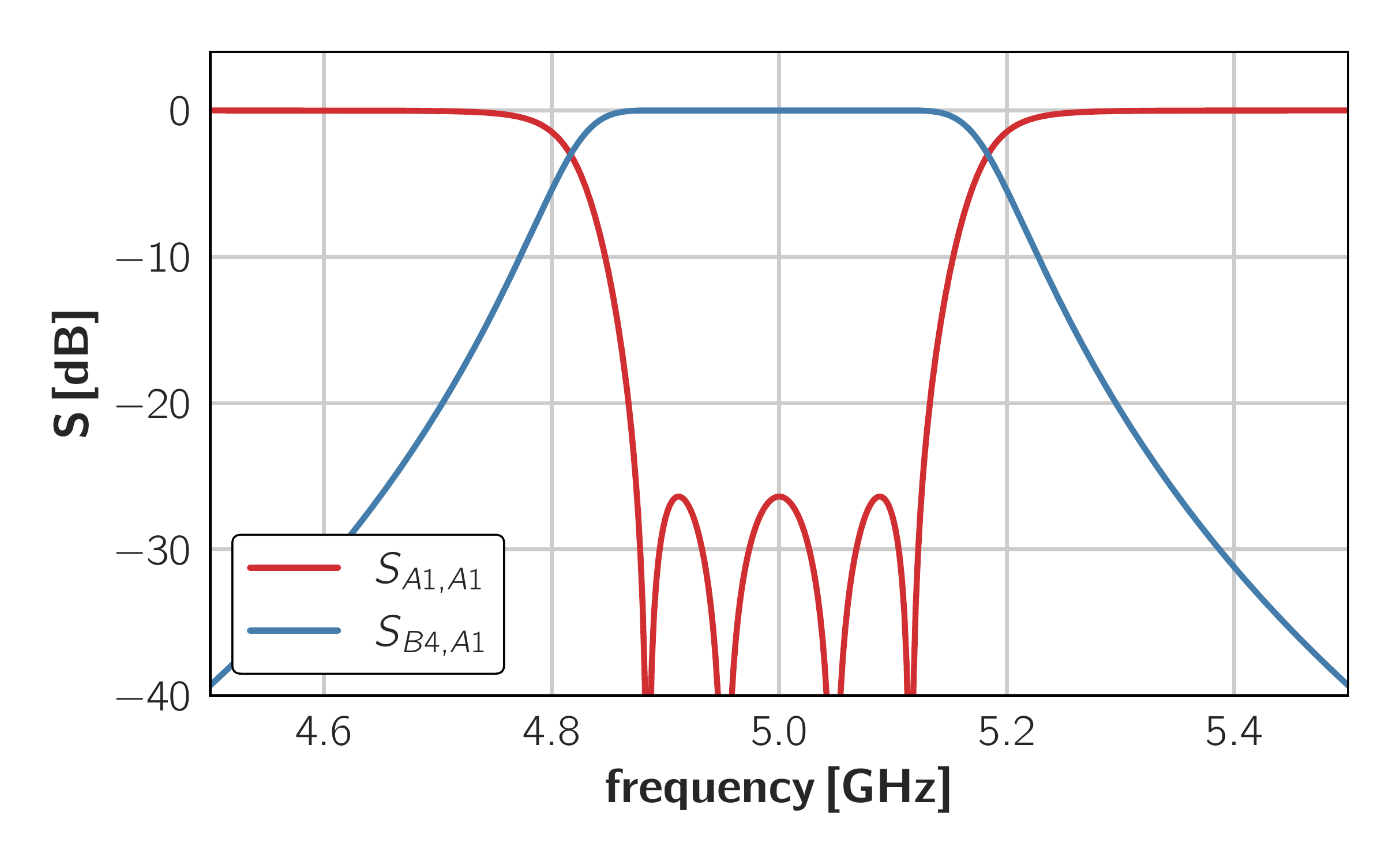}
    \caption{S-parameters of a matched parametric converter, calculated using Eq.~(\ref{eq:s_params_components}) and the coupling matrix Eq.~(\ref{eq:broadband_FC_matrix}). The frequency axis is referenced to the signal port.\label{fig:broadband_FC_sparams}}
\end{figure}

Figure \ref{fig:broadband_FC_sparams} shows the resulting S-parameters, calculated using Eq.~(\ref{eq:s_params_components}) for the coupling matrix Eq.~(\ref{eq:broadband_FC_matrix}) with the parameter values above. The trace labeled $S_{A1,A1}$ is the reflection off of mode $A_1$, and the trace labeled $S_{B4,A1}$ is the transmission through the circuit, equivalent to the conversion gain. Since this is a Chebyshev network, the bandwidth parameter corresponds to the extent of the ripple (0.01~dB in this case) rather than to the 3~dB points~\cite{pozar2009microwave}.

A circuit implementation for a Josephson parametric frequency converter with the above characteristics is shown schematically in Fig.~\ref{fig:broadbandFC}(b). The parametric coupler is a dc-SQUID with a total critical current of $I_0=2.74\;\mu$A ($L_J=120$~pH). The inductors $L_2$ and $L_3$ in Fig.~\ref{fig:broadbandFC}(b) are both 1.2~nH, resulting in resonator impedances $Z_2=44.8\;\Omega$ and $Z_3=58.9\;\Omega$ when the SQUID is biased to its operating point. We choose the impedances of the other resonators $Z_1$ and $Z_4$ to equal $35\;\Omega$ and $45\;\Omega$ respectively, so that $L_1=Z_1/\omega_A=1.11\;$nH and $L_4=Z_4/\omega_B=1.02\;$nH. Remember that we have quite a bit of flexibility in choosing the resonator impedances\textemdash these should be chosen in accordance with the fabrication process capabilities, and the values here are merely for demonstration purposes.  

With the resonator impedances now set, we use Eqs.~(\ref{eq:J_01})-(\ref{eq:J_nn1}) to calculate the values of the admittance inverters. $J_{23}$ is implemented as the parametric coupler and its value determines the pump amplitude. The remaining inverters are implemented as capacitive $\pi$-sections, Fig.~\ref{fig:inverters_pi}(b), and from their values we calculate the capacitances $C_{01}=212\;$fF, $C_{12}=43\;$fF, $C_{34}=17\;$fF, and $C_{45}=110\;$fF. Finally, capacitors $C_1=0.675\;$pF, $C_2=0.667$~pF, $C_3=0.369$~pF, and $C_4=0.384\;$pF are calculated similarly to Sec.~\ref{sec:filter_example}.

\begin{figure}[th]
    \centering
    \includegraphics[width=\columnwidth]{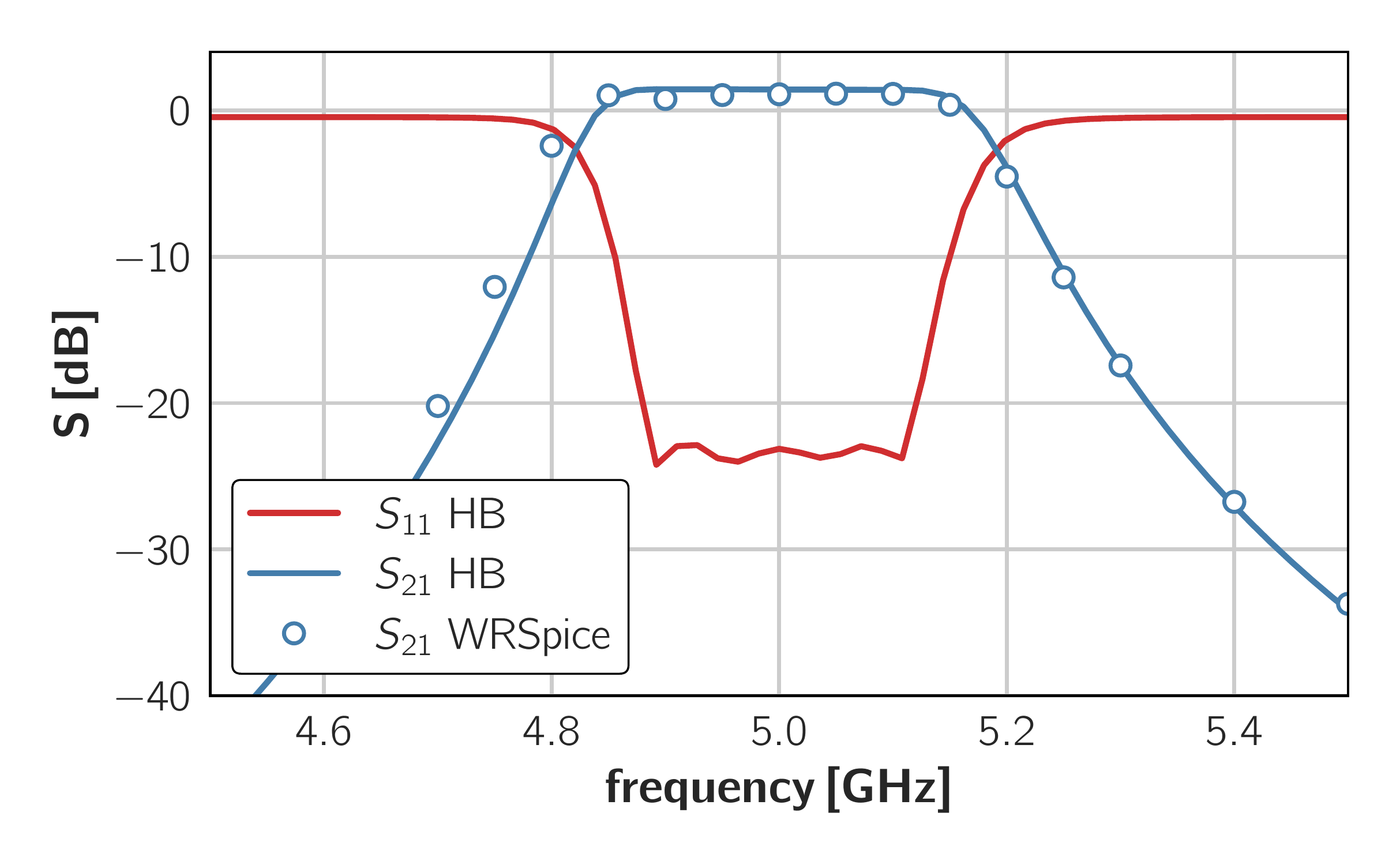}
    \caption{Simulated S-parameters of a matched parametric converter vs frequency referenced to the signal frequency. Solid lines are harmonic-balance simulations in Keysight ADS, circles are data from simulation in WRSpice. S$_{21}>0$~dB in the figure is a consequence of the Manley-Rowe relations \cite{manley1956some}, requiring power gain in an up-converting process that has unity photon-gain.\label{fig:broadband_FC_spice}}
\end{figure}

Figure~\ref{fig:broadband_FC_spice} shows S-parameters from circuit simulations of Fig.~\ref{fig:broadbandFC}(b) with the component values given above. The solid lines represent $S_{11}$ (red) and $S_{21}$ (blue) from harmonic balance simulation in Keysight ADS, where the dc-SQUID parametric coupler is modeled as a numerically-pumped inductance $L=L_J/\cos{\pi\phi}$ (see Appendix \ref{apx:JJSQUID}), with $\phi=\phi_{dc}+\phi_{ac}\cos(\omega_Pt)$, $\phi_{dc}=0.224$, $\phi_{ac}=0.18$, and $\omega_P/2\pi=2\;$GHz. Circles represent $S_{21}$ data from a simulation in WRSpice \cite{wrspice} that includes a full model of the Josephson junctions. In this simulation the SQUID is biased with $\Phi_{dc}=0.224\Phi_0$ as above, and pumped at  $\omega_P/2\pi=2\;$GHz with $\Phi_{ac}=0.2\Phi_0$.

The simulations are in good mutual agreement, and both are in reasonable agreement with the calculated S-parameters in Fig.~\ref{fig:broadband_FC_sparams}. Notably, both circuit simulations (Fig.~\ref{fig:broadband_FC_spice}) show conversion gain of $\approx 1.1\;$dB, consistent with a predicted $G=10\log_{10}(\omega_B/\omega_A)=1.46\;$dB based on the Manley-Rowe relations~\cite{Matthaei1961, manley1956some}, whereas S-parameters calculated using Eq.~(\ref{eq:s_params_components}) (Fig.~\ref{fig:broadband_FC_sparams}), yield a maximum $G = 0$ dB. The latter result reflects the expected conservation of photon number under ideal frequency conversion.

\subsection{Broadband parametric circulator}\label{sec:all_toghether_circ}
The parametric circulator as described by Ref.~\onlinecite{lecocq2017nonreciprocal} consists of three modes connected pairwise via parametric frequency conversion processes. One of the pumps can have a phase of $\pi/2$ with respect to the others, giving rise to circulation. The circuit was already discussed in Sec.~\ref{sec:graph_examples}.

\subsubsection{Matching strategy}
To broadband match the parametric circulator we will adopt the ideas of Anderson~\cite{anderson1967}, which were originally applied to ferrite circulators. We will approximate the working circulator as a single effective $LCR$ resonator, whose quality factor depends on parametric coupling strengths $\beta$ in the 3-mode `core', Fig.~\ref{fig:cir_graph_matrix}(a). We will then design a network to match this effective resonated load following Sec.~\ref{sec:resonated_load}, by finding the values of $\beta$ that satisfy the decrement condition Eq.~(\ref{eq:decrement}) for our choice of network prototype. 

In general, each port of the 3-mode core circuit can be outfitted with a different matching network (having a different prototype, number of sections, or bandwidth), however, here we choose to illustrate the design concepts with a simple, symmetric design, which can be fully worked-out analytically. If all ports have equal coupling to the environment $\gamma_A=\gamma_B=\gamma_C$ and all parametric coupling strengths are equal, the circulator circuit has cyclic symmetry, suggesting that the analysis that follows can be done equivalently on either of the bare circulator modes, and that the same matching network parameters (bandwidth and prototype coefficients) can be used on all ports.

The coupling matrix of the bare circulator, Fig.~\ref{fig:cir_graph_matrix}, with all coupling strengths set to equal $\beta_c$ (as in Sec.~\ref{sec:graph_example_circulator}, we factor out the phase of the coupling and write it out explicitly, therefore $\beta_c$ is real below) and equal port dissipation rates, is:
\begin{equation}\label{eq:bare_Mc}
    \mathbf{M}_c=
    \begin{bmatrix}
        \Delta_{A1} & i\beta_c & \beta_c \\
        -i\beta_c & \Delta_{B1} & \beta_c \\
        \beta_c & \beta_c & \Delta_{C1}
    \end{bmatrix}.
\end{equation}

\begin{figure}[th]
    \centering
    \includegraphics[width=\columnwidth]{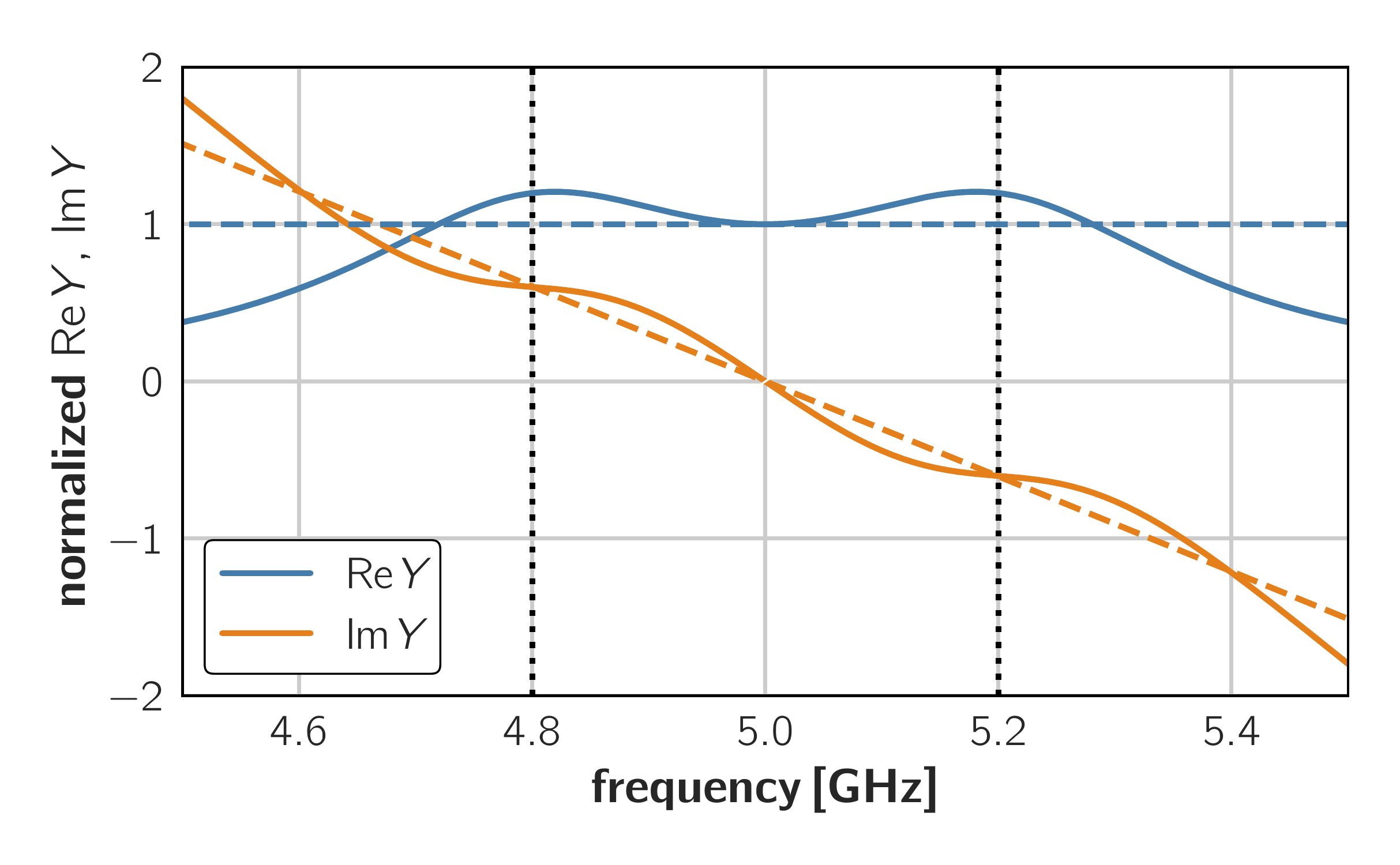}
    \caption{\label{fig:circ_RLC_apporx}Real and imaginary parts of the normalized port admittance (solid), along with their approximation (dashed) for the purpose of matching circuit design. Vertical dashed lines indicate detunings of $\pm\gamma_0/2$.}
\end{figure}

Using Eq.~(\ref{eq:port_admittance}) we can calculate the admittance seen from any of the ports. For example the admittance seen from port A is:
\begin{equation}
    \frac{Y_A}{Y_0}=-2i\frac{\Delta_{A1}\Delta_{B1}\Delta_{C1}-\beta^2_c\left(\Delta_{A1}+\Delta_{B1}+\Delta_{C1}\right)}{\Delta_{B1}\Delta_{C1}-\beta^2_c}-1,
\end{equation}
whose real part at the center of the band $\omega=\omega_A$ is: 
\begin{equation}
    \mathrm{Re}\left\{\frac{Y_A}{Y_0}\right\}=\frac{8\beta^2_c}{1+4\beta^2_c}.
\end{equation}
We approximate the slope of the imaginary part of the admittance using the line connecting the points $\mathrm{Im}\{\frac{Y_A(\omega_A\pm\gamma_0/2)}{Y_0}\}$, \ie, the average slope over the target frequency band (see Fig.~\ref{fig:circ_RLC_apporx}). From here, using Eq.~(\ref{eq:loaded_res_Q}) we find the quality factor of the effective resonated load presented at the signal port as a function of the coupling strength $\beta_c$:
\begin{equation}\label{eq:circ_effective_Q}
    Q=\frac{\omega_A}{\gamma_0}\frac{12\beta^3_c-6\beta^2_c+2\beta_c+1}{8\beta^2_c}
\end{equation}
If the network that we choose to match the circulator with is symmetric, such that its prototype coefficients satisfy $g_0g_1=g_Ng_{N+1}$, then Eq.~(\ref{eq:circ_effective_Q}), together with Eq.~(\ref{eq:correspond_source}) for $\gamma_0$, and the decrement relation Eq.~(\ref{eq:decrement}), result in $12\beta^3_c-14\beta^2_c+2\beta_c+1=0$, of which $\beta_c=0.5$ is a root. If the network is not symmetric then $\beta_c$ can be found numerically in a similar fashion. The normalized admittance of the bare circulator with $\beta_c=0.5$ is shown in Fig.~\ref{fig:circ_RLC_apporx} (solid lines) along with the resonated load approximation (dashed lines), for $\gamma_0/2\pi=400\;$MHz.

We are now ready to embed the bare circulator in a matching network. We will choose a 3-pole Chebyshev prototype with 0.01 dB ripple, with coefficients~\cite{MYJ} $g_i=\{1.0,~0.6291,~0.9702,~0.6291,~1.0\}$, and build the network by adding two additional, passively-coupled modes to each port of the bare circulator. The resulting graph is shown in Figure \ref{fig:circ_circuit}(a), and the corresponding $9\times9$ EoM matrix can be written in the mode basis $\vec{v}=\left[A_3,~B_3,~C_3,~A_2,~B_2,~C_2,~A_1,~B_1,~C_1\right]$ in block form:
\begin{equation}\label{eq:circ_full}
    \mathbf{M}=
    \begin{bmatrix}
        \hat{\Delta}_3 & \hat{\beta}_{23} & \hat{0} \\
        \hat{\beta}_{23} & \hat{\Delta}_2 & \hat{\beta}_{12} \\
        \hat{0} & \hat{\beta}_{12} & M_c
    \end{bmatrix},
\end{equation}
where the $3\times3$ blocks are $\hat{\Delta}_k=\mathrm{diag}(\Delta_{Ak},~\Delta_{Bk},~\Delta_{Ck})$, $\hat{\beta}_{jk}=\beta_{jk}\mathbb{I}$, $M_c$ is the bare circulator matrix Eq.~(\ref{eq:bare_Mc}), and $\hat{0}$ is a $3\times3$ block of zeros. Choosing a bandwidth of 250 MHz and a center frequency of $\omega_A/2\pi=5\;$GHz, and using Eqs.~(\ref{eq:correspond_load}) and~(\ref{eq:correspond_beta}) we obtain $\gamma_0/2\pi=400\;$MHz, $\beta_{12}=\beta_{23}=0.403$, and $\beta_c=0.5$ was found above.

\begin{figure}[tbh]
    \centering
    \includegraphics[width=\columnwidth]{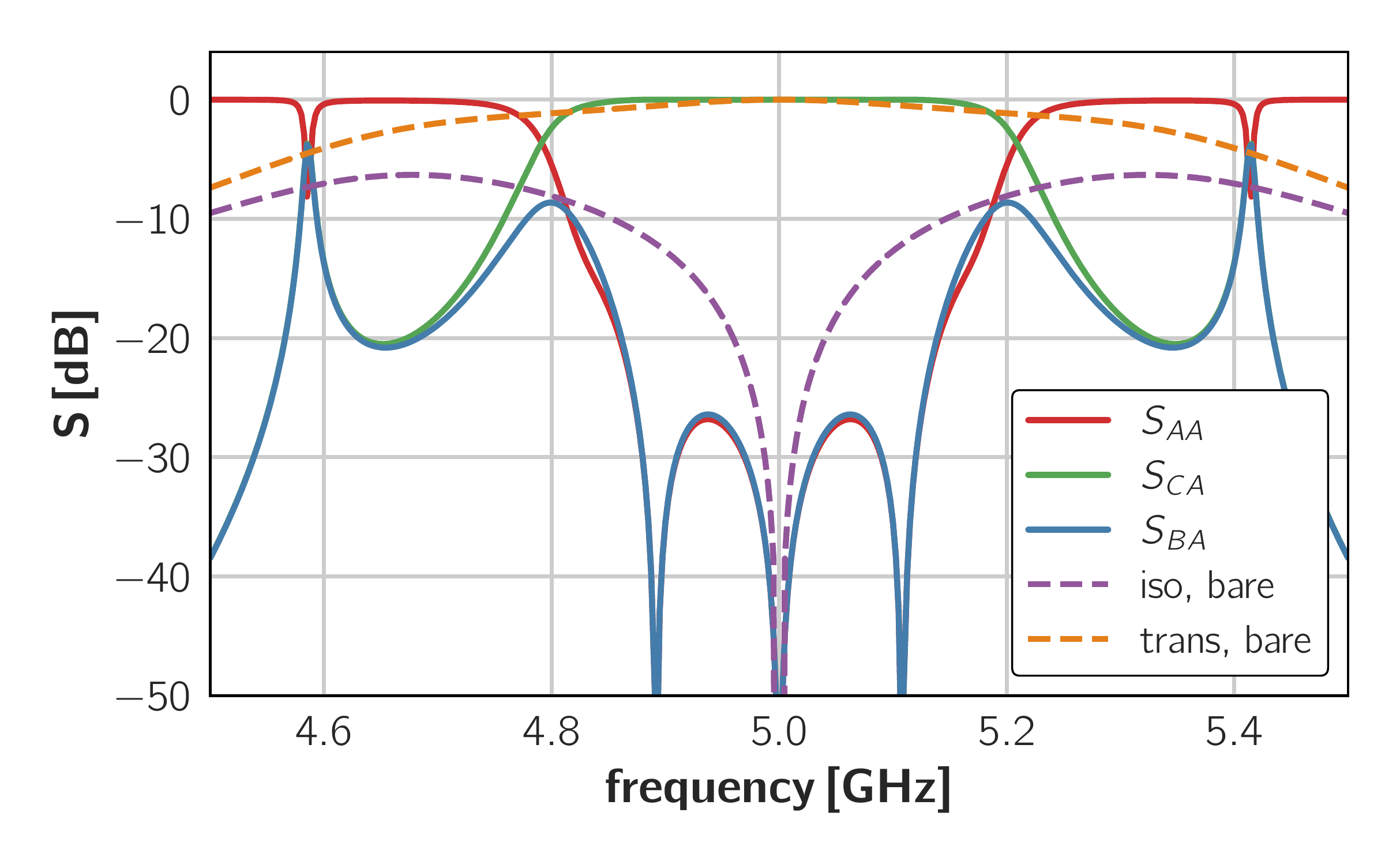}
    \caption{\label{fig:circ_sparams}S-parameters of a broadband matched parametric circulator. $S_{AA}$ (red) is the reflection from port A, $S_{CA}$ (green) and $S_{BA}$ (blue) are forward and reverse transmission, respectively. Dashed lines show the forward transmission and reverse isolation of a bare 3-mode circulator having the same $\gamma_0$}
\end{figure}

Figure \ref{fig:circ_sparams} shows an S-parameter calculation using Eq.~(\ref{eq:s_params_components}) with the matrix Eq.~(\ref{eq:circ_full}). The trace labeled $S_{AA}$ (red) is the reflection from port $A$, $S_{CA}$ (green) is the forward transmission at the direction of the circulation, from port $A$ to port $C$, and $S_{BA}$ (blue) is the reverse transmission from port $A$ to port $B$. We see that over the bandwidth of the device we get flat transmission with low insertion loss in the forward direction, and better than 20 dB reverse isolation. Fig.~\ref{fig:circ_sparams} also shows the transmission and reverse isolation (dashed lines) of a bare 3-mode circulator having the same $\gamma_0$, demonstrating the benefit of broadband-matching the circulator in improving the bandwidth of usable isolation. 

\subsubsection{Circuit implementation}
In Section~\ref{sec:graph_example_circulator}, we mentioned that the minimal construction of a parametric circulator requires only one of the modes to be at a different frequency than the others. For the circuit implementation of the matched circulator, we will take advantage of this feature, and set the frequencies of both the $A$ and $C$ modes to 5 GHz, while the $B$ modes frequency is set to 7 GHz. We will design with the same prototype as above, and with the same bandwidth of 250 MHz.

\begin{figure*}[htb]
    \centering
    \includegraphics[width=\textwidth]{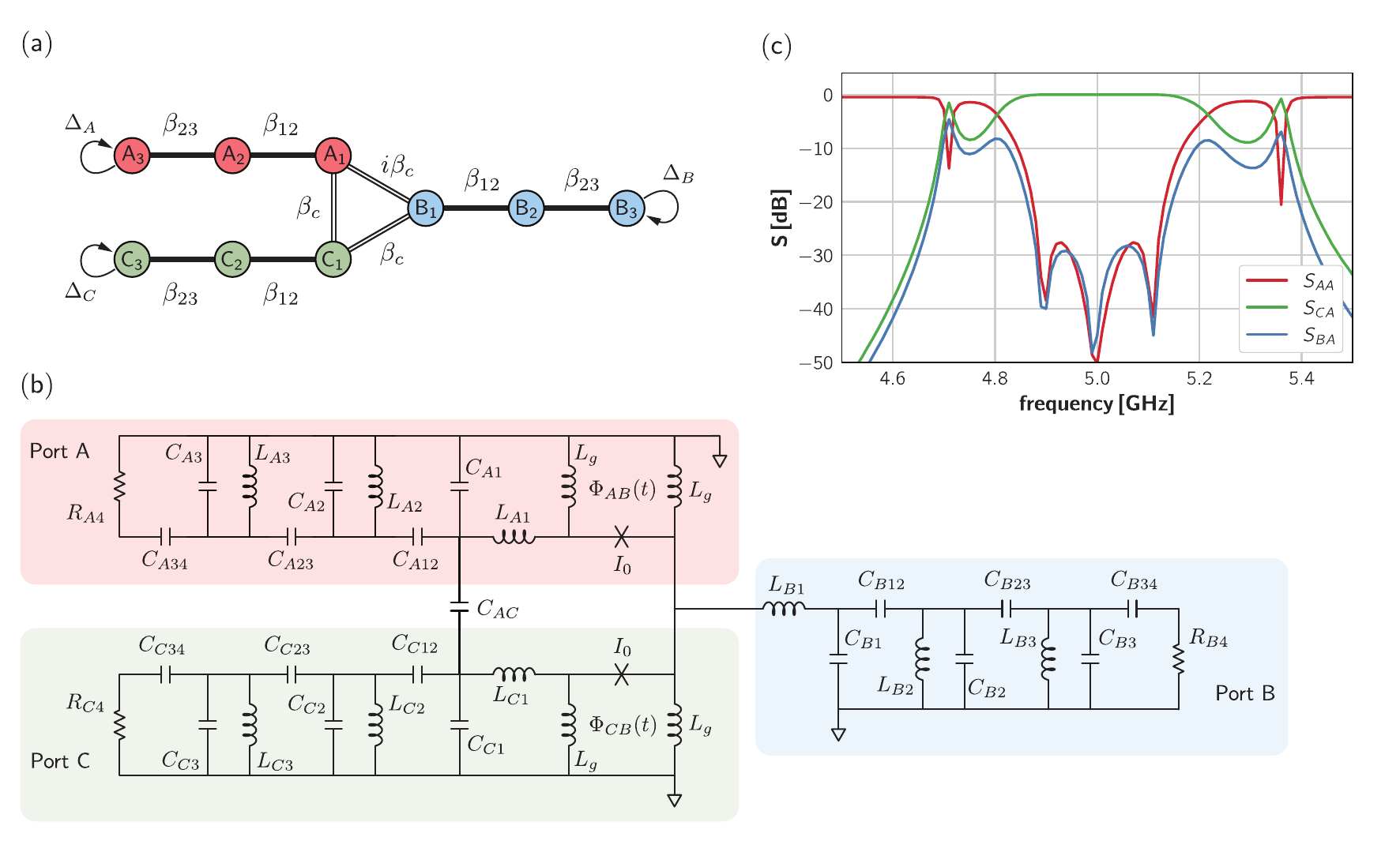}
    \caption{\label{fig:circ_circuit} (a) Coupled mode graph of a matched parametric circulator. (b) Circuit implementation of a Josephson parametric circulator with ports A and C operating at 5 GHz, and port B operating at 7 GHz. Component values are given in Table~\ref{tab:circ_components}. (c) S-parameters from harmonic balance simulation. $S_{AA}$ (red) is the reflection from port A, $S_{CA}$ (green) and $S_{BA}$ (blue) are forward and reverse transmission, respectively.}
\end{figure*}

\begin{table}[hbt]
\caption{Values of circuit components appearing in Fig.~\ref{fig:circ_circuit}(b), and used in the simulation shown in Fig.~\ref{fig:circ_circuit}(c) \label{tab:circ_components}}
\begin{ruledtabular}
\begin{tabular}{cdcd}
\textrm{Capacitors}&
\multicolumn{1}{c}{\textrm{Value (pF)}}&
\textrm{Inductors}&
\multicolumn{1}{c}{\textrm{Value (nH)}}\\
\colrule
$C_{A1},~C_{C1}$ & 4.515 & $L_{A1},~L_{C1}$ &  0.05 \\
$C_{A2},~C_{C2}$ & 0.845 & $L_{A2},~L_{C2}$ & 0.955\\
$C_{A3},~C_{C3}$ & 0.777 & $L_{A3},~L_{C3}$ & 0.955 \\
$C_{B1}$ &2.886 & $L_{B1}$ & 0.100 \\
$C_{B2}$ &0.655 & $L_{B2}$ & 0.682 \\
$C_{B3}$ &0.590 & $L_{B3}$ & 0.682 \\
$C_{AC}$ & 0.403 & $L_g$ & 0.150 \\
$C_{A12},~C_{C12}$ & 0.148 & &\\
$C_{A23},~C_{C23}$ & 0.068 & &\\
$C_{A34},~C_{C34}$ & 0.249 & &\\
$C_{B12}$ & 0.068 & &\\
$C_{B23}$ & 0.035 & &\\
$C_{B34}$ & 0.147 & & 
\end{tabular}
\end{ruledtabular}
\end{table}

The circuit schematic is shown in Fig.~\ref{fig:circ_circuit}(b). The parametric couplings between modes $A1$ and $B1$, and between modes $C1$ and $B1$, are implemented here using rf-SQUID couplers~\cite{chen2014gmon, naaman2016on-chip}. The coupling between modes $A1$ and $C1$ is now passive since these modes are resonant, and can be implemented using a capacitive admittance inverter whose value can be calculated from the $\beta_c = 0.5$ we found in the previous section, using Eq.~(\ref{eq:beta_to_J}):
\begin{equation}
    C_{AC}=\frac{J_c}{\omega_A}=\frac{2\gamma_0\beta_c}{\omega^2_A\sqrt{Z_{A1}Z_{C1}}}.
\end{equation}
For the circuit in Fig.~\ref{fig:circ_circuit}(b), we have $Z_{A1}=Z_{C1}=6.283\;\Omega$, giving $C_{AC}=0.403\;$pF. The rest of the elements are calculated in a similar fashion to what we have done in Sec.~\ref{sec:all_together_converter}. All component values are listed in Table.~\ref{tab:circ_components}. 

Fig.~\ref{fig:circ_circuit}(c) shows the results of a harmonic balance simulation with the rf-SQUIDs modeled as numerically pumped mutuals $M=M_0+\delta M\cos(\omega_Pt)$. Both rf-SQUIDs are assumed to be biased such that they provide zero passive mutual coupling, $M_0=0$, and both are pumped at $\omega_P/2\pi=2\;$GHz so that the mutual is modulated by $\delta M=20\;$pH, and the phase of the drive to the $C-B$ coupler is $87^\circ$ with respect to that of the $A-B$ coupler. In the simulation, an adjustment of the $C-B$ coupler pump phase away from the ideal $90^\circ$ was used to compensate for parasitic frequency-dependence of the capacitor networks implementing the admittance inverters. 

The S-parameters from circuit simulations are in reasonable agreement with the calculated S-parameters in Fig.~\ref{fig:circ_sparams}, which is remarkable considering the vastly different methods and tools used to obtain these results. This agreement gives us confidence in the foundational relations that we established here between the coupling matrix language and that of band-pass network synthesis.

\subsection{Matched Josephson parametric amplifiers}\label{sec:all_together_paramp}
A matched degenerate Josephson parametric amplifier based on filter design techniques was already demonstrated in Ref.~\onlinecite{naaman2019high}. In Ref.~\onlinecite{roy2015broadband}, Roy \textit{et al.} derived a matching circuit for a Josephson parametric amplifier from first principles rather than using techniques from microwave engineering; in Section~\ref{sec:vijay} we show that their result is equivalent to a 2-pole Butterworth matching circuit design.

The Josephson parametric amplifier, like its varactor-based predecessors~\cite{Matthaei1961}, is a negative-resistance amplifier in which the pumped nonlinear element\textemdash typically a SQUID\textemdash presents the signal port with a negative effective resistance \cite{sundqvist2014negative}: this is the admittance of the idler circuit as transformed through the anti-conjugating admittance inverter of Eq.~(\ref{eq:pa_admittance_inverter}), as illustrated by Eq.~(\ref{eq:pumpistor_Y}). The stray linear shunt inductance of the SQUID, $L'_0$ in Fig.~\ref{fig:LequivInverters}(b), is resonated by adding a shunt capacitance. Therefore the amplifier can be matched using the techniques of Sec.~\ref{sec:resonated_load} for matching resonated loads. Observe, however~\cite{Matthaei1961, MYJ}, that the reflection (voltage) gain off of a negative resistance $-|R_{PA}|$, $\sqrt{G}=\frac{-|R_{PA}|-Z_0}{-|R_{PA}|+Z_0}$, is equivalent to the inverse of the usual reflection coefficient off of a positive resistance $|R_{PA}|$, $\rho=\frac{|R_{PA}|-Z_0}{|R_{PA}|+Z_0}$, so that $\sqrt{G}=1/\rho$. Therefore, when we design the matching circuit we will design for a finite reflection\textemdash an engineered mismatch\textemdash that is either flat or has specified ripple over the band when the network is terminated by $|R_{PA}|$. This is a different requirement than what is used in typical filter and matching networks (where the goal is to approximate $\rho=0$), so we cannot use the usual network prototypes designed for those situations. Appendix~\ref{apx:cauer_synthesis} shows how to calculate suitable prototypes based on Butterworth and Chebyshev response characteristics. Tables of Chebyshev prototypes for matching negative-resistance amplifiers can also be found in Ref.~\onlinecite{Getsinger1963}.

As already mentioned, physical realizations of immittance inverters carry their own parasitic frequency dependence; this is true as well for the pumped nonlinear element, \ie, $|R_{PA}|$ is some function of frequency, see for example Eq.~(\ref{eq:pumpistor_Y}). We proceed here to use the value of $|R_{PA}|$ at the center of the band, with the understanding that these methods are best suitable for designs with fractional bandwidths of up to $\approx20$\%.

\subsubsection{Broadband non-degenerate amplifier}\label{sec:broadband_paramp}
Here we will design a non-degenerate parametric amplifier with a bandwidth of 500 MHz, a signal band centered at 5 GHz, an idler band centered at 7 GHz, and using a 3-pole Chebyshev prototype designed for 20 dB signal gain and 0.5 dB gain ripple. From Appendix~\ref{apx:cauer_synthesis} (Table~\ref{tab:cheby_amp_coefs}) we calculate the prototype coefficients:
\begin{equation}
    \{g_0\dots g_4\}=\{1.0,~0.5899,~0.6681,~0.3753,~0.9045\},\nonumber
\end{equation}
where $g_0$ correspond to the amplifier's negative resistance and $g_4$ is the $50\;\Omega$ port. Both signal and idler circuits will use the same prototype\textemdash this is not necessary, but is convenient.

\begin{figure}
    \centering
    \includegraphics[width=\columnwidth]{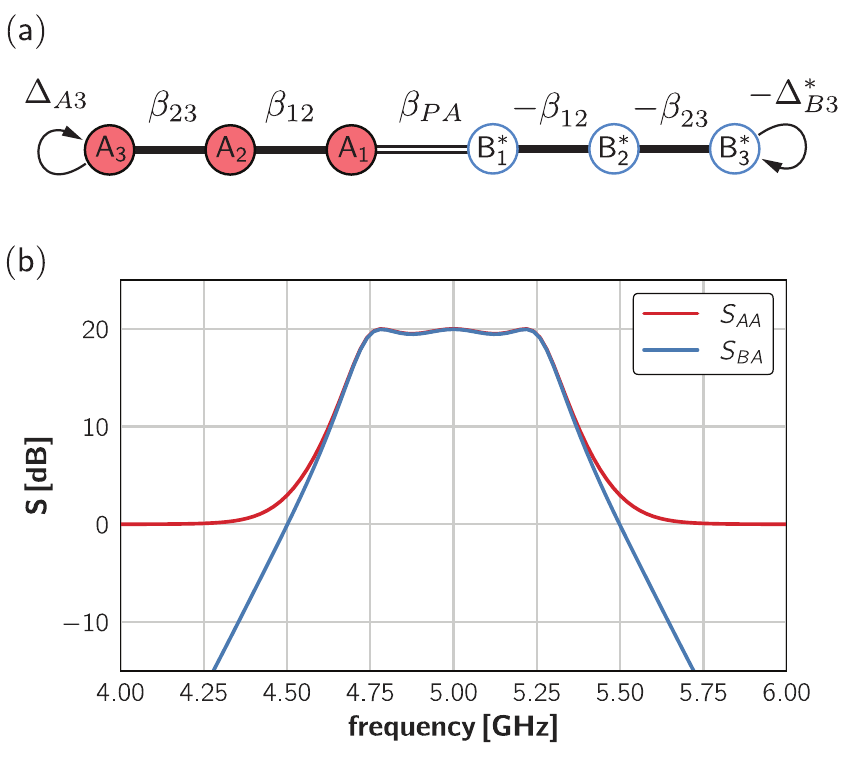}
    \caption{(a) Coupled-mode graph for a matched non-degenerate parametric amplifier, (b) S-parameters for a matched parametric amplifier. $S_{AA}$ is the signal reflection gain, and $S_{BA}$ is the signal-idler trans gain.}
    \label{fig:NDPA_graph}
\end{figure}

Figure \ref{fig:NDPA_graph}(a) shows the coupled mode graph for the amplifier, where the conjugated $B^*$ modes are distinguished by their open face shading. Modes $A_1$ and $B_1^*$ are coupled via a parametric coupler pumped at the sum frequency $\omega_P=\omega_A+\omega_B=12\;$GHz. All other couplings between same-frequency modes are passive, and modes $A_3$ and $B^*_3$ are connected to ports.

The $6\times6$ EoM matrix is tri-diagonal in the mode basis $\vec{v}=\left[A_3,~A_2,~A_1,~B_1^*,~B_2^*,~B_3^*\right]$, and can be written by inspection using the anti-conjugation rules in Fig.~\ref{fig:matrix_rules} for the $B$ (idler) modes: 
\begin{equation}\label{eq:PA_matrix}
    \mathbf{M}_{PA}=
    \begin{bmatrix}
        \Delta_{A3} & \beta_{23} & 0 & 0 & 0 & 0\\
        \beta_{23} & \Delta_{A2} & \beta_{12} & 0 & 0 & 0\\
        0 & \beta_{12} & \Delta_{A1} & \beta_{PA} & 0 & 0\\
        0 & 0 & -\beta^*_{PA} & -\Delta^*_{B1} & -\beta_{12} & 0\\
        0 & 0 & 0 & -\beta_{12} & -\Delta^*_{B2} & -\beta_{23}\\
        0 & 0 & 0 & 0 & -\beta_{23} & -\Delta^*_{B3}
    \end{bmatrix}
\end{equation}

From the network prototype coefficients $g_i$ we calculate the coupling terms for the passive elements, $\beta_{23}=0.339$ and $\beta_{12}=0.270$ using Eq.~(\ref{eq:correspond_beta}). The parametric coupling term $|\beta_{PA}|=0.287$ can be found using Eq.~(\ref{eq:beta_pump}) in Section~\ref{sec:pump_amplitude}. The port coupling rates are $\gamma_0/2\pi=1.476\;$GHz.

Fig.~\ref{fig:NDPA_graph}(b) shows S-parameters calculated with Eq.~(\ref{eq:s_params_components}) and the matrix in Eq.~(\ref{eq:PA_matrix}) with the parameters above. In the figure, $S_{AA}$ is the signal reflection gain from port $A_3$ and $S_{BA}$ is the idler trans-gain at port $B_3^*$.

Circuit implementations will vary based on the coupler architecture (see Appendix~\ref{apx:para_inverter}), but after fixing the impedances of the resonators embedding the coupler, all other circuit components can be readily calculated using the network prototypes and Eqs.~(\ref{eq:J_jk}) and~(\ref{eq:J_nn1}), as was demonstrated in detail in the preceding sections.

\subsubsection{Analysis of Roy \textit{et al.}}\label{sec:vijay}
\begin{figure}
    \centering
    \includegraphics[width=\columnwidth]{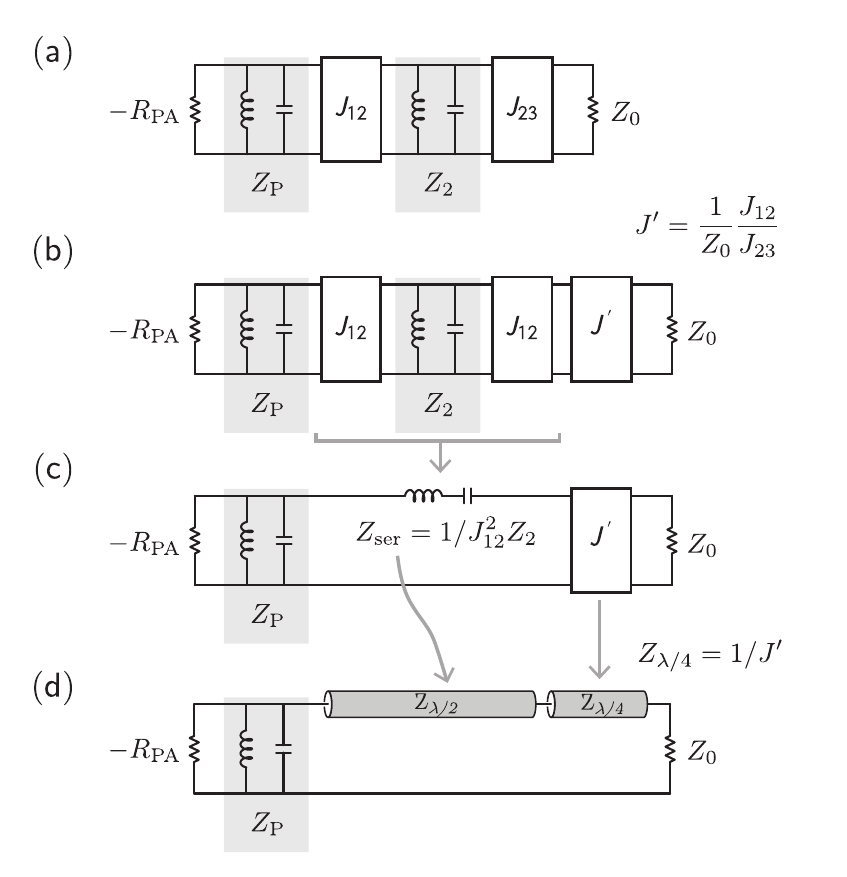}
    \caption{\label{fig:transformation}Transformation used to go from shunt coupled resonators to ladder configuration with series auxiliary mode (a-c), and finally to the transmission-line topology of Ref.~\onlinecite{roy2015broadband} (d). The pumped junction is represented by a negative resistance $-R_{PA}$.}
\end{figure}

Roy \textit{et al.}~\cite{roy2015broadband} were the first to describe a Josephson parametric amplifier with a synthesized impedance matching network. A schematic of the circuit is shown in Figure~\ref{fig:transformation}(d), where the pumped capacitively-shunted SQUID is represented by the negative resistance $-R_{PA}$ in parallel with the resonator $Z_p$. Broadband impedance matching of the device is accomplished via a series combination of two transmission line resonators, as shown in the figure. The simplicity of this particular implementation makes it amenable to fabrication with modest resources, while producing relatively wide bandwidth. The basic design has been recently reproduced in Refs.~\cite{grebel2021fluxpumped, duan2021broadband} albeit with small modifications. Despite its approachability, the design parameters were derived using a physics approach, and modifications to the gain, bandwidth and center frequency are less straightforward to produce \cite{grebel2021fluxpumped}.

Here, we will show that the design of Ref.~\onlinecite{roy2015broadband} can be derived from a synthesized 2-pole network with Butterworth characteristics. The plan is to start with a 2-pole matching network prototype suitable for negative resistance amplifiers. One of the poles is the resonated SQUID, the other pole is what Ref.~\onlinecite{roy2015broadband} calls the ``auxiliary resonator." We will first show how to obtain design parameters for a ladder circuit as shown in Fig.~\ref{fig:transformation}(c), and then transform it to the circuit topology of Ref.~\onlinecite{roy2015broadband}, Fig.~\ref{fig:transformation}(d).

For a given center frequency, $\omega_0$, and bandwidth, $\Delta\omega$, the design is fully constrained by the inductance $L_p$ of the Josephson circuit, in this case a dc SQUID, at the operating flux bias point. This is resonated out with a shunt capacitor $C_p$, to match a target frequency $\omega_0$. The design in~\cite{roy2015broadband} had $\omega_0/2\pi=6$ GHz, $w=\Delta\omega/\omega_0\approx0.1$, and the capacitance shunting the SQUID was $C_p=3.4$ pF. From this, $Z_p=1/\omega_0C_p=7.8\,\Omega$.

We use a 2-pole max-flat (Butterworth) prototype with 20 dB gain, calculated according to Appendix~\ref{apx:cauer_synthesis} (Table~\ref{tab:butterworth_amp_coefs})
\begin{equation}\label{eq:roy_gs}
    \{g_0,\dots, g_3\}=\{1.0,\,0.408,\, 0.234,\, 1.106\},
\end{equation}
where $g_0$ is the pumped junction and $g_3$ is the $50\,\Omega$ load. To obtain the desired ladder topology of Fig.~\ref{fig:transformation}(c), it is natural to use the method of Sec.~\ref{sec:denorm}. From Eq.~(\ref{eq:denorm_parallel}) we have:
\begin{equation}
    Y_p=g_1\frac{\omega_0}{\Delta\omega}Y_\mathrm{ref},
\end{equation}
where $Y_\mathrm{ref}$ is the reference admittance. Usually we will have $Y_\mathrm{ref}=Y_0=(50\;\Omega)^{-1}$, however here, since $Y_p=1/Z_p$ is constrained, we will use this equation instead to find the appropriate reference impedance for the network,
\begin{equation}\label{vijay_ref_z}
    Z_\mathrm{ref}=\frac{1}{Y_\mathrm{ref}}=g_1\frac{\omega_0}{\Delta\omega}Z_p=31.8\;\Omega.
\end{equation}
Using $Z_\mathrm{ref}$ above in Eq.~(\ref{eq:denorm_series}), we can now find the impedance of the series ``auxiliary" resonator:
\begin{equation}\label{eq:roy_zres}
    Z_\mathrm{ser}=g_2\frac{\omega_0}{\Delta\omega}Z_\mathrm{ref}=74.4\;\Omega.
\end{equation}
Finally the load impedance, Eq.~(\ref{eq:denorm_load}), is denormalized to $Z_\mathrm{L}=Z_\mathrm{ref}/g_3=28.8\;\Omega$. To transform $Z_\mathrm{L}$ to the $Z_0=50\;\Omega$ environment, we insert an admittance inverter, $J'$ in Fig.~\ref{fig:transformation}(c), whose value is 
\begin{equation}\label{eq:roy_Jprime}
    J'=1/\sqrt{Z_\mathrm{L}Z_0}=0.0264\;\Omega^{-1}.
\end{equation}

For completeness, before proceeding to the implementation of the circuit shown in Fig.~\ref{fig:transformation}(d), we will show how the same circuit parameters can be derived using an approach based on the method in Sec.~\ref{sec:design_method}. We start from the same prototype coefficients and the constrained value of $Z_p$ above, and construct the circuit of Fig.~\ref{fig:transformation}(a), where the inverter values can be calculated in the usual way Eqs.~(\ref{eq:J_01}-\ref{eq:J_nn1}) and assuming some arbitrary value for $Z_2$. Next, we factor $J_{23}$ into a series combination of $J_{12}$ and $J'=\frac{1}{Z_0}\frac{J_{12}}{J_{23}}$, as shown in Fig.~\ref{fig:transformation}(b), and transform the shunt resonator $Z_2$ and the adjacent inverters into a series resonator, $Z_\mathrm{ser}=1/J^2_{12}Z_2=g_1g_2Z_p/w^2$, to arrive again at Fig.~\ref{fig:transformation}(c). The values of $Z_\mathrm{ser}$ and $J'$ found here are identical to those found above using the direct ladder synthesis.

Proceeding with the circuit implementation of Ref.~\onlinecite{roy2015broadband}, the series resonator was implemented as a half-wave transmission line resonator with impedance $Z_{\lambda/2}$, and the inverter $J'$ is implemented as a quarter wave transmission line with impedance $Z_{\lambda/4}$. The two transmission lines are in series with each other as shown in Fig.~\ref{fig:transformation}(d). 

The impedance $Z$ seen looking through the series resonator $Z_\mathrm{ser}$ towards a load $Z_\mathrm{L}=1/J'^2Z_0$ (Fig.~\ref{fig:transformation}(c)) near the resonance frequency, can be approximated as (writing $\omega=\omega_0+\delta\omega$):
\begin{equation}\label{eq:roy_series_res}
    Z=Z_\mathrm{L}+2jZ_\mathrm{ser}\frac{\delta\omega}{\omega_0}.
\end{equation}
Likewise, the impedance $Z'$ seen through a half wave transmission line with $Z_{\lambda/2}$ in series with a quarter-wave transmission line with $Z_{\lambda/4}$ (Fig.~\ref{fig:transformation}(d)) near the resonance frequency $\omega_0$ is
\begin{widetext}
    \begin{equation}\label{eq:roy_tlines}
        Z' \simeq \frac{Z_{\lambda/r}^2}{Z_0}+j\frac{\pi}{2}\frac{\delta\omega}{\omega_0}\left[Z_{\lambda/4}\left(1-\frac{Z_{\lambda/4}^2}{Z_0^2}\right)+2Z_{\lambda/2}\left(1-\frac{Z_{\lambda/4}^4}{Z_0^2Z_{\lambda/2}^2}\right)\right],
    \end{equation}
\end{widetext}
this is the basis for Eq.~(S40) in Ref.~\onlinecite{roy2015broadband} and Eq.~(2) in Ref.~\onlinecite{duan2021broadband}.

Equating Eq.~(\ref{eq:roy_series_res}) and (\ref{eq:roy_tlines}), we get the following quadratic equation for $Z_{\lambda/2}$ which we can solve to get the impedance value:
\begin{equation}\label{eq:roy_half_wave}
    Z_{\lambda/2}^2-Z_{\lambda/2}\left[\frac{2}{\pi}Z_\mathrm{ser}-\frac{1}{2}Z_{\lambda/4}\left(1-\frac{Z_{\lambda/4}^2}{Z_0^2}\right)\right]-\frac{Z_{\lambda/4}^4}{Z_0^2}=0
\end{equation}
where $Z_{\lambda/4} = 1/J'$ (Eq.~\ref{eq:roy_Jprime}), and $Z_\mathrm{ser}$ found in Eq.~(\ref{eq:roy_zres}).

Using the numerical values in Eq.~(\ref{eq:roy_zres}) and (\ref{eq:roy_Jprime}), we finally get $Z_{\lambda/2}=54.55\,\Omega$ and $Z_{\lambda/4}=37.93\,\Omega$. This agrees with the values of $Z_{\lambda/2}\approx 58\,\Omega$ and $Z_{\lambda/4}\approx 40\,\Omega$, reported in Ref.~\onlinecite{roy2015broadband}.

\begin{figure}
    \centering
    \includegraphics[width=\columnwidth]{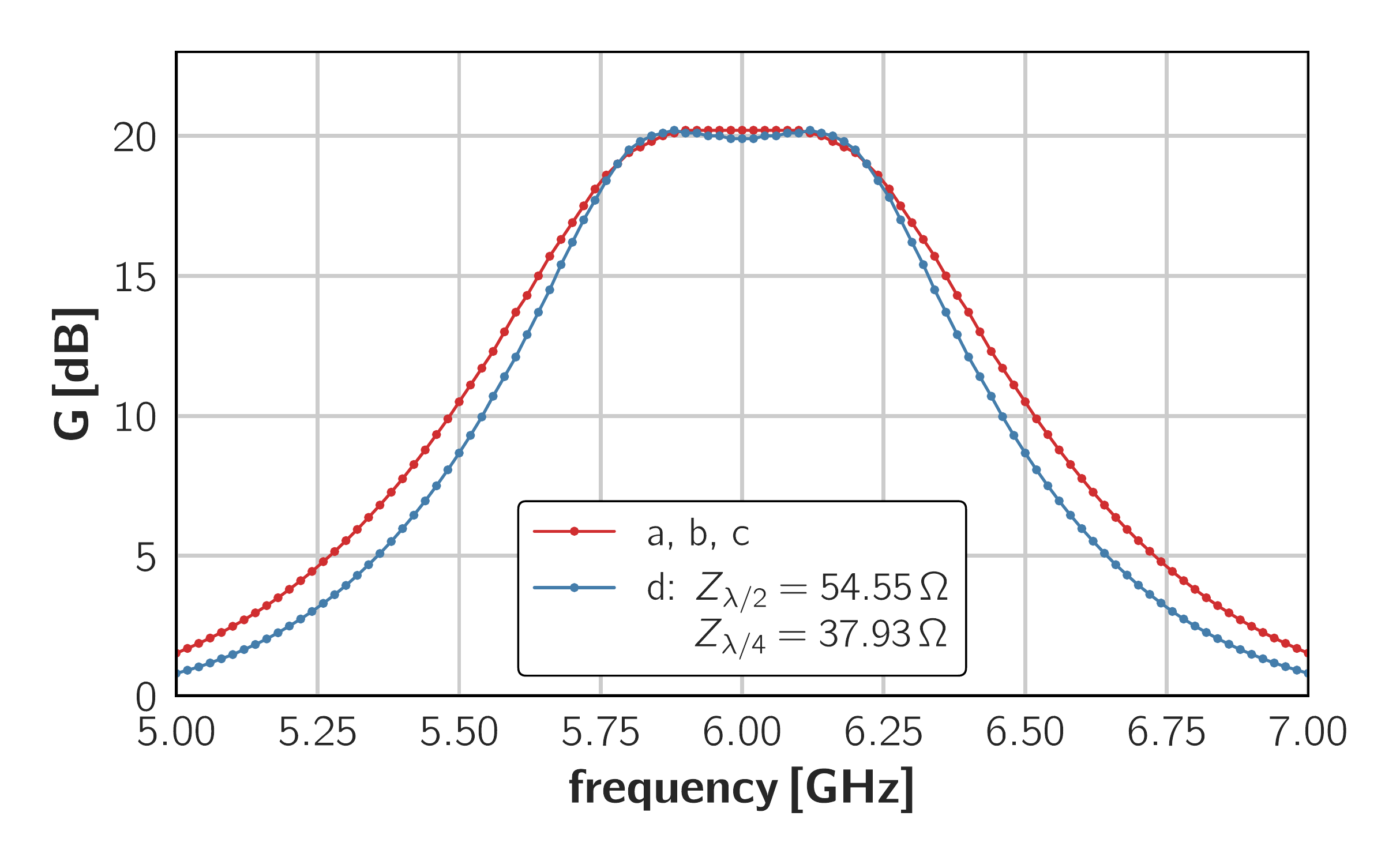}
    \caption{\label{fig:roy_sims}ADS harmonic balance simulations of the circuits in Fig.~\ref{fig:transformation}(a-c), and the transmission line implementations, Fig.~\ref{fig:transformation}(d), with the parameters indicated in the legend.}
\end{figure}

Figure \ref{fig:roy_sims} shows harmonic balance simulations using a symbolic nonlinear device model in Keysight ADS for the pumped SQUID. All simulation curves use the same flux bias $\Phi_\mathrm{dc}=0.3075\,\Phi_0$ (chosen to get the shunted SQUID resonance frequency to 6 GHz), and pump amplitude $\Phi_\mathrm{ac}=0.111\,\Phi_0$. The curve labeled (a-c) (red) corresponds to the schematics in Fig.~\ref{fig:transformation}(a)-(c), where the inverters are implemented as ideal Y-matrices. These circuits produce identical results showing that the transformations in Fig.~\ref{fig:transformation}(a)-(c) are equivalent. The blue trace (d) represents a half-wave/quarter-wave implementation as in Ref.~\onlinecite{roy2015broadband} corresponding to the schematic in Fig.~\ref{fig:transformation}(d), with the transmission line impedances as obtained in this section. The two traces should ideally be equivalent, however, it is not surprising that the trace (d) shows slightly narrower response than the transformations in (a-c), since it is using approximations that become worse with detuning away from $\omega_0$. The synthesized parameters and simulated response agree with what was reported in Ref.~\onlinecite{roy2015broadband}, showing that their matched broadband parametric amplifier design can be understood in terms of simple network synthesis methods known from microwave engineering.

\subsubsection{Calculating the parametric amplifier pump amplitude}\label{sec:pump_amplitude}
In Section \ref{sec:immittance_paramp} we saw that sum-frequency parametric coupling (amplification process) presents the input port with an effective negative resistance, resulting in gain. This negative resistance, as we have seen in Eqs.~(\ref{eq:pa_impedance_inverter}-\ref{eq:pumpistor_Y}) and Fig.~\ref{fig:LequivInverters}, depends on the pump amplitude.  Earlier in this Section, we designed matching networks for parametric amplifiers, assuming that the pump amplitude (or alternatively, the parametric coupling strength $\beta_p$) has been set correctly so that the negative impedance $-|R_{PA}|$ presented to the matching network satisfies the so-called decrement condition, Eq.~(\ref{eq:decrement}):
\begin{equation}\label{eq:apx_decrement}
    Q=|R_{PA}|/Z_1=g_0g_1/w,
\end{equation} 
where $g_0$ and $g_1$ are the prototype coefficients, $w$ is the fractional bandwidth, and $Z_1$ the impedance of the resonance embedding the modulated inductor. 

In the coupled-mode EoM matrix picture, the pump amplitude is expressed by the dimensionless parametric coupling strength $\beta_p$. We can calculate it as the normalized dissipation rate of resonator $Z_1$ due to the effective resistance $|R_{PA}|$: 
\begin{equation}\label{eq:beta_pump}
    \beta_p=\frac{\omega_0}{2\gamma_0}\frac{Z_1}{|R_{PA}|}=\frac{1}{2}\frac{g_Ng_{N+1}}{g_0g_1},
\end{equation}
where we have used $\gamma_0=\Delta\omega/g_Ng_{N+1}$, Eq.~(\ref{eq:correspond_load}), and the decrement relation, Eq.~(\ref{eq:apx_decrement}). 

As a simple check, for the 2-mode parametric amplifier of Fig.~\ref{fig:FCPA_graph_matrix}(c) with 20 dB gain, $N=1$ and, using $g_0=1.0$ and $g_2=0.9045$ calculated according to Appendix~\ref{apx:cauer_synthesis}, we get from Eq.~(\ref{eq:beta_pump}) that $\beta_p=\frac{g_2}{2g_0}=0.452$, which recovers the result of Eq.~(\ref{eq:paramp_graph_gain}).

In a circuit implementation, the pump amplitude is expressed via the parameter $\alpha$, defined in Figs.~\ref{fig:LequivInverters}(b) and~\ref{fig:MequivInverters}(b). Using the shunt representation in these figures, we can calculate $\alpha$ for a given prototype, signal power gain $G$, and fractional bandwidth $w$:
\begin{equation}\label{eq:alpha_pump}
    \alpha=\frac{\sqrt{G}-1}{\sqrt{G}+1}\left(\frac{w}{g_1}\right)^2\times \left\{
        \begin{aligned}
            g^2_{N+1};\;\;&N\in \mathrm{even}\\
            g^{-2}_{N+1};\;\;&N \in \mathrm{odd}
        \end{aligned},
    \right.
\end{equation}
as shown in Appendix~\ref{apx:para_inverter}.

Eqs.~\ref{eq:beta_pump} and~\ref{eq:alpha_pump} are useful in circuit design and simulation. Eq.~\ref{eq:alpha_pump} in particular should be used to ensure that the chosen physical implementation of the parametric coupler can deliver the required modulation strength. Knowledge of this quantity can also inform the design of the experimental setup, with respect to \eg, choice of pump generator, attenuation in the microwave cable delivering the pump, or on-chip coupling architecture. It can also serve as a starting point for experimentally finding the optimal pump power during the amplifier bring-up and calibration procedure.

\subsection{Design flow}\label{sec:design_flow}
As a summary, we outline a general design flow that we have found useful in prototyping and simulating parametrically coupled networks.
\begin{enumerate}
\item \textit{Draw the network graph.} \label{par:draw_graph} For a given required functionality, we find it useful to first sketch the coupling graph, Sec.~\ref{sec:graphs}. Doing so we can readily identify the required coupling processes, plan the mode basis, and assign mode frequencies and ports.
\item \textit{Calculate network prototype coefficients.}\label{par:coefs} Based on the desired response characteristics and the topology of the circuit, find the appropriate prototype coefficients $\{g_i\}$, either from tables when appropriate \cite{pozar2009microwave, MYJ}, or by direct calculation as outlined in Appendix~\ref{apx:cauer_synthesis}.
\item \textit{Assign coupling rates.}\label{par:rates} Given a bandwidth requirement for the circuit, calculate the port dissipation rates $\gamma_k$ using the prototype coefficients and Eqs.~(\ref{eq:correspond_source}) and~(\ref{eq:correspond_load}). Calculate the normalization rate $\gamma_0$ from Eq.~(\ref{eq:gamma0_def}), and the coupling terms $\beta_{jk}$ from Eq.~(\ref{eq:correspond_beta}). For networks involving parametric amplification, use Eq.~(\ref{eq:beta_pump}) to calculate $\beta_{PA}$ that corresponds to the sum-frequency pump, as was done in Sec.~\ref{sec:all_together_paramp}. 
\item \textit{Write down the EoM matrix.} From the network graph, write down the EoM matrix using the rules in Fig.~\ref{fig:matrix_rules}. This step, and indeed steps \ref{par:coefs} and \ref{par:rates} as well, are amenable to automation.
\item \textit{Calculate S-parameters.} The S-parameters of the design can now be calculated from the EoM matrix using Eq.~(\ref{eq:s_params_components}) to verify the desired behavior of the circuit. It is easy at this step to explore, for example, the effects of different pump phases and amplitudes, or different network prototype coefficients.
\item \textit{Determine parametric coupler implementation.} \label{par:implement_coupler} The configuration of the parametric coupler will constrain the rest of the circuit. In a Josephson junction circuit, we have several options for the coupler, including a shunt-connected dc-SQUID (Sec.~\ref{sec:all_together_converter}), an rf-SQUID coupler (Sec.~\ref{sec:all_toghether_circ}), or a balanced bridge \cite{abdo2013nondegenerate, naaman2017josephson}, each with their own advantages and ability to incorporate advanced nonlinear devices such as the SNAIL \cite{frattini2018optimizing} or SQUID arrays \cite{naaman2019high}. With the implementation fixed, construct resonators to embed the coupler, \eg~by shunting the Josephson element with capacitors to achieve the specified resonance frequency. The impedances of the resonators embedding the coupler will be important parameters in the following steps.  Use Eq.~(\ref{eq:alpha_pump}) or~(\ref{eq:alpha2}) to ensure the required modulation strength is consistent with the coupler design. 
\item \textit{Determine resonator implementation.}\label{par:implement_resonators} With the resonators embedding the parametric coupler determined in step \ref{par:implement_coupler}, next choose the implementation (\eg~lumped-element, quarter-wave, cavity) and impedances of the remaining resonators in the circuit based on the available technology. 
\item \textit{Calculate immittance inverters.} With all resonator impedances now known, use Eqs.~(\ref{eq:J_01})-(\ref{eq:J_nn1}) to calculate the values of the admittance $J$ inverters. If the resonators are not lumped, or the implementation calls for $K$ inverters instead, refer to \cite{MYJ} for alternative expressions for inverter values. 
\item \textit{Determine inverter implementation.}\label{par:implement_inverters} For $J$ inverters in a lumped-element circuit, choose either a capacitive or inductive circuit (Fig.~\ref{fig:inverters_pi}) for each of the inverters, and calculate the component values. Capacitive implementations are favorable in Josephson circuits as they do not introduce additional superconducting loops. Inductive implementations can be advantageous in circuits that require transition between single-ended and balanced sections \cite{naaman2017josephson}.  Other implementations are more appropriate for $K$ inverters or for distributed components; these can be found in \cite{pozar2009microwave, collin2007foundations, MYJ}.
\item \textit{Calculate component values.} Calculate resonator inductance and capacitance values based on the resonator impedance and frequency, and absorb the negative reactances of the adjacent inverters, as demonstrated in Sec.~\ref{sec:filter_example}. It is important to check at this point that all component values are realizable in the given technology; iterate through steps \ref{par:implement_resonators}-\ref{par:implement_inverters} to fix any issues, as we have flexibility to adjust the resonator impedances.
\item \textit{Perform circuit simulations.} With all component values now known, simulate the circuit in a tool such as Keysight ADS, AWR Microwave Office, or WRSpice using harmonic balance or transient analyses. The parametric couplers can be approximated as a numerically modulated inductance or mutual inductance, varying at the pump frequency. This step is important as it will reveal any issues stemming from, \eg, the fact that the inverters are not ideal and their reactances are frequency-dependent. Most tools allow for manual trimming of component values in real-time, and this functionality can be used to manually compensate for nonidealities in components. The designer should also explore here the effects of component variations and parasitics on the performance of the device.
\end{enumerate}

At the completion of the above steps, the design should be ready for layout and manufacture. As noted, many of these steps can be automated based on the rules and equations given in this tutorial, allowing the designer to focus their creative effort on steps \ref{par:draw_graph}, \ref{par:implement_coupler}, and the physical design.

\section{Conclusion and Outlook}
The design of parametrically coupled networks and that of microwave filters both boil down to engineering the mutual coupling rates between resonant modes, and between these modes and the environment. These disciplines, however, approach this problem with seemingly different language, tools, and methods. In this tutorial we have shown that, in fact, a common language can be used to describe both electrical filter networks and parametric networks, regardless of their physical implementation. 

We have shown that given prescribed response characteristics for a 1D-connected system of $N$ modes, which can be fully described by $N+2$ prototype coefficients, the coupling rates in the system can be calculated from these coefficients to realize the desired response. The correspondence we established between the $\beta$ (coupling efficiency) and $\gamma$ (dissipation rate) of the coupled-modes language, to the admittance (impedance) inverters $J$ $(K)$ of the filter synthesis language, offers agility in designing systems that may contain a mix of both passive and parametric couplings. Techniques developed for matching more complicated electrical systems, like ferrite circulators and amplifiers, as well as techniques for designing with internally lossy components, can now be readily used in the context of parametrically coupled devices. The principles outlined here also open the door for microwave engineers and filter designers to have significant impact on key technologies, components, and subsystems in the quantum computer stack.

We presented a range of example designs based on simple filter networks, however, more sophisticated multiply-connected structures have been researched, both in the microwave filter arena \cite{cameron2003advanced, cameron2018microwave} and in Josephson parametric amplifiers \cite{liu2021minimal}. By bridging the gap between these disciplines we hope to spur further research that utilizes advanced filter synthesis methods to produce new and performant parametric amplifiers, transduction networks \cite{wang2022generalized},  and non-reciprocal device designs. Further research is also needed to expand on the methods outlined here to include systematic analysis and optimization of quantum noise performance of these networks.

\begin{acknowledgments}
We thank A.\,Bengtsson, J.\,Estrada, P.\,Ravindran, J.D.\,Teufel, and T.C.\,White for critical comments on the manuscript.
Commercial software is identified by name for readers interested in accurately reproducing simulation results presented here. Such identification is not intended to imply recommendation or endorsement by NIST, nor is it intended to imply that this software is necessarily the best available for this purpose.
\end{acknowledgments}

\appendix
\section{$Z,~Y$, and $ABCD$ matrix representations for modulated inductors}
\label{apx:ZY_ABCD}
In Section~\ref{sec:immittance} we wrote the $\mathbf{Z}$ matrices for parametrically pumped circuit elements. Here we will show how to transform those to $ABCD$ ($\mathbf{T}$) matrices and factor the component into passive and purely-parametric parts, as we have done in Figs.~\ref{fig:LequivInverters} and ~\ref{fig:MequivInverters}. 

\subsection{Equivalent series inverter circuits} 
The $Z$-matrix form is especially useful in helping to identify a convenient circuit representation. This becomes clear if we convert an arbitrary $Z$-matrix to a transmission ($T$- or $ABCD$) matrix ~\cite{pozar2009microwave}. One can show that the resulting transformation can be factored into three separate matrices, 
\begin{equation}
\mathbf{T}(\mathbf{Z}) = 
    \begin{bmatrix}
        1 & z_{11}\\0 & 1
    \end{bmatrix}
    \begin{bmatrix}
        0 & -z_{12}\\z_{21}^{-1} & 0
    \end{bmatrix}
    \begin{bmatrix}
        1 & z_{22}\\0 & 1
    \end{bmatrix}.
    \label{eq:Z2T}
\end{equation}

\begin{figure}
    \centering
    \includegraphics[width=2.5in]{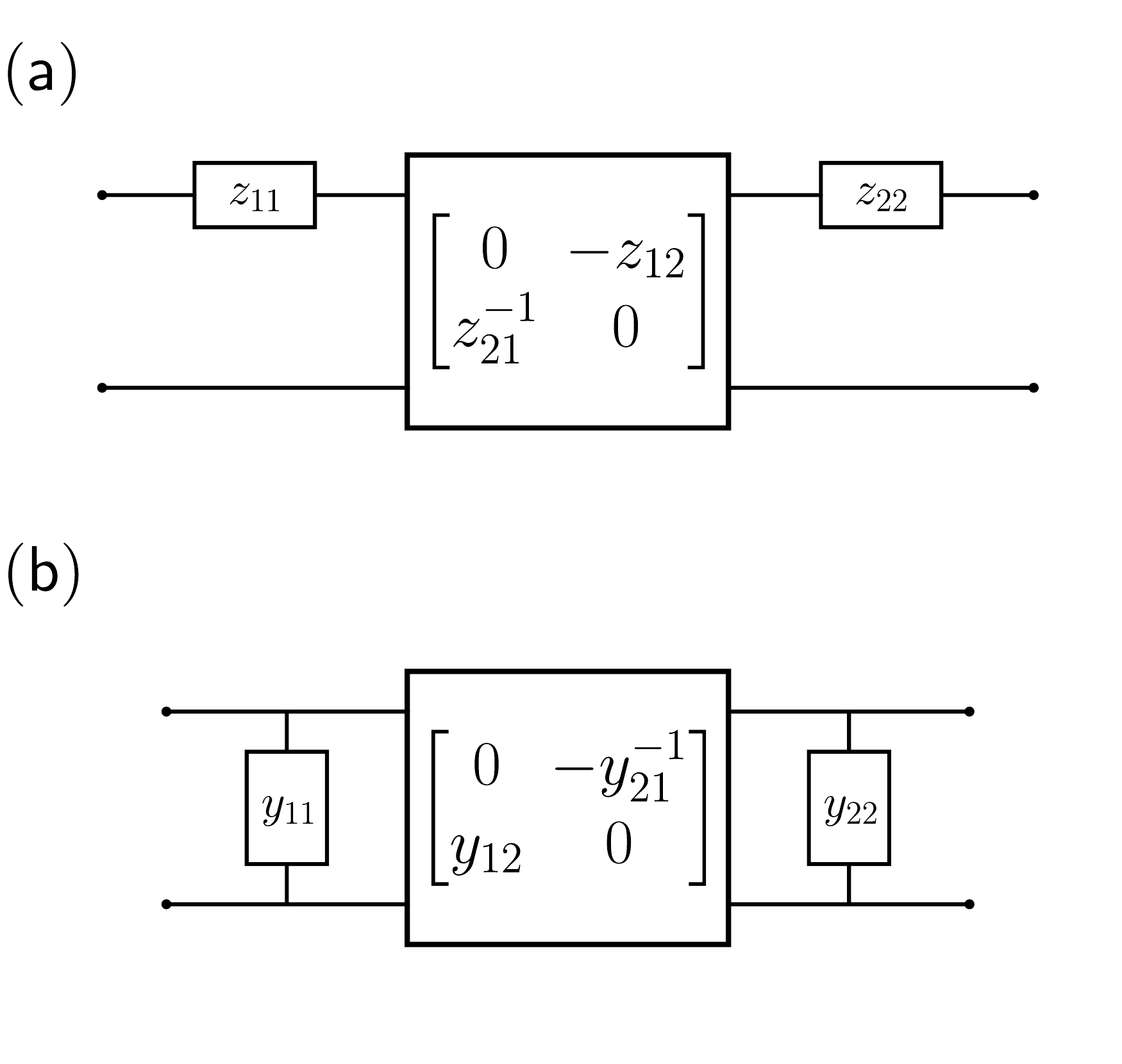}
    \caption{Equivalent 2-port circuit networks for an arbitrary (a) $Z$-matrix and (b) $Y$-matrix. The center element is represented by the equivalent inverter-like transmission matrix.}
    \label{fig:ZY2T}
\end{figure}

From this form, we see that diagonal elements of the $Z$-matrix represent series impedances, while off-diagonal elements correspond to an inverter-like object sandwiched between these impedances, Fig.~\ref{fig:ZY2T}.

If we substitute into Eq.~(\ref{eq:Z2T}) the $Z$-matrices we derived for the modulated inductor, Eq.~\eqref{eq:ZLFC} for parametric frequency conversion, and~Eq.~\eqref{eq:ZLPA} for parametric amplification, we have the following transmission matrix equivalents,
\begin{eqnarray}
    \mathbf{T}_{L,\mathrm{FC}}^\mathrm{ser} &=& \nonumber \\     
    && \begin{bmatrix}
        1 & j\omega_1L_0\\0 & 1
    \end{bmatrix}
    \begin{bmatrix}
        0 & -jK_1^*\\
        -j/K_2& 0
    \end{bmatrix}
    \begin{bmatrix}
        1 & j\omega_2L_0\\0 & 1
    \end{bmatrix},
    \label{eq:TLFC}
\end{eqnarray}
and
\begin{eqnarray}
    \mathbf{T}_{L,\mathrm{PA}}^\mathrm{ser} &=& \nonumber \\    
    & &\begin{bmatrix}
        1 & j\omega_1L_0\\0 & 1
    \end{bmatrix}
    \begin{bmatrix}
        0 & -jK_1\\
        j/K_2^*& 0
    \end{bmatrix}
    \begin{bmatrix}
        1 & -j\omega_2L_0\\0 & 1
    \end{bmatrix},
    \label{eq:TLPAseries}
\end{eqnarray}
where we have defined a complex inverter constant,
\begin{equation}
    K_\ell \equiv \omega_\ell \frac{\delta L}{2}.
\end{equation}

From both of these expressions, we see that the time-invariant part of the inductance, $L_0$, looks like a series inductor, with its impedance reckoned at both the signal ($\omega_1$) and idler frequencies ($\pm\omega_2$). Meanwhile, the inverter's role is to convert current at the signal/idler frequencies to voltage at the idler/signal frequencies. The strength of this conversion process is determined by the pump amplitude $|\delta L|$. In Figure \ref{fig:LequivInverters}(b), we denote this as the \textit{series representation} for the parametrically modulated inductor.

\subsection{Equivalent shunt inverter circuits} 
In building a frequency converter or amplifier circuit, $L_0$ should be regarded as part of the core resonator and corresponding matching sections, which we term the \textit{signal} and \textit{idler matching networks}, each evaluated at their respective frequencies. In the series representation, $L_0$ is used to build the last series $LC$ section in a multipole matching network, up to the inverter. However, it is often the case that parallel $LC$ sections are easier to implement in a given technology~\cite{pozar2009microwave,collin2007foundations}. To convert to a \textit{parallel representation}, one can write down a transmission matrix factorization, similar to \eqref{eq:Z2T}, but for the admittance matrix $\mathbf{Y}=\mathbf{Z}^{-1}$,
\begin{equation}
    \mathbf{T}(\mathbf{Y}) =
    \begin{bmatrix}
        1 & 0\\y_{11} & 1    
    \end{bmatrix}
    \begin{bmatrix}
        0 & -y_{21}^{-1}\\y_{12} & 0 
    \end{bmatrix}
    \begin{bmatrix}
        1 & 0\\y_{22} & 1
    \end{bmatrix}.
    \label{eq:Y2T}
\end{equation}

Applying Eq.~(\ref{eq:Y2T}) with the inverted $Z$-matrices from Sec.~\ref{sec:immittance} yields an equivalent \textit{shunt representation} (see Figure \ref{fig:LequivInverters}(b)),
\begin{eqnarray}
    \mathbf{T}_{L,\mathrm{FC}}^\mathrm{shunt} =\hspace{1.3cm}& \nonumber \\ 
    \begin{bmatrix}
        1 & 0\\(j\omega_1L_0')^{-1} & 1
    \end{bmatrix}&
    \begin{bmatrix}
        0 & j/J_1'\\
        jJ_2'^*& 0
    \end{bmatrix}
    \begin{bmatrix}
        1 & 0\\(j\omega_2L_0')^{-1} & 1
    \end{bmatrix}\hspace{1.0cm}
    \label{eq:TLFCshunt}
\end{eqnarray}
and
\begin{eqnarray}
    \mathbf{T}_{L,\mathrm{PA}}^\mathrm{shunt} =\hspace{1.3cm}& \nonumber \\ 
    \begin{bmatrix}
        1 & 0\\(j\omega_1L_0')^{-1} & 1
    \end{bmatrix}&
    \begin{bmatrix}
        0 & j/J_1'^*\\
        -jJ_2'& 0
    \end{bmatrix}
    \begin{bmatrix}
        1 & 0\\(-j\omega_2L_0')^{-1} & 1
    \end{bmatrix},\hspace{0.5cm}
    \label{eq:TLPAshunt}
\end{eqnarray}
where we have defined an effective shunt inductance
\begin{equation}
    L_0' \equiv L_0\left(1 - \frac{|\delta L|^2}{4L_0^2}\right)
\end{equation}
and inverter constant
\begin{equation}
    K_\ell' = J_\ell'^{-1} \equiv \omega_\ell L_0 \left(\frac{2L_0}{\delta L} - \frac{\delta L^*}{2 L_0}\right).
\end{equation}

The key results of this Appendix are summarized, in compact form, in Fig.~\ref{fig:LequivInverters}(b). Similar arguments can be used to derive the relations summarized in Fig.~\ref{fig:MequivInverters}(b) for the modulated mutual inductance.

\section{Josephson Junctions and SQUIDs as modulated inductors}\label{apx:JJSQUID}
The role of the parametric coupler element in Section\,\ref{sec:immittance} is played by a modulated inductor, which in Josephson parametric devices, can be performed by any of a variety of single and multi-junction devices \cite{frattini2018optimizing, abdo2013nondegenerate, naaman2019high, chen2014gmon}. Here we give a primer on two very simple choices, a single Josephson junction and the dc-SQUID.

\subsection{Single Josephson junction}
The Josephson equations \cite{josephson1962possible,clarke2006squid} yield a simple expression for the inductance of a single junction as a function of bias current $I$,
\begin{equation}
    L_J(I)  =\frac{\Phi_0}{2\pi I_0}\frac{1}{\sqrt{1 - I^2/I_0^2}},
    \label{eq:LJvsI}
\end{equation}
where $\Phi_0\approx 2.068\,\mu\mathrm{A}\cdot \mathrm{nH}$ is the single flux quantum, and $I_0$ is the junction critical current. This is a nonlinear function of the bias current and, if expanded around $I=0$, in the small current limit this can be approximated as 
\begin{equation}
    L_J(I) \simeq L_{J0} \left(1 + \frac{1}{2}\frac{I^2}{I_0^2}\right),
\label{eq:LJquadratic}
\end{equation}
where we define the Josephson inductance, $L_{J0}\equiv \Phi_0/2\pi I_0$. Modulation of the inductance is achieved by injecting a pump current, $I_P(t) = |I_P|\cos(\omega_P t + \phi_P)$, yielding a time dependent inductance,
\begin{eqnarray}
    L_J^\mathrm{4w}(t) &\simeq L_{J0}\left[1 + \frac{1}{2}\frac{|I_P|^2}{I_0^2}\left(1 + \cos\left(2\omega_Pt + 2\phi_P\right)\right)\right]\nonumber\\
    &\simeq L_{J0}' + \delta L \cos(2\omega_Pt + 2\phi_P),
    \label{eq:LJ4wave}
\end{eqnarray}  
with an effective base inductance
\begin{equation}
    L_{J0}'\equiv L_{J0}\left(1 + \frac{1}{2}\frac{|I_P|^2}{I_0^2} \right)
\end{equation}
and inductance modulation amplitude
\begin{equation}
    \delta L \equiv \frac{L_{J0}}{2}\frac{|I_P|^2}{I_0^2}.
\end{equation}

\paragraph{4-wave mixing.} Eq.~(\ref{eq:LJ4wave}) differs slightly from the expression in Eq.~(\ref{eq:body_vflux}) having a modulation frequency $2\omega_P$ instead of $\omega_P$. In this configuration, where we're using a quadratic nonlinearity in the current, the inductance modulation is doubled in frequency. In other words, the nonlinear element is modulated by a sub-harmonic drive. Since, the purpose of this element is to achieve mixing between a signal and idler, we require only that the element modulate at the difference frequency between the two. Therefore, for frequency conversion we have $2\omega_P = |\omega_1 - \omega_2|$ (connecting $\omega_1$ and $\omega_2$), and for amplification we have $2\omega_P = \omega_1 + \omega_2$ (connecting $\omega_1$ and $-\omega_2$). This approach is often termed \textit{4-wave mixing}, but can be interpreted as a consequence of using a quadratic nonlinearity to achieve the required mixing product frequencies for parametric coupling.

\paragraph{3-wave mixing.} Alternatively, one can choose to add a dc-bias to the pump current, $I_P(t) = I_\mathrm{dc} + |I_P|\cos(\omega_P t + \phi_P)$, in which case the lowest order expansion for the inductance is
\begin{equation}
    L_J^\mathrm{3w}(t) \simeq L_{J0}' + \delta L \cos{(\omega_P t + \phi_P)},
    \label{eq:LJ3wave}
\end{equation}
with
\begin{equation} 
L_{J0}'\equiv \frac{L_{J0}}{\sqrt{1 - I_\mathrm{dc}^2/I_0^2}}
\end{equation}
and
\begin{equation}
    \delta L \equiv L_{J0}\frac{I_\mathrm{dc}}{I_0} \frac{1}{\left(1 - I_\mathrm{dc}^2/I_0^2 \right)^{3/2}} |I_P|.
\end{equation}

Eq.~(\ref{eq:LJ3wave}) corresponds to `direct' modulation of the inductance at $\omega_P$ as in Eq.~(\ref{eq:body_vflux}), a consequence of the dominant linear dependence of $\delta L$ on current when expanded around a bias current.

While the current nonlinearity above is convenient to use, it requires some care in handling the large pump tone required to modulate the inductance, which can be close to the signal frequency in the case of 4-wave mixing, necessitating additional components in the signal path to cancel or filter it out. In practice, these extra components can impact the overall noise performance and complicate operation and bringup.

\subsection{dc SQUID}
As an alternative to using the current nonlinearity, one can utilize a loop topology that interferes two supercurrent paths. The result is an effective Josephson inductance that can be tuned by the magnetic flux threading the loop. There are several variants of these SQUID (superconducting quantum interference device) configurations, but in this work, for simplicity we utilize an ideal dc-SQUID. This is comprised of two Josephson junctions in parallel, with no stray inductance and a means to modulate its magnetic flux \cite{clarke2006squid}. In the limit where stray inductance is neglible, the inductance has a simple form,
\begin{equation}\label{eq:LSQ_flux}
    L_\mathrm{SQ}(I,\Phi) = \frac{L_{J0}(I)}{2|\cos{(\pi \Phi_A/\Phi_0)}|},
\end{equation}
where $\Phi_A$ is the magnetic flux induced by an adjacent flux line, and can include both dc and ac components. $L_{J0}(I)$ is the single junction current-dependent inductance, Eq.~(\ref{eq:LJvsI}).

Eq.~(\ref{eq:LSQ_flux}) is a nonlinear, periodic function in the applied flux. As with the Josephson current nonlinearity, one can likewise choose to bias such that the dominant first order expansion term in flux is either quadratic/4-wave ($\Phi_\mathrm{dc} = 0$) or linear/3-wave ($|\Phi_\mathrm{dc}|\neq0$ modulo $(n+1/2)\Phi_0$ for integer $n$). Just as we did for the current-pumped junction, we can likewise expand the inductance for a given flux pump modulation, $\Phi_A(t) = \Phi_{dc} + \Phi_P\cos{(\omega_Pt + \phi_P)}$. One can then generate both 4-wave (Eq.~(\ref{eq:LJ4wave})) and 3-wave (Eq.~(\ref{eq:LJ3wave})) inductance modulation forms as we did above \cite{sundqvist2014negative}.

\section{Using the $J$/$K$-inverter models of parametric coupling in amplifier design}\label{apx:para_inverter}

In Figs.~\ref{fig:LequivInverters} and~\ref{fig:MequivInverters}, we presented equivalent models for the parametrically-modulated inductor and mutual as impedance ($K$) or admittance ($J$) inverters. Here we will show how to use these `parametric inverter' models as aids in the synthesis of wideband amplifier circuits, as well as in the systematic calculation of the full 2-port response response of these circuits. In addition, we will show how these inverter models can be used to set the parametric modulation strength, or pump amplitude, based on the bandwidth and response characteristics in a multi-pole matched amplifier. 

\begin{figure*}[htb]
    \centering
    \includegraphics[width=\textwidth]{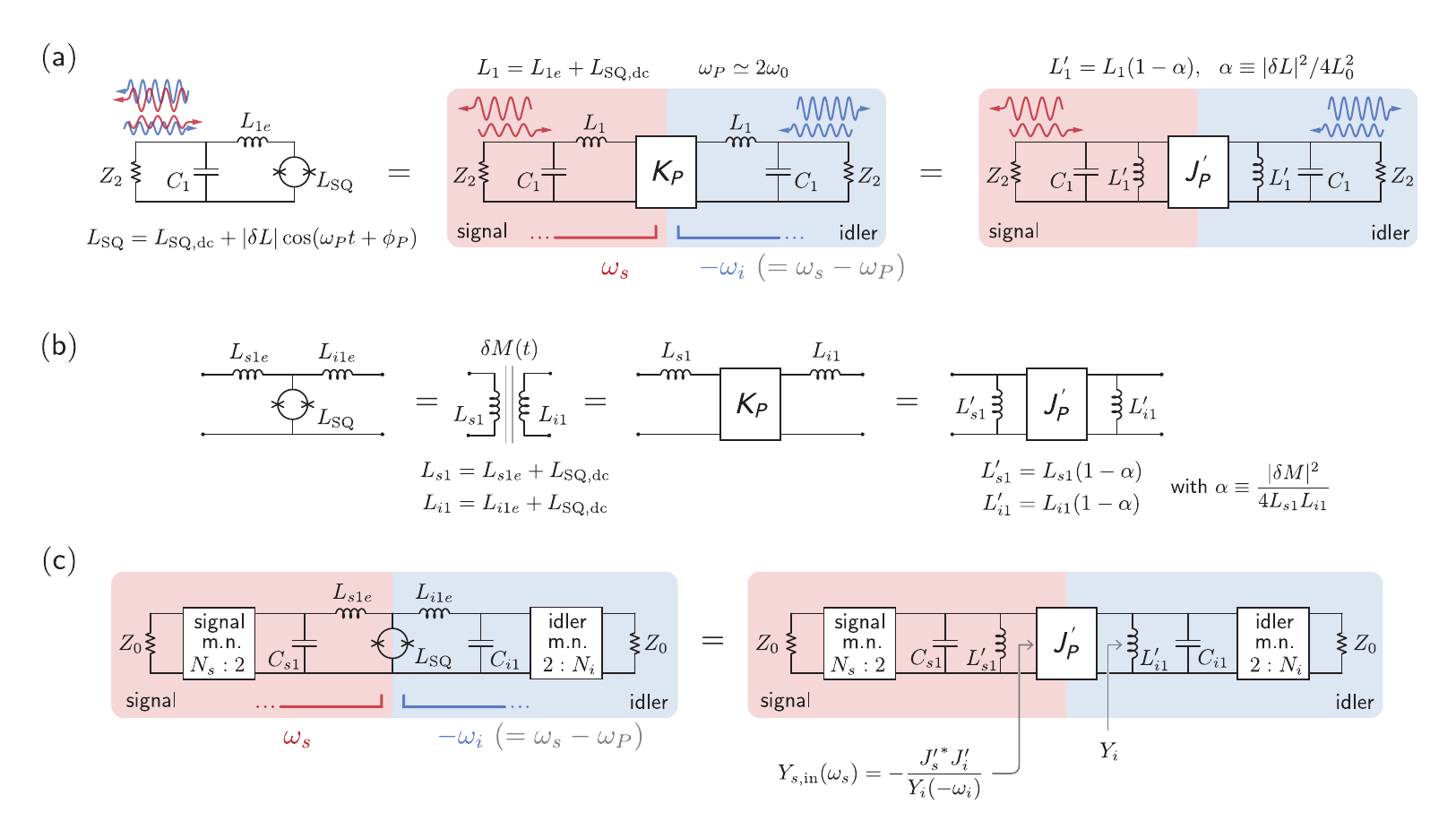}
    \caption{
    (a) Transformation of single pole degenerate Josephson parametric amplifier (left) to a 2-port (signal/idler) series parametric inverter representation ($K_P$, middle), and a shunt parametric inverter representation ($J'_P$, right). (b) Equivalence between the parametrically modulated inductor tee and modulated mutual inductance circuits, and their  $K$- and $J$- inverters counterparts. The inverter formulae are given in Fig.~\ref{fig:MequivInverters}. (c) Example multipole matching strategy for a nondegenerate parametric amplifier using a parametric $J$-inverter. The tee-equivalent $J$ inverter transforms the SQUID-embedded double-resonator circuit into effective parallel $LC$ sections of multipole signal/idler matching networks (m.n. in the figure). The input admittance looking into the inverter from the signal circuit, $Y_{s,\mathrm{in}}(\omega_s)$ is the nondegenerate equivalent to the pumpistor model admittance \cite{sundqvist2014negative} since $Y_i^*(\omega_i) = Y_i(-\omega_i)$.
    \label{fig:inverter_examples}}
\end{figure*}

As noted in Sec.~\ref{sec:immittance_paramp}, the immitance inverter models reduce to the `pumpistor model' \cite{sundqvist2014negative} for reflection JPAs; however, they are more general in that they allow one to obtain full 2-port scattering parameters, since they treat the signal and idler ports equally. To be more concrete, in the degenerate parametric amplifier case, the incoming signal and idler tones are driven from the same \emph{physical} port and then both re-emitted from the same \emph{physical} port (see Fig.\,\ref{fig:inverter_examples}(a, left)). This can still be described as a 2-port scattering problem in the signal/idler basis, where the signal is the `left' \emph{frequency} port and the idler is the `right' \emph{frequency} port, as illustrated in Fig.~\ref{fig:inverter_examples}(a, middle). Likewise, any matching network (\eg, Refs.~\cite{naaman2019high, roy2015broadband}) would be mirrored on either side, but evaluated at the respective frequencies: $\omega_s$ on the signal side and $-\omega_i\;(=\omega_s - \omega_P)$ on the idler side. In a nondegenerate amplifier, the distinction between physical and frequency ports is not necessary, and the signal/idler matching networks can be physically distinct with different numbers of poles and filter types, allowing for more flexibility in design.

\subsection{Degenerate parametric amplifier}
In Figs.~\ref{fig:LequivInverters} and~\ref{fig:MequivInverters}, we labeled the inverter equivalents as $K$-type or $J$-type, depending on the topology of the circuit in which they are embedded. When designing a degenerate parametric amplifier around a flux- or current-pumped Josephson circuit, it is natural to resonate the Josephson circuit out using a parallel capacitor, Fig.~\ref{fig:inverter_examples}(a, left). In this case, since the parametrically modulated component is embedded in a parallel resonator, it is more convenient to use the `shunt representation' in Figs.~\ref{fig:LequivInverters} and~\ref{fig:MequivInverters}, and model the parametric coupler as an admittance $J$-inverter, Fig.~\ref{fig:inverter_examples}(a, right). 
The full circuit can be analyzed by virtually mirroring it and evaluating each element at the signal and idler frequencies (Fig.\,\ref{fig:inverter_examples}(b-c)). Note that we have included the possibility of stray inductance $L_{1e}$ in the transformation to a 2-frequency-port circuit. We can then write out the total transmission ($ABCD$) matrix for this example by cascading the $ABCD$ matrices of the constituent components. For the single-pole amplifier in Fig.~\ref{fig:inverter_examples}(a) we have,
\begin{widetext}
\begin{equation}
\mathbf{T}_\Sigma =
    \begin{bmatrix}
        1 & 0\\
        j\omega_s C_1 & 1
    \end{bmatrix}
    \begin{bmatrix}
        1 & 0\\
        (j\omega_s L_1^\prime)^{-1} & 1    
    \end{bmatrix}
    \begin{bmatrix}
        0 & j/{J_s^\prime}^*\\
        -j{J_i^\prime} & 0
    \end{bmatrix}
    \begin{bmatrix}
        1 & 0\\
        (-j\omega_i L_1^\prime)^{-1} & 1
    \end{bmatrix}
    \begin{bmatrix}
        1 & 0\\
        -j\omega_i C_1 & 1
    \end{bmatrix},
    \label{eq:TtotalJPA}
\end{equation}
\end{widetext}
where the inverter constants $J_{s/i}$ are defined in Fig.~\ref{fig:LequivInverters}. From Eq.~(\ref{eq:TtotalJPA}), and assuming a port impedance $Z_2$ as in Fig.~\ref{fig:inverter_examples}(a), one can compute the scattering matrix   \cite{pozar2009microwave}, whose elements express the reflection (\textit{cis}-)gains and conversion (\textit{trans}-)gains \textit{vs.} frequency.

\subsection{Nondegenerate amplifier}

The single-pole design in the above section is simple, and can be easily extended to a nondegenerate amplifier. One starts with a modulated inductance with a value that is constrained by its physical realization (\eg, available critical current densities in a given process), and the center frequency of the amplifier can then be set by the value of $C_1$. The value of the terminating impedance $Z_2$ can be chosen to yield the desired bandwidth, $\Delta\omega = 1/2\pi Z_2 C_1$. In turn, $Z_2$ can be realized by transforming the $50\,\Omega$ source impedance using an immitance inverter (see \Cref{apx:inverter_termination}).
In designs using multi-pole matching networks, the relative immittances of each of the sections can be systematically determined by the filter prototypes. In Section~\ref{sec:broadband_paramp} we designed a coupled-mode network for a matched nondegenerate amplifier, but abstracted the physical details of the parametric coupler. Here we demonstrate a physical circuit synthesis procedure using the method of Sec.~\ref{sec:denorm} and the parametric inverter model from Sec.~\ref{sec:immittance}. Furthermore, whereas in Sec.~\ref{sec:broadband_paramp} we assumed the same prototype and bandwidth was used for both signal and idler ports, here we treat a more general synthesis problem that does not make this assumption.

We start with the circuit shown in Fig.~\ref{fig:inverter_examples}(b), left, as the core of our design: an inductor tee where the center inductance is the modulated inductor (\eg, a Josephson circuit) and the arm inductances are required to form distinct tank circuits that make up the first resonator sections of arbitrary matching networks on the signal and idler sides. As we saw in \Cref{sec:denorm}, filter sections are either parallel or series resonators yet the tee looks like neither. To begin our synthesis procedure, we transform the physical topology of this core element into an effective circuit comprised of two parallel $LC$ resonators, sandwiching a pure parametric $J$-type inverter. The steps for this transformation are summarized in Fig.\,\ref{fig:inverter_examples}(b). Indexing the filter sections from the core outwards, these resonators form the respective $k=1$ sections of extended matching networks on the signal (labeled with subscript $s$) and idler (labeled with $i$) sides. Following \Cref{sec:denorm}, we can relate the impedance of these sections to the reference impedances of the filter networks, $Z_{m,0}$:
\begin{eqnarray}
Z_{m,1} &=& \omega_m L_{m,1}'\label{eq:sec1parallel_impedance}\\ 
&=&\dfrac{\omega_m}{\Delta \omega_m}\dfrac{Z_{m,0}}{g_{m,1}}, \label{eq:sec1_charimpedance}
\end{eqnarray}
where $g_{m,k}$ is the prototype coefficient corresponding to the $k$th section in the $m$th circuit, and with $m\in\{s,i\}$. Note that the reference impedances for the respective signal and idler circuits, $Z_{m,0}$, as extracted from Eq.~\ref{eq:sec1_charimpedance} \emph{are not} generally going to be $50\,\Omega$.

With the reference impedance found from Eq.~(\ref{eq:sec1_charimpedance}) and the prototype coefficients $\{g_{m,k}\}$, we can proceed to derive a circuit with alternating series/parallel resonators (\Cref{sec:denorm}). As in the degenerate amplifier example above, the 2-port scattering parameters can be calculated simply by computing the total transmission matrix product, however in this case, the signal and idler modes correspond to separate physical circuits and ports. 

\begin{figure}
    \centering
    \includegraphics[width=\columnwidth]{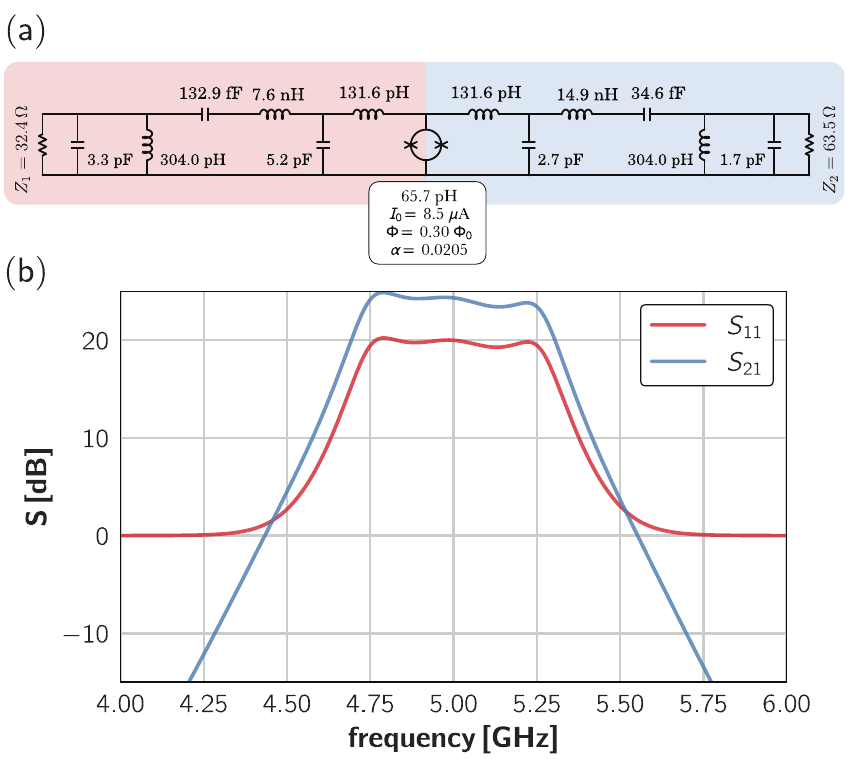}
    \caption{\label{fig:PA_sparams}(a) Example nondegenerate amplifier example circuit based on 3-pole Chebyshev graph in \Cref{fig:NDPA_graph},  centered at 5 GHz (signal) and 7 GHz (idler). (b) Calculated $S$-parameters for a matched parametric amplifier (referred to the port impedances $Z_1$ and $Z_2$, not to 50\,$\Omega$). $S_{11}$ is the signal reflection \textit{cis}-gain, and $S_{21}$ is the signal-idler trans-gain. In contrast to \Cref{fig:NDPA_graph}, these represent the power gain rather than the photon number gain, therefore $S_{21}$ has an extra overall relative scaling factor $f_2 Z_2/f_1 Z_1$ (4.4\,dB at 5.0\,GHz).}
    \label{fig:NDPA_circuit}
\end{figure}

\Cref{fig:NDPA_circuit}(a) shows a denormalized ladder circuit (alternating series/parallel resonators) computed for the 3-pole (0.5 dB ripple) Chebyshev prototype we have already used in Sec.~\ref{sec:broadband_paramp}. Component values are indicated in the figure. Note that the terminating impedances of the signal port $Z_1$ and the idler port $Z_2$ are not $50\,\Omega$; these can be transformed to $50\,\Omega$ by inserting admittance inverters as in, \eg, Sec.~\ref{sec:vijay}. Note also that the impedances of the series resonators in the circuit are rather high, producing inconvenient component values. These sections could be transformed to parallel resonators by use of additional inverters following Sec.~\ref{sec:design_method}. The addition of physical admittance inverters, however, will invariably introduce additional frequency dependence in the circuit \cite{MYJ} that will become more pronounced in designs with either higher bandwidth, or greater frequency separation between the signal and the idler.

\Cref{fig:NDPA_circuit}(b) shows reflection and transmission gain for the circuit in panel (a), computed from its cascaded $ABCD$ matrix. In contrast to \Cref{fig:NDPA_graph}(b), these represent the power gain rather than the photon number gain, therefore $S_{21}$ has an extra overall relative scaling factor $f_2 Z_2/f_1 Z_1$ (4.4\,dB at 5.0\,GHz). Also visible is a sloped gain profile, stemming from a frequency dependence of the parametric inverter itself; this effect is inherent to all physical implementations of nondegenerate parametric amplifiers \cite{Matthaei1961}, but is absent in calculations based on ideal (frequency-independent) coupling rates, such as in Fig.~\ref{fig:NDPA_graph}. Methods for equalizing the gain slope can be found in, \eg, Ref.~\cite{mellor1975synthesis}. Alternatively one can slightly detune the $k=1$ resonators \cite{Matthaei1961}. 

\subsection{Modulation amplitude, $|\delta M|$} 
Here we show how to calculate the modulation amplitude in a general multi-pole parametric amplifier circuit. Knowledge of this parameter is useful for a few reasons. First, it gives the designer a basis for setting requirements on physical parameters such as the participation ratio of the modulated inductance to the total inductance, or the on-chip coupling strength of the ac bias line to the modulated component. Second, it helps determine whether the required inverter constant (in this example $J_m^\prime$, as defined in Sec.~\ref{sec:immittance_paramp} and Figs.~\ref{fig:LequivInverters} and ~\ref{fig:MequivInverters}) is even physically tenable for a given Josephson circuit. 

In \Cref{fig:LequivInverters,fig:MequivInverters}, we parameterize the inverter constants and effective inductances in terms of a normalized modulation parameter $\alpha$. An expression for $\alpha$ can be derived by requiring that the parametric inverter, when looking from either its signal or idler side, transforms the opposite port's immittance to the correct value to yield the required reflection gain on resonance. For instance, looking into the $J^\prime$ inverter from the signal side on resonance, as in Fig.~\ref{fig:inverter_examples}(c) right, we see an input admittance, Eq.~(\ref{eq:pa_admittance_inverter}), can be written as
\begin{eqnarray}
Y_{s,\mathrm{in}} (\omega = \omega_{s0})
&=& - \dfrac{{J_s^\prime}^* J_i^\prime}{Y_{i,N_i + 1}}\\
&=& - \dfrac{\alpha}{\omega_{s0} L_s^\prime \omega_{i0} L_i^\prime Y_{i,N_i + 1}}.
\label{eq:Jprimeinverter_inputadmittance}
\end{eqnarray}
Here, $\omega_{s0}$ and $\omega_{i0}$ are the signal and idler center frequencies, $Y_{i,N_i + 1}$ is the idler termination admittance, and we have used the definitions in Fig.~\ref{fig:MequivInverters}(b). Note that only the terminating admittance $Y_{i,N_i + 1}$ appears in Eq.~(\ref{eq:Jprimeinverter_inputadmittance}) because at the center frequency of the network all shunt resonators are effectively open-circuit and all series resonators are short-circuit, so that the network is effectively `transparent' and does not transform the value of the terminating admittance in any way.  Setting $Y_{s,\mathrm{in}}$ equal to the required value to obtain power gain $G$ in reflection, Sec.~\ref{sec:all_together_paramp} ($Y_{s,N_s + 1}$ is the signal port termination admittance),
\begin{equation}
Y_{s,\mathrm{in}}(\omega = \omega_{s0}) = Y_{s,N_s + 1} \dfrac{\sqrt{G} - 1}{\sqrt{G} + 1},
\end{equation}
we can, after recognizing the terms in the denominator of Eq.~(\ref{eq:Jprimeinverter_inputadmittance}) as denormalized filter section immittances, derive a general form for the modulation amplitude,
\begin{eqnarray}
\alpha 
&=& \dfrac{|\delta M|^2}{4 L_{s,1} L_{i,1}} \label{eq:alpha1}\\
&=& 
\left[\dfrac{\sqrt{G} + 1}{\sqrt{G} - 1}\right]^{\pm 1} \dfrac{\Delta\omega_{s}}{\omega_{s0}}\dfrac{\Delta\omega_i}{\omega_{i,0}}
\dfrac{g_{s,N_s + 1}^{p_s}g_{i,N_i + 1}^{p_i}}{g_{s,1}g_{i,1}}.\label{eq:alpha2}
\end{eqnarray}
The positive and negative signs in the exponent of the gain factor correspond to the use of series/$K$-type and shunt/$J$-type inverters, respectively. Likewise, Eq.~(\ref{eq:alpha2}) is generalized to arbitrary filter orders for the signal/idler circuits, $N_{s/i}$, different filter prototype coefficients, and differing fractional bandwidth specifications $\Delta\omega_m/\omega_{m0}$. The $p_{s/i}$ are parity exponents that evaluate to $p_m = 1$ when $N_m$ is even, and $p_m=-1$ when $N_m$ is odd. This result expands on the examples we have given in the main text, where the same bandwidth and prototype was used for both signal and idler. 

Since $\alpha$ is physically bounded by the $k=1$ mode inductances, $\alpha\leq 1$, Eq.~(\ref{eq:alpha2}) sets limits to the available bandwidth. Likewise, for a fixed bandwidth and gain, one finds that the normalized modulation amplitude $\sqrt{\alpha}$ gets smaller at increasing prototype order. In practical terms, this means that it gets easier to achieve the target gain as one increases the order of the matching network.

\section{Inverters at terminations}\label{apx:inverter_termination}

\begin{figure}[htbp]
    \begin{center}
    \includegraphics[width=\columnwidth]{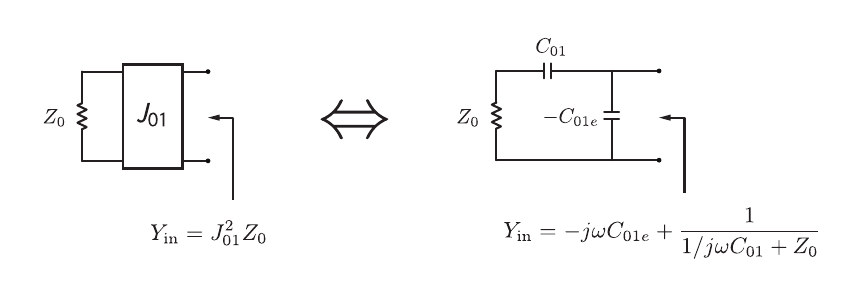}
    \caption{
    (left) Inverter at termination, (right) Capacitive inverter equivalent.
    }
    \label{fig:inverterterm}
    \end{center}
\end{figure}

The canonical inverter realization consists of either a $\pi$ network (Fig.~\ref{fig:inverters_pi}) or a tee network where the central element (a series or shunt reactance) is sandwiched between (shunt or series) elements with identical reactance values, except with the opposite sign \cite{collin2007foundations,pozar2009microwave}. When placed between two resonator circuits, these negative reactances can be absorbed into the reactances forming adjacent resonator sections. However, when an inverter is placed at a real termination, $Z_0$ (Figure \ref{fig:inverterterm}), there is no reactance at the termination to absorb the negative reactance. In this tutorial, we follow the standard practice and use an asymmetric 2-element topology that places the negative reactance on one side \cite{MYJ}. One can equate the real and imaginary parts of the input admittance of the ideal inverted termination (Fig.~\ref{fig:inverterterm}, left) to the capacitive 2-element topology (Fig.~\ref{fig:inverterterm}, right), and derive for the series coupling capacitance,
\begin{equation}
C_{01} = \frac{J_{01}}{\omega_0}\frac{1}{\sqrt{1 - Z_0^2 J_{01}^2}}
\end{equation}
shunted by a negative reactance, $-C_{01e}$ where
\begin{equation}\label{eq:c01e}
C_{01e} = \frac{J_{01}}{\omega_0}\sqrt{1 - Z_0^2 J_{01}^2}.
\end{equation}
Here we have set $\omega = \omega_0$ so that this transformation is exact at the center frequency.

This particular topology is similar to the capacitive $\pi$ inverter, with a leg removed. The expressions are similar to the symmetric $\pi$ version, Fig.~\ref{fig:inverters_pi}(b), but with a scaling factor that is dependent on the desired transformation ratio, since $Z_0^2 J_{01}^2 = Y_\mathrm{in}/Y_0$ (with $Z_0 \equiv 1/Y_0$).  One can derive a similar variant for the tee topology using the same approach. As in the canonical $\pi$ and tee topologies, this inverter is constructed with frequency-dependent reactances, so the inversion is only exactly correct at the center frequency $\omega_0$.  

\section{Calculating prototype coefficients by Cauer synthesis}\label{apx:cauer_synthesis}
Here, we briefly review the procedure for calculating prototype coefficients using the well-known \textit{insertion loss} technique (\textit{c.f.} Ref.~\onlinecite{pozar2009microwave}) and Cauer synthesis \cite{aatre1986network}. The technique is based on the fact that the input impedance of a ladder circuit, as a function of frequency, can be expressed as a continued-fraction, Eq.~(\ref{eq:continued_fraction}). Since prototype tables for passively-terminated filter networks are commonly available~\cite{pozar2009microwave,MYJ}, we will focus here on prototypes designed specifically for active, negative resistance loads, such as presented by parametric amplifiers. 

We start with a power loss function, $P_L(\omega)$, which is an even function of frequency, normalized to a cutoff frequency of $2\pi\times$1 Hz. This function is usually expressed as a polynomial of order $2N$, where $N$ usually corresponds to the number of reactances (for low-pass networks) or resonant modes (for band-pass networks) in the circuit. The reflected power at the input of the network is given by:
\begin{equation}\label{eq:reflection_from_pil}
    |\Gamma(\omega)|^2 = 1 - \frac{1}{P_L(\omega)}=\frac{P_L(\omega)-1}{P_L(\omega)}.
\end{equation}
For Butterworth networks, $P_L$ can be written as:
\begin{equation}\label{eq:butter_pil}
    P_L(\omega) = A\left(1+k^2\omega^{2N}\right),
\end{equation}
and for Chebyshev networks, it can be written as:
\begin{equation}\label{eq:cheby_pil}
    P_L(\omega) = A\left(1+k^2T^2_N(\omega)\right),
\end{equation}
where $T_N(\omega)$ is a Chebyshev polynomial of order $N$. In both expressions, $A\in[0,1]$ controls the match ($P_L=1$ means zero reflection), and $k^2$ controls the reflection at the cutoff frequency where $\omega=1$. 

\subsection{Cauer synthesis}\label{apx:factoring}
Our task is to factor Eq.~(\ref{eq:reflection_from_pil}) into  $|\Gamma(s)|^2=\Gamma(s)\Gamma(-s)$, where $s=j\omega$. Note that we have a choice of sign for $\Gamma$: if we choose a positive sign, then when $\omega\rightarrow\infty$ the low-pass prototype looks like an open circuit, and we may use $\Gamma(s)$ to write the input impedance of the network (normalized to 1 $\Omega$) $Z(s)=\frac{1+\Gamma(s)}{1-\Gamma(s)}$. If we choose the negative sign, then when $\omega\rightarrow\infty$ the low-pass prototype looks like a short circuit, and we may use $\Gamma(s)$ instead to write the input admittance of the network (normalized to 1 $\Omega^{-1}$)  $Y(s)=\frac{1-\Gamma(s)}{1+\Gamma(s)}$.

Since $|\Gamma(s)|^2$ is a ratio of polynomials of order $2N$, $\Gamma(s)$ is a ratio of polynomials of order $N$, $\Gamma(s)=R(s)/D(s)$, where:
\begin{align}
    R(s) &= \prod^N_{i=1}\left(s-z_i\right)\label{eq:cauer_zeros}\\
    D(s) &= \prod^N_{i=1}\left(s-p_i\right)\label{eq:cauer_poles},
\end{align}
where $p_i$ are the complex roots of the denominator of $|\Gamma(s)|^2$ that are in the left-half-plane (LHP), $\mathrm{Re}\{p_i\}<0$. $z_i$ are the complex roots of the numerator of $|\Gamma(s)|^2$, which we pick either from the LHP or from the imaginary axis in conjugate pairs, or from the origin, until we have exactly $N$ of them.

We can then write (for positive sign of $\Gamma$)
\begin{equation}\label{eq:z_polynomials}
    Z(s)=\frac{1+\Gamma(s)}{1-\Gamma(s)}=\frac{D(s)+R(s)}{D(s)-R(s)} 
\end{equation}
or $Y(s)=1/Z(s)$ (for a negative sign choice). Expanding out Eqs.~(\ref{eq:cauer_zeros}) and~(\ref{eq:cauer_poles}) to polynomial form and plugging into Eq.~(\ref{eq:z_polynomials}), we will get a ratio of polynomials where the order of the numerator is greater than that of the denominator by 1. 

From here, we proceed to use polynomial long division to write $Z(s)$, or $Y(s)$, as a continued fraction expansion (Cauer form):
\begin{equation}
    Z=g_1s+\cfrac{1}{g_2s+\cfrac{1}{g_3s+\dots}}.
    \label{eq:continued_fraction}
\end{equation}
The coefficients $\{g_i\}$ of the Cauer expansion are the filter prototype coefficients. The procedure for writing the continued-fraction expansion for a fraction of polynomials and extracting $\{g_i\}$ can be automated using \textit{e.g.} the \texttt{numpy} python package, where the inputs \texttt{n} and \texttt{d} to the function below are the numerator and denominator polynomials of the impedance (or the admittance), Eq.~(\ref{eq:z_polynomials}), respectively:
\begin{verbatim}
import numpy as np
def cauer(n: np.poly1d, d: np.poly1d):
    assert n.order == d.order + 1
    g = [1.0]
    while n.order > 0:
        quot, rem = np.polydiv(n, d)
        g.append(quot.coef[0])
        n = d
        d = rem
    g.append(1 / quot.coef[-1])
    return g
\end{verbatim}

\subsection{Prototype synthesis for negative resistance amplifiers}
In Section~\ref{sec:all_together_paramp} we saw that the reflection gain from an active negative-resistance load is equivalent to the return loss from a passive termination having the same absolute value of resistance. However, gain in the signal mode also necessitates gain in the idler mode, and because the amplified signals are correlated~\cite{hatridge2011dispersive}, the power loss function that we need to use when synthesizing a matching network should be calculated based on the gain parameter:
\begin{equation}\label{eq:g_pl}
    G_{PL} = \left(\sqrt{G} + \sqrt{G-1}\right)^2 = 4G-2+{\cal O}\left(\frac{1}{G}\right),
\end{equation}
where $G$ is the \emph{signal} power gain. This is in contrast to Ref.~\onlinecite{Getsinger1963}, which assumes that the amplified signal and idler are not correlated.

In other words, if $P_\textrm{s}$ is the input signal power, then $G_{PL}P_\textrm{s}$ is the power extracted from the pump. The matching network must be designed to allow this power to be dissipated in both the signal and idler terminations.

\subsubsection{Butterworth response}
For the Butterworth response type, we set $A$ in Eq.~(\ref{eq:butter_pil}) so that $|\Gamma(0)|^2=1/G_{PL}$, and $k^2$ in Eq.~(\ref{eq:butter_pil}) so that at the cutoff frequency $|\Gamma(1)|^2=2/G_{PL}$, to define the amplifier bandwidth as the half-gain points (3 dB bandwidth).

The power loss function we use to generate normalized Butterworth low-pass prototypes of order $N$ for a desired signal power gain $G$ is therefore given by:
\begin{equation}\label{eq:butter_amp_pil}
    P_L(\omega)=\frac{G_{PL}}{G_{PL}-1}\left[1+\left(\frac{1}{G_{PL}-2}\right)\omega^{2N}\right],
\end{equation}
where $G_{PL}$ is defined in Eq.~(\ref{eq:g_pl}). Calculated prototype coefficients for a practical selection of gains are listed in Table.~\ref{tab:butterworth_amp_coefs}.

\begin{table*}[h!]
\caption{Butterworth coefficients calculated according to Appendix~\ref{apx:cauer_synthesis} for several amplifier gains, and for 2-, 3-, and 4-section networks, using the power loss function Eq.~(\ref{eq:butter_amp_pil}). Coefficient $g_0$ refers to the active load.\label{tab:butterworth_amp_coefs}}
\begin{ruledtabular}
\begin{tabular}{lcdddddd}
\textrm{Signal gain}&
\textrm{order}&
\multicolumn{1}{c}{\textrm{$g_0$}}&
\multicolumn{1}{c}{\textrm{$g_1$}}&
\multicolumn{1}{c}{\textrm{$g_2$}}&
\multicolumn{1}{c}{\textrm{$g_3$}}&
\multicolumn{1}{c}{\textrm{$g_4$}}&
\multicolumn{1}{c}{\textrm{$g_5$}}\\
\colrule
$G = 15$~dB & 2 &1.0&  0.6067 & 0.2733 & 1.1969 & & \\
&3 & 1.0&  0.8122 & 0.6587 & 0.3710 & 0.8355 & \\
&4 & 1.0& 0.9266 & 0.8729 & 0.9049 & 0.2266 & 1.1969\\
\colrule
$G=17$~dB & 2 & 1.0& 0.5149 & 0.2587 & 1.1528 & &\\
& 3  & 1.0& 0.7078 & 0.6417 & 0.3381 & 0.8674 &\\
& 4 & 1.0&  0.8175 & 0.8613 & 0.8361 & 0.2263 & 1.1528\\
\colrule
$G=20$~dB & 2 & 1.0& 0.4085 & 0.2343 & 1.1055 & &\\
& 3 & 1.0& 0.5846 & 0.6073 & 0.2981 & 0.9045 &\\
& 4 & 1.0& 0.6878 & 0.8309 & 0.7527 & 0.2225 & 1.1055\\
\colrule
$G=25$~dB & 2 &  1.0 & 0.2852 & 0.1921 & 1.0579 & &\\
& 3 &  1.0 & 0.4372 & 0.5371 & 0.2468 & 0.9453 &\\
& 4 &  1.0 & 0.5310 & 0.7594 & 0.6461 & 0.2103 & 1.0579\\
\colrule
$G=30$~dB & 2 &  1.0 & 0.2035 & 0.1531 & 1.0321 & &\\
& 3 &  1.0 & 0.3352 & 0.4631 & 0.2071 & 0.9689 &\\
& 4 &  1.0 & 0.4206 & 0.6775 & 0.5630 & 0.1941 & 1.0321
\end{tabular}
\end{ruledtabular}
\end{table*}

\subsubsection{Chebyshev response}
For the Chebyshev response type, for a given order $N$, and specified signal gain $G_0$ in dB, and gain ripple $R$ in dB, we define:
\begin{equation}
    G^\mathrm{dB}_\mathrm{min}=\left\{
    \begin{aligned}
        G_0;&\;N\in\mathrm{even}\\
        G_0-R;&\;N\in\mathrm{odd}
    \end{aligned}
    ,\right.
\end{equation}
and:
\begin{equation}
    G^\mathrm{dB}_\mathrm{max}=\left\{
    \begin{aligned}
        G_0+R;&\;N\in\mathrm{even}\\
        G_0;&\;N\in\mathrm{odd}
    \end{aligned}
    .\right.
\end{equation}

If $G_\mathrm{max}$ and $G_\mathrm{min}$ are the corresponding signal power gains in \emph{linear} power units, then we should calculate the power loss function in terms of $G^{PL}_\mathrm{max}=4G_\mathrm{max}-2$ and $G^{PL}_\mathrm{min}=4G_\mathrm{min}-2$, similar to Eq.~(\ref{eq:g_pl}). The power loss function can then be written as:
\begin{equation}\label{eq:cheby_amp_pil}
    P_L(\omega) = \frac{G^{PL}_\mathrm{max}}{G^{PL}_\mathrm{max}-1}\left[1+\left(\frac{1}{G^{PL}_\mathrm{min}}-\frac{1}{G^{PL}_\mathrm{max}}\right)T^2_N(\omega)\right].
\end{equation}
Note that for Chebyshev response, the parameter $k^2$ in Eq.~(\ref{eq:cheby_pil}) does not correspond to the 3 dB (half gain) points, but rather, it controls the bandwidth over which the gain oscillates by the ripple amount~\cite{pozar2009microwave}. Therefore the 3 dB bandwidth of the amplifier will generally be greater than the specified bandwidth parameter $\Delta\omega$ that appears in Sections~\ref{sec:bandpass} and~\ref{sec:all_together}. Calculated prototype coefficients for a practical selection of gain and ripple specifications are listed in Table.~\ref{tab:cheby_amp_coefs}. Additional listings are given in Refs.~\onlinecite{MYJ, Getsinger1963}, however the gain specified therein is the total gain in both signal and idler bands, leading roughly to a 6 dB difference from our definition: what we denote as $G=20$~dB, would appear as $G\approx26$~dB in the convention of Refs.~\onlinecite{MYJ, Getsinger1963}.

\begin{table*}[h!]
\caption{Chebyshev coefficients calculated according to Appendix~\ref{apx:cauer_synthesis} for several values of amplifier gain and gain ripple, and for 2-, 3-, and 4-section networks, using the power loss function Eq.~(\ref{eq:cheby_amp_pil}). Coefficient $g_0$ refers to the active load.\label{tab:cheby_amp_coefs}}
\begin{ruledtabular}
\begin{tabular}{llcdddddd}
\textrm{Signal gain}&
\textrm{ripple}&
\textrm{order}&
\multicolumn{1}{c}{\textrm{$g_0$}}&
\multicolumn{1}{c}{\textrm{$g_1$}}&
\multicolumn{1}{c}{\textrm{$g_2$}}&
\multicolumn{1}{c}{\textrm{$g_3$}}&
\multicolumn{1}{c}{\textrm{$g_4$}}&
\multicolumn{1}{c}{\textrm{$g_5$}}\\
\colrule

$G=17$~dB& $R=0.1$~dB & 2 &  1.0  &  0.2769 & 0.1451 & 1.1528 & &\\
& & 3 &  1.0  &  0.5595 & 0.5410 & 0.3098 & 0.8674 &\\
& & 4 &  1.0  &  0.7489 & 0.8450 & 0.9181 & 0.2767 & 1.1528\\

& $R=0.5$~dB &2 &  1.0  & 0.3981 & 0.2206 & 1.1528 & &\\
& &3 &  1.0  & 0.7062 & 0.7029 & 0.4326 & 0.8674 &\\
& & 4 &  1.0  &  0.8533 & 0.9943 & 1.1289 & 0.3667 & 1.1528\\

& $R=1.0$~dB &2 &  1.0  &  0.4567 & 0.2642 & 1.1527 & &\\
& & 3 &  1.0  &  0.7822 & 0.7854 & 0.5095 & 0.8674 &\\
& & 4 &  1.0  &  0.8892 & 1.0592 & 1.2252 & 0.4182 & 1.1527\\
\colrule

$G=20$~dB & $R=0.1$~dB & 2 &  1.0  &  0.2204 & 0.1310 & 1.1055 & &\\
& & 3 &  1.0  &  0.4656 & 0.5126 & 0.2707 & 0.9045 &\\
& & 4 &  1.0  &  0.6370 & 0.8200 & 0.8243 & 0.2683 & 1.1055\\

& $R=0.5$~dB &2 &  1.0  &  0.3184 & 0.1982 & 1.1055 & &\\
&&3 &  1.0  &  0.5899 & 0.6681 & 0.3753 & 0.9045 &\\
&&4 &  1.0  &  0.7296 & 0.9671 & 1.0147 & 0.3525 & 1.1055\\

& $R=1.0$~dB & 2 &  1.0  &  0.3666 & 0.2366 & 1.1055 & &\\
&& 3 &  1.0  &  0.6545 & 0.7488 & 0.4397 & 0.9045 &\\
&& 4 &  1.0  &  0.7629 & 1.0310 & 1.1032 & 0.3999 & 1.1055\\
\colrule

$G=25$~dB & $R=0.1$~dB & 2 &  1.0  &  0.1546 & 0.1069 & 1.0579 & &\\
&& 3 &  1.0  &  0.3520 & 0.4541 & 0.2214 & 0.9453 &\\
&& 4 &  1.0  &  0.4997 & 0.7559 & 0.7044 & 0.2485 & 1.0579\\

& $R=0.5$~dB & 2 &  1.0  &  0.2246 & 0.1608 & 1.0579 & &\\
&& 3 &  1.0  &  0.4487 & 0.5939 & 0.3039 & 0.9453 &\\
&& 4 &  1.0  &  0.5768 & 0.8950 & 0.8679 & 0.3226 & 1.0579\\

& $R=1.0$~dB & 2 &  1.0  &  0.2599 & 0.1912 & 1.0579 & &\\
&& 3 &  1.0  &  0.4992 & 0.6682 & 0.3537 & 0.9453 &\\
&& 4 &  1.0  &  0.6060 & 0.9554 & 0.9450 & 0.3632 & 1.0579\\

\colrule
$G=30$~dB & $R=0.1$~dB & 2 &  1.0  &  0.1107 & 0.0849 & 1.0321 & &\\
&& 3 &  1.0  & 0.2722 & 0.3920 & 0.1839 & 0.9689 &\\
&& 4 &  1.0  &  0.4013 & 0.6794 & 0.6112 & 0.2256 & 1.0321\\

& $R=0.5$~dB & 2 &  1.0  &  0.1615 & 0.1272 & 1.0321 & &\\
&& 3 &  1.0  &  0.3488 & 0.5139 & 0.2506 & 0.9689 &\\
&& 4 &  1.0  &  0.4663 & 0.8071 & 0.7530 & 0.2899 & 1.0321\\

& $R=1.0$~dB & 2 &  1.0  &  0.1875 & 0.1507 & 1.0321 & &\\
&& 3 &  1.0  &  0.3892 & 0.5796 & 0.2901 & 0.9689 &\\
&& 4 &  1.0  &  0.4918 & 0.8630 & 0.8204 & 0.3245 & 1.0321\\
\end{tabular}
\end{ruledtabular}
\end{table*}

\section{Cauer form of a 1D array of nearest-neighbor coupled modes using graph reduction}\label{apx:kron_to_cauer}
We start by looking at the example of Fig.~\ref{fig:reduction_example} for a four-mode circuit and then generalize. Eq.~(\ref{eq:kron_cfrac}) writes the reduced-mode detuning term for the system as:
\begin{equation}\label{eq:apx_kron_cfrac}
    \Delta'_1=\Delta_1-\cfrac{\beta^2_{12}}{\Delta_2-\cfrac{\beta^2_{23}}{\Delta_3-\cfrac{\beta^2_{34}}{\Delta_4}}}.
\end{equation}
For a 1D simply-connected system, we can take $\beta_{jk}$ to be real, so we drop the absolute value in $|\beta_{jk}|^2$ compared to Eq.~(\ref{eq:kron_cfrac}). 

We will make the resonant pump assumption, and change variable to a normalized detuning, as common in microwave engineering:
\begin{equation}\label{s_def}
    s=-2i\frac{\left(\omega-\omega_0\right)}{\Delta\omega},
\end{equation}
where $\omega$ is the signal frequency, $\omega_0$ is the mode frequency, and $\Delta\omega$ is the bandwidth. With this substitution we can write the detuning terms appearing in Eq.~(\ref{eq:apx_kron_cfrac}) as follows:
\begin{align}
    \Delta_1&=i\frac{\Delta\omega}{2\gamma_0}s+i\frac{\gamma_{01}}{2\gamma_0}\\
    \Delta_2&=\Delta_3=i\frac{\Delta\omega}{2\gamma_0}s\\
    \Delta_4&=i\frac{\Delta\omega}{2\gamma_0}s+i\frac{\gamma_{45}}{2\gamma_0}
\end{align}

Plugging this into Eq.~(\ref{eq:apx_kron_cfrac}) and using Eq.~(\ref{eq:admittance_reduced}), we can write the admittance seen from the port (after some algebra) as:
\begin{equation}\label{eq:admittance_cfrac}
    \frac{Y}{Y_0}=2\left(\frac{\beta_{12}\beta_{34}}{\beta_{23}}\right)^2\frac{\gamma_0}{\gamma_{01}}\left[g_1s+\cfrac{1}{g_2s+\cfrac{1}{g_3s+\cfrac{1}{g_4s+\cfrac{1}{g_5}}}}\right]
\end{equation}
where we have defined:
\begin{align*}
    g_1&=\left(\frac{\beta_{23}}{\beta_{12}\beta_{34}}\right)^2\frac{\Delta\omega}{2\gamma_0},\\
    g_2&=\left(\frac{\beta_{34}}{\beta_{23}}\right)^2\frac{\Delta\omega}{2\gamma_0},\\
    g_3&=\left(\frac{1}{\beta_{34}}\right)^2\frac{\Delta\omega}{2\gamma_0},\\
    g_4&=\frac{\Delta\omega}{2\gamma_0}, \\
    g_5&=\frac{2\gamma_0}{\gamma_{45}}.
\end{align*}
We further define $g_0$ as the prefactor in Eq.~(\ref{eq:admittance_cfrac}). This brings the continued-fraction expression for the admittance, Eq.~(\ref{eq:admittance_cfrac}), to exact equivalence with the Cauer expansion in Appendix.~\ref{apx:cauer_synthesis}. The coefficients $\{g_i\}$ are the prototype coefficients of the normalized low-pass network (\textit{c.f.} Sec.~\ref{sec:bandpass}) that represents the circuit.

Looking at products of the form $g_jg_{j+1}$, and generalizing to a 1D coupled system of $N$ modes:
\begin{align*}
    g_0g_1&=\frac{\Delta\omega}{\gamma_{01}},\\
    g_jg_{j+1}&=\frac{1}{\beta^2_{j,j+1}}\left(\frac{\Delta\omega}{2\gamma_0}\right)^2,\\
    g_Ng_{N+1}&=\frac{\Delta\omega}{\gamma_{N,N+1}},
\end{align*}
which are equivalent to Eqs.~(\ref{eq:correspond_source})-(\ref{eq:correspond_beta}).

%

\clearpage
\end{document}